\documentclass[%
reprint,
superscriptaddress,
 amsmath,amssymb,
 aps,
floatfix,
]{revtex4-1}

\usepackage{graphicx}% Include figure files
\usepackage{epstopdf}
\usepackage{dcolumn}% Align table columns on decimal point
\usepackage{bm}% bold math
\usepackage{color}
\usepackage[caption=false]{subfig}
\usepackage{xspace}
\usepackage{epsfig}
\usepackage{multirow}
\usepackage[]{units}
\usepackage{cleveref}
\usepackage[mathlines]{lineno}% Enable numbering of text and display math
\usepackage{afterpage} 
\input symbols

\begin{document}

\preprint{XXX}

\title{Letter of Intent to Construct a nuPRISM Detector in the J-PARC Neutrino Beamline}% Force line breaks with \\

 %%%%%%%%%%%%%%%%%%%%%%%%%%%%%%%%%%%%%%%%%%%%%%%%%%%%%%%%%%%%%%
% T2K author list generated on Mon, 07 Apr 2014 23:39:50 +0900
% setting: extra = 0 revtex = 1 yearrule = 1 shiftrule = 1
%         shift rules based on: period1 = 1213 period2 = 1314
% Number of authors = 313
%%%%%%%%%%%%%%%%%%%%%%%%%%%%%%%%%%%%%%%%%%%%%%%%%%%%%%%%%%%%%%

\newcommand{\INSTC}{\affiliation{University of Alberta, Centre for Particle Physics, Department of Physics, Edmonton, Alberta, Canada}}
\newcommand{\INSTEE}{\affiliation{University of Bern, Albert Einstein Center for Fundamental Physics, Laboratory for High Energy Physics (LHEP), Bern, Switzerland}}
\newcommand{\INSTFE}{\affiliation{Boston University, Department of Physics, Boston, Massachusetts, U.S.A.}}
\newcommand{\INSTD}{\affiliation{University of British Columbia, Department of Physics and Astronomy, Vancouver, British Columbia, Canada}}
\newcommand{\INSTGA}{\affiliation{University of California, Irvine, Department of Physics and Astronomy, Irvine, California, U.S.A.}}
\newcommand{\INSTI}{\affiliation{IRFU, CEA Saclay, Gif-sur-Yvette, France}}
\newcommand{\INSTGB}{\affiliation{University of Colorado at Boulder, Department of Physics, Boulder, Colorado, U.S.A.}}
\newcommand{\INSTFG}{\affiliation{Colorado State University, Department of Physics, Fort Collins, Colorado, U.S.A.}}
\newcommand{\INSTFH}{\affiliation{Duke University, Department of Physics, Durham, North Carolina, U.S.A.}}
\newcommand{\INSTBA}{\affiliation{Ecole Polytechnique, IN2P3-CNRS, Laboratoire Leprince-Ringuet, Palaiseau, France }}
\newcommand{\INSTEF}{\affiliation{ETH Zurich, Institute for Particle Physics, Zurich, Switzerland}}
\newcommand{\INSTEG}{\affiliation{University of Geneva, Section de Physique, DPNC, Geneva, Switzerland}}
\newcommand{\INSTDG}{\affiliation{H. Niewodniczanski Institute of Nuclear Physics PAN, Cracow, Poland}}
\newcommand{\INSTCB}{\affiliation{High Energy Accelerator Research Organization (KEK), Tsukuba, Ibaraki, Japan}}
\newcommand{\INSTED}{\affiliation{Institut de Fisica d'Altes Energies (IFAE), Bellaterra (Barcelona), Spain}}
\newcommand{\INSTEC}{\affiliation{IFIC (CSIC \& University of Valencia), Valencia, Spain}}
\newcommand{\INSTEI}{\affiliation{Imperial College London, Department of Physics, London, United Kingdom}}
\newcommand{\INSTGF}{\affiliation{INFN Sezione di Bari and Universit\`a e Politecnico di Bari, Dipartimento Interuniversitario di Fisica, Bari, Italy}}
\newcommand{\INSTBE}{\affiliation{INFN Sezione di Napoli and Universit\`a di Napoli, Dipartimento di Fisica, Napoli, Italy}}
\newcommand{\INSTBF}{\affiliation{INFN Sezione di Padova and Universit\`a di Padova, Dipartimento di Fisica, Padova, Italy}}
\newcommand{\INSTBD}{\affiliation{INFN Sezione di Roma and Universit\`a di Roma ``La Sapienza'', Roma, Italy}}
\newcommand{\INSTEB}{\affiliation{Institute for Nuclear Research of the Russian Academy of Sciences, Moscow, Russia}}
\newcommand{\INSTHA}{\affiliation{Kavli Institute for the Physics and Mathematics of the Universe (WPI), Todai Institutes for Advanced Study, University of Tokyo, Kashiwa, Chiba, Japan}}
\newcommand{\INSTCC}{\affiliation{Kobe University, Kobe, Japan}}
\newcommand{\INSTCD}{\affiliation{Kyoto University, Department of Physics, Kyoto, Japan}}
\newcommand{\INSTEJ}{\affiliation{Lancaster University, Physics Department, Lancaster, United Kingdom}}
\newcommand{\INSTFC}{\affiliation{University of Liverpool, Department of Physics, Liverpool, United Kingdom}}
\newcommand{\INSTFI}{\affiliation{Louisiana State University, Department of Physics and Astronomy, Baton Rouge, Louisiana, U.S.A.}}
\newcommand{\INSTJ}{\affiliation{Universit\'e de Lyon, Universit\'e Claude Bernard Lyon 1, IPN Lyon (IN2P3), Villeurbanne, France}}
\newcommand{\INSTM}{\affiliation{Michigan State University, Department of Physics and Astronomy, East Lansing, Michigan, U.S.A.}}
\newcommand{\INSTCE}{\affiliation{Miyagi University of Education, Department of Physics, Sendai, Japan}}
\newcommand{\INSTDF}{\affiliation{National Centre for Nuclear Research, Warsaw, Poland}}
\newcommand{\INSTFJ}{\affiliation{State University of New York at Stony Brook, Department of Physics and Astronomy, Stony Brook, New York, U.S.A.}}
\newcommand{\INSTGJ}{\affiliation{Okayama University, Department of Physics, Okayama, Japan}}
\newcommand{\INSTCF}{\affiliation{Osaka City University, Department of Physics, Osaka, Japan}}
\newcommand{\INSTCI}{\affiliation{Osaka University, Department of Physics, Osaka, Toyonaka, Japan}}
\newcommand{\INSTCJ}{\affiliation{Osaka University, Research Center for Nuclear Physics(RCNP), Ibaraki, Osaka, Japan}}
\newcommand{\INSTGG}{\affiliation{Oxford University, Department of Physics, Oxford, United Kingdom}}
\newcommand{\INSTBB}{\affiliation{UPMC, Universit\'e Paris Diderot, CNRS/IN2P3, Laboratoire de Physique Nucl\'eaire et de Hautes Energies (LPNHE), Paris, France}}
\newcommand{\INSTGC}{\affiliation{University of Pittsburgh, Department of Physics and Astronomy, Pittsburgh, Pennsylvania, U.S.A.}}
\newcommand{\INSTFA}{\affiliation{Queen Mary University of London, School of Physics and Astronomy, London, United Kingdom}}
\newcommand{\INSTE}{\affiliation{University of Regina, Department of Physics, Regina, Saskatchewan, Canada}}
\newcommand{\INSTGD}{\affiliation{University of Rochester, Department of Physics and Astronomy, Rochester, New York, U.S.A.}}
\newcommand{\INSTBC}{\affiliation{RWTH Aachen University, III. Physikalisches Institut, Aachen, Germany}}
\newcommand{\INSTFB}{\affiliation{University of Sheffield, Department of Physics and Astronomy, Sheffield, United Kingdom}}
\newcommand{\INSTDI}{\affiliation{University of Silesia, Institute of Physics, Katowice, Poland}}
\newcommand{\INSTEH}{\affiliation{STFC, Rutherford Appleton Laboratory, Harwell Oxford,  and  Daresbury Laboratory, Warrington, United Kingdom}}
\newcommand{\INSTCH}{\affiliation{University of Tokyo, Department of Physics, Tokyo, Japan}}
\newcommand{\INSTBJ}{\affiliation{University of Tokyo, Institute for Cosmic Ray Research, Kamioka Observatory, Kamioka, Japan}}
\newcommand{\INSTCG}{\affiliation{University of Tokyo, Institute for Cosmic Ray Research, Research Center for Cosmic Neutrinos, Kashiwa, Japan}}
\newcommand{\INSTGK}{\affiliation{Tokyo Institute of Technology, Department of Physics, Tokyo, Japan}}
\newcommand{\INSTGI}{\affiliation{Tokyo Metropolitan University, Department of Physics, Tokyo, Japan}}
\newcommand{\INSTF}{\affiliation{University of Toronto, Department of Physics, Toronto, Ontario, Canada}}
\newcommand{\INSTB}{\affiliation{TRIUMF, Vancouver, British Columbia, Canada}}
\newcommand{\INSTG}{\affiliation{University of Victoria, Department of Physics and Astronomy, Victoria, British Columbia, Canada}}
\newcommand{\INSTDJ}{\affiliation{University of Warsaw, Faculty of Physics, Warsaw, Poland}}
\newcommand{\INSTDH}{\affiliation{Warsaw University of Technology, Institute of Radioelectronics, Warsaw, Poland}}
\newcommand{\INSTFD}{\affiliation{University of Warwick, Department of Physics, Coventry, United Kingdom}}
\newcommand{\INSTGE}{\affiliation{University of Washington, Department of Physics, Seattle, Washington, U.S.A.}}
\newcommand{\INSTGH}{\affiliation{University of Winnipeg, Department of Physics, Winnipeg, Manitoba, Canada}}
\newcommand{\INSTEA}{\affiliation{Wroclaw University, Faculty of Physics and Astronomy, Wroclaw, Poland}}
\newcommand{\INSTH}{\affiliation{York University, Department of Physics and Astronomy, Toronto, Ontario, Canada}}

%\INSTC
%\INSTEE
%\INSTFE
\INSTD
\INSTGA
%\INSTI
%\INSTGB
%\INSTFG
%\INSTFH
%\INSTBA
%\INSTEF
\INSTEG
%\INSTDG
\INSTCB
\INSTED
%\INSTEC
\INSTEI
%\INSTGF
%\INSTBE
%\INSTBF
%\INSTBD
\INSTEB
\INSTHA
%\INSTCC
\INSTCD
%\INSTEJ
%\INSTFC
%\INSTFI
%\INSTJ
\INSTM
%\INSTCE
%\INSTDF
\INSTFJ
%\INSTGJ
%\INSTCF
\INSTCI
\INSTCJ
%\INSTGG
%\INSTBB
%\INSTGC
%\INSTFA
\INSTE
\INSTGD
%\INSTBC
%\INSTFB
%\INSTDI
\INSTEH
\INSTCH
\INSTBJ
\INSTCG
\INSTGK
%\INSTGI
\INSTF
\INSTB
%\INSTG
%\INSTDJ
\INSTDH
%\INSTFD
%\INSTGE
%\INSTGH
%\INSTEA
\INSTH

\author{S.\,Bhadra}\INSTH
\author{A.\,Blondel}\INSTEG
\author{S.\,Bordoni }\INSTED
\author{A.\,Bravar}\INSTEG
\author{C.\,Bronner}\INSTCD
%\author{R.G.\,Calland}\INSTFC
\author{J.\,Caravaca Rodr\'iguez}\INSTED
\author{M.\,Dziewiecki}\INSTDH
\author{T.\,Feusels}\INSTD
\author{G.A.\,Fiorentini\,Aguirre}\INSTH
\author{M.\,Friend}\thanks{also at J-PARC, Tokai, Japan}\INSTCB
\author{L.\,Haegel}\INSTEG
\author{M.\,Hartz}\INSTHA\INSTB
\author{R.\,Henderson}\INSTB
\author{T.\,Ishida}\thanks{also at J-PARC, Tokai, Japan}\INSTCB
\author{M.\,Ishitsuka}\INSTGK
\author{C.K.\,Jung}\thanks{affiliated member at Kavli IPMU (WPI), the University of Tokyo, Japan}\INSTFJ
\author{A.C.\,Kaboth}\INSTEI
\author{H.\,Kakuno}\INSTGI
\author{H.\,Kamano}\INSTCJ
\author{A.\,Konaka}\INSTB
\author{Y.\,Kudenko}\thanks{also at Moscow Institute of Physics and Technology and National Research Nuclear University "MEPhI", Moscow, Russia}\INSTEB
\author{M.\,Kuze}\INSTGK
\author{T.\,Lindner}\INSTB
\author{K.\,Mahn}\INSTM
\author{J.F.\,Martin}\INSTF
\author{J.\,Marzec}\INSTDH
\author{K.S.\,McFarland}\INSTGD
\author{S.\,Nakayama}\thanks{affiliated member at Kavli IPMU (WPI), the University of Tokyo, Japan}\INSTBJ
\author{T.\,Nakaya}\INSTCD\INSTHA
\author{S.\,Nakamura}\INSTCI
\author{Y.\,Nishimura}\INSTCG
\author{A.\,Rychter}\INSTDH
\author{F.\,S\'anchez}\INSTED
\author{T.\,Sato}\INSTCI
\author{M.\,Scott}\INSTB
\author{T.\,Sekiguchi}\thanks{also at J-PARC, Tokai, Japan}\INSTCB
\author{M.\,Shiozawa}\INSTBJ\INSTHA
\author{T.\,Sumiyoshi}\INSTGI
\author{R.\,Tacik}\INSTE\INSTB
\author{H.K.\,Tanaka}\thanks{affiliated member at Kavli IPMU (WPI), the University of Tokyo, Japan}\INSTBJ
\author{H.A.\,Tanaka}\thanks{also at Institute of Particle Physics, Canada}\INSTD
\author{S.\,Tobayama}\INSTD
\author{M.\,Vagins}\INSTHA\INSTGA
\author{J.\,Vo}\INSTED
\author{D.\,Wark}\INSTEH
\author{M.O.\,Wascko}\INSTEI
\author{M.J.\,Wilking}\INSTFJ
\author{S.\,Yen}\INSTB
\author{M.\,Yokoyama}\thanks{affiliated member at Kavli IPMU (WPI), the University of Tokyo, Japan}\INSTCH
\author{M.\,Ziembicki}\INSTDH

\collaboration{The nuPRISM Collaboration}\noaffiliation

\date{ \today }  % It is always \today, today,

\maketitle

%\begin{abstract}

\onecolumngrid
\vspace{-8mm}
{\small
As long-baseline neutrino experiments enter the precision era, the difficulties associated with understanding neutrino interaction cross sections on atomic nuclei are expected to limit experimental sensitivities to neutrino oscillation parameters. In particular, the ability to relate experimental observables to the incident neutrino energy in all previous experiments has relied solely on theoretical models of neutrino-nucleus interactions, which currently suffer from very large theoretical uncertainties.

By observing charged current \numu interactions over a continuous range of off-axis angles from 1$^\circ$ to 4$^\circ$, the \nuprism water Cherenkov detector can provide a direct measurement of the far detector lepton kinematics for any given set of oscillation parameters, which largely removes neutrino interaction modeling uncertainties from T2K oscillation measurements. This naturally provides a direct constraint on the relationship between lepton kinematics and neutrino energy. In addition, \nuprism is a sensitive probe of sterile neutrino oscillations with multiple energy spectra, which provides unique constraints on possible background-related explanations of the MiniBooNE anomaly. Finally, high-precision measurements of neutrino cross sections on water are possible, including electron neutrino measurements and the first ever measurements of neutral current interactions as a function of neutrino energy.

The \nuprism detector also provides significant benefits to the proposed Hyper-Kamiokande project. A demonstration that neutrino interaction uncertainties can be controlled will be important to understanding the physics reach of Hyper-K. In addition, \nuprism will provide an easily accessible prototype detector for many of the new hardware components currently under consideration for Hyper-K. The following document presents the configuration, physics impact, and preliminary cost estimates for a \nuprism detector in the J-PARC neutrino beamline.\\ \\ \\
}
%\end{abstract}
\twocolumngrid


%\maketitle

\vspace{1cm}

{\large
\tableofcontents}

\clearpage

\section{Introduction}
\label{sec:intro}

With the publications of the first ever observation of $\nu_e$ appearance, and the world's most precise measurement of $\theta_{23}$, T2K has achieved its initial experimental goals with only 8.5\% of the approved protons on target (POT). The next phase of the experiment will make even more precise measurements of $\nu_e$ appearance and $\nu_\mu$ disappearance using both neutrinos and anti-neutrinos in order to probe the value of $\delta_{CP}$, the $\theta_{23}$ octant, and $\left|\Delta m^2_{32}\right|$. In conjunction with measurements from NO$\nu$A, these measurements may also provide a constraint on the neutrino mass hierarchy.

In order to achieve these goals, a more precise understanding of neutrino interaction cross sections is required. Currently, T2K is forced to rely on neutrino interaction generators to translate experimental observables into constraints on the neutrino energy spectrum, which depends on the value of the oscillation parameters. Measurements of very forward-going muons on the carbon target employed by the existing near detector, ND280, are translated into constraints on the 4$\pi$ muon angular distribution on a water target seen at the far detector. The interactions of final-state hadrons both within the nucleus and within each detector medium can have a significantly different impacts on the near and far detector analyses. Some of the backgrounds at far detector, Super-Kamiokande (Super-K), are poorly constrained at ND280. This is particularly true of NC$\pi^+$ events, which are problematic both because the cross section is not well measured, and because $\pi^+$ reconstruction at Super-K is not well understood. This results in a contamination of both the $\nu_\mu$ and $\nu_e$ samples that produces large systematic uncertainties.

It is also necessary to measure events with single, electron-like rings in order to constrain any differences in the \nue and \numu cross sections. These events can be caused by a variety of sources, such as beam $\nu_e$, single $\gamma$ production, $\pi^0$ production, external $\gamma$ background, sterile neutrino oscillations and radiative muon production. An excess of such events has been observed by MiniBooNE. It is important to confirm whether a similar excess exists on a water target, ideally with a water Cherenkov detector, and if found, the cause must be understood in order to perform precision \nue appearance measurements.

The least constrained component of these neutrino interaction models, however, is the relationship between the experimentally observable lepton kinematics and the energy of the incident neutrino. At present, there is an experimentally-unconstrained and potentially large bias in the ability to translate lepton kinematics to neutrino energy. Current estimates, based solely on new, recently developed models, suggest that this bias may be one of T2K's largest systematic uncertainties, and no existing dataset can provide a constraint on this uncertainty in a manner that does not rely on neutrino interaction models. Had neutrino interaction models been trusted to provide this relationship just 5 years ago, current models suggest that 20 to 30\% of events where only the final state lepton was observed would have been reconstructed with an incorrect neutrino energy in a way that would not have been constrained or even detectable. Even a high-performance near detector, capable of precisely measuring all charged particles in the final state, would be forced to rely on models that relate lepton kinematics to hadronic final states, and no modern theoretical models offer a prediction for such a relationship within a nuclear environment.

%The obvious solution to constrain the final state particle kinematics for interactions at a fixed neutrino energy is the use of a monochromatic neutrino beam.  In principle, such beams could be made using the two-body decays from a monochromatic charged pion beam, but such a beam is inefficient and technically challenging to build.  In a conventional neutrino beam, the peak neutrino energy varies with the angle between the neutrino direction and average pion direction, off-axis angle, and the neutrino spectrum becomes narrower at larger off-axis angles.  This energy dependence with off-axis angle can be used to over-constrain the relationship between observed final state particle kinematics and predicted neutrino spectra in a manner not possible with measurements at a single off-axis angle.

The nuPRISM water-Cherenkov detector takes advantage of the energy dependence of the neutrino flux with off-axis angle by spanning a continuous range of 1 to 4 degrees in off-axis angle.
%The \nuprism neutrino detection principle provides a mechanism whereby the far detector response for any given oscillation hypothesis can be directly measured at the near detector. 
%The \nuprism concept involves a 50-60 m tall water-Cherenkov detector located about 1 km from the T2K neutrino production point, covering an off-axis angle range of 1-4$^{\circ}$.
This technique has the potential to significantly reduce uncertainties from neutrino interaction modeling in T2K oscillation analyses, as is demonstrated for the muon neutrino disappearance measurement described in Section~\ref{sec:physics}. In particular, these measurements will provide the first direct experimental constraint on the relationship between lepton kinematics and neutrino energy using measurements of final state muons at many different off-axis angles.
 In order to construct a more cost-effective detector that can reasonably be built on a timescale that is applicable to T2K, this document proposes to instrument a subset of the full water volume on a frame that moves vertically within the water tank, which sequentially samples the full off-axis range of the shaft in 5-6 separate running periods.

The construction of a \nuprism detector in the next 3-5 years can also provide significant benefits to Hyper-Kamiokande (Hyper-K). The problems with understanding neutrino interactions can have a larger impact on Hyper-K, since Hyper-K will have much smaller statistical errors, and a demonstration that these uncertainties can be managed with a \nuprism near detector will significantly enhance the physics case for Hyper-K.
In addition, \nuprism is an easily accessible water Cherenkov detector that provides an ideal environment to test Hyper-K technology. Hyper-K proposes to use new, in-water electronics, new solid state hybrid-PMTs (HPDs), and a new tank and liner construction to prevent leaks, all of which require extensive testing in a prototype detector.
Finally, \nuprism will provide an intermediate physics program that bridges the gap from T2K phase I to Hyper-K, which can provide continuity within the Japanese physics community while Hyper-K is being designed and constructed.

The remainder of this document provides an overview of the detector components and physics potential of \nuprism. The results for a full T2K $\nu_\mu$ disappearance analysis are provided, and a variety of additional \nuprism neutrino energy spectrum fits are presented to demonstrate how the \nuprism technique can constrain \nue cross sections, which will be important for measurements of $\nu_e$ appearance and $\delta_{CP}$, as well as several different oscillation backgrounds. Cost estimates have been obtained for the items that are expected to dominate the cost of the project, in particular photomultiplier tubes (PMTs) and civil construction. For the additional less expensive items, cost estimates from a very similar project proposed in 2005, the T2K 2~km water Cherenkov detector, are used to guide expectations for the full \nuprism project cost.

\subsection{\label{sec:enu_determine}Uncertainties in Neutrino Energy Determination}

Prior to 2009, neutrino interaction models assumed that neutrinos, when encountering a nuclear target, interact a single nucleon. The initial state of the nucleon was characterized by a binding energy and Fermi momentum, which were drawn from either a Fermi gas~\cite{LlewellynSmith:1971zm,Smith:1972xh} or a more specialized spectral function treatment~\cite{Benhar:1989aw}. In this paradigm, all the remaining dynamics of charge-current quasi-elastic (CCQE) interactions, in which the target neutron is converted into an outgoing proton, are encapsulated in a set of three vector and three axial-vector form factors. Most of these form factors are tightly constrained from external electron and pion scattering experiments (for a detailed discussion, see Ref.~\cite{nue-numu-cross-section}). The largest remaining uncertainty is on the axial vector form factor, which is assumed to take a dipole form,
\begin{equation}
F_A(Q^2)=\frac{F_A(0)}{(1+\frac{Q^2}{M_A^2})^2}.
\end{equation}
The parameter $F_A(0)$ is precisely known from nuclear beta decay, which leaves $M_A$ as the remaining uncertain parameter. Modifying $M_A$ simultaneously alters both the overall CCQE cross section and the shape of the $Q^2$ distribution.

\begin{figure*}[htpb]
     \begin{center}
       \includegraphics[width=15cm]{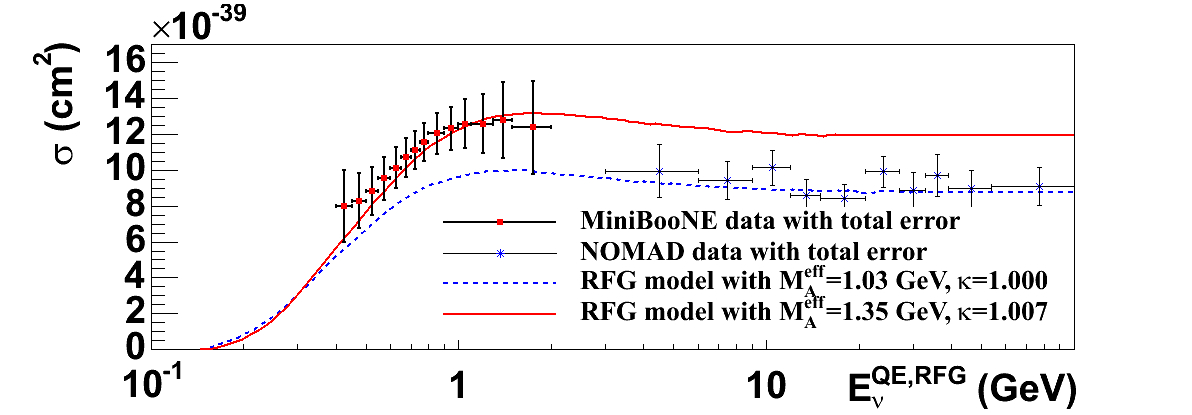}
       \caption{The CCQE cross section measurements are shown for MiniBooNE and NOMAD. The data show significant differences between measurements made at low and high energies.}
     \end{center}
\label{fig:miniboonenomad}
\end{figure*}

In 2009, the first comparison of MiniBooNE CCQE-like data at neutrino energies around 1~GeV and NOMAD data at higher energies was released. A reproduction of that comparison 
%with LSND and SciBooNE data included 
is shown in Figure~\ref{fig:miniboonenomad}. The MiniBooNE data are consistent with an $M_A$ value of 1.35~GeV (an additional empirical parameter, $\kappa$ is consistent with no modification at $1~\sigma$), while the NOMAD data prefer an $M_A$ of 1.03~GeV. 
This discrepancy is currently unexplained by neutrino-nucleus interaction models and is an outstanding experimental question that can be addressed by nuPRISM (see Section~\ref{sec:cc0pi}).
%This discrepancy could not be explained using the neutrino-nucleus interaction models available at that time.

Later in 2009, the Marteau~\cite{Marteau:1999kt} formalism for the treatment of neutrino scattering on nucleon 
pairs in nuclei was resurrected by Martini {\it et al.}~\cite{Martini:2009uj,Martini:2010ex,Martini:2011wp} to explain the
 higher event rate and muon kinematic distributions observed by MiniBooNE.  If this explanation of the MiniBooNE event 
excess were correct, it would imply that neutrino energy reconstruction for all previous neutrino experiments on nuclei
 at the GeV scale could have significant biases for 20-30\% of CCQE-like events. In the past few years, the models of Martini {\it et al.} and Nieves {\it et al.}~\cite{Nieves:2011pp} have begun to incorporate these effects, but such calculations are very difficult and the predictions of just these two models produce significantly different results when applied to T2K oscillation analyses~\cite{TN172}.
 
There exists circumstantial experimental evidence for multinucleon interaction mechanisms in both neutrino and electron scattering, but nothing that allows us to conclusively solve the problem or even to down-select among the various calculations.  In electron scattering, the reaction mechanism is different due to the absence of an axial-vector current component.  In neutrino scattering experiments with broadband beams, the evidence is only circumstantial, since we must rely on the predictions of the models themselves to extract the neutrino energy for any given event. Other approaches, such as making precise measurements of the hadronic final state, are limited by a lack of theoretical understanding of the expected hadron kinematics for multinucleon events. Even the final state hadron spectra for CCQE events are modified by nuclear effects which are also not well understood.

\begin{figure}[htpb]
     \begin{center}
       \includegraphics[width=7cm]{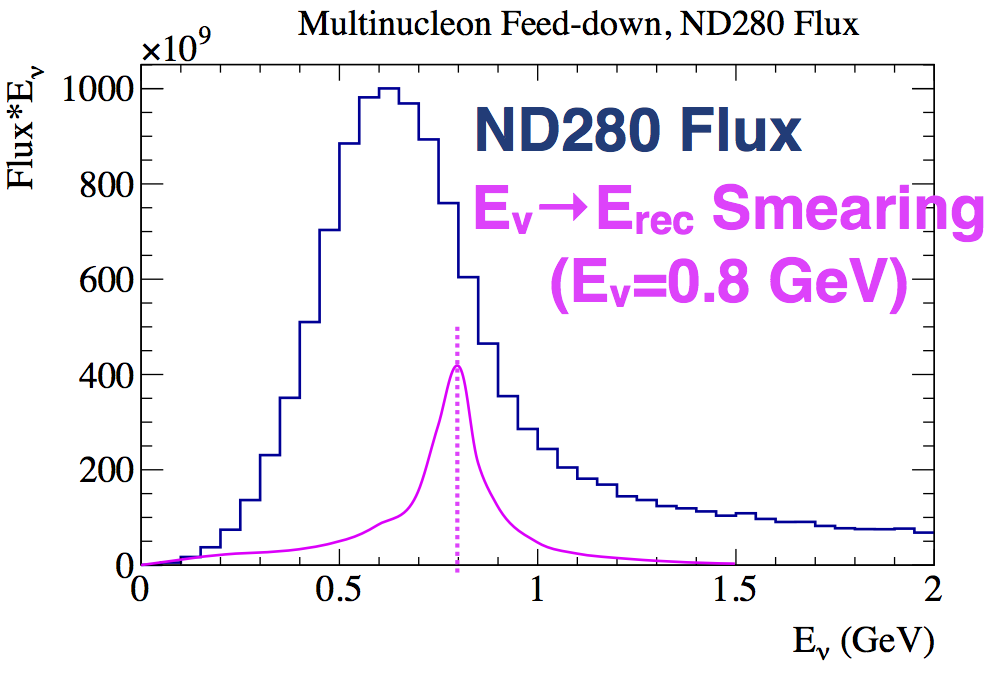}
       \includegraphics[width=7cm]{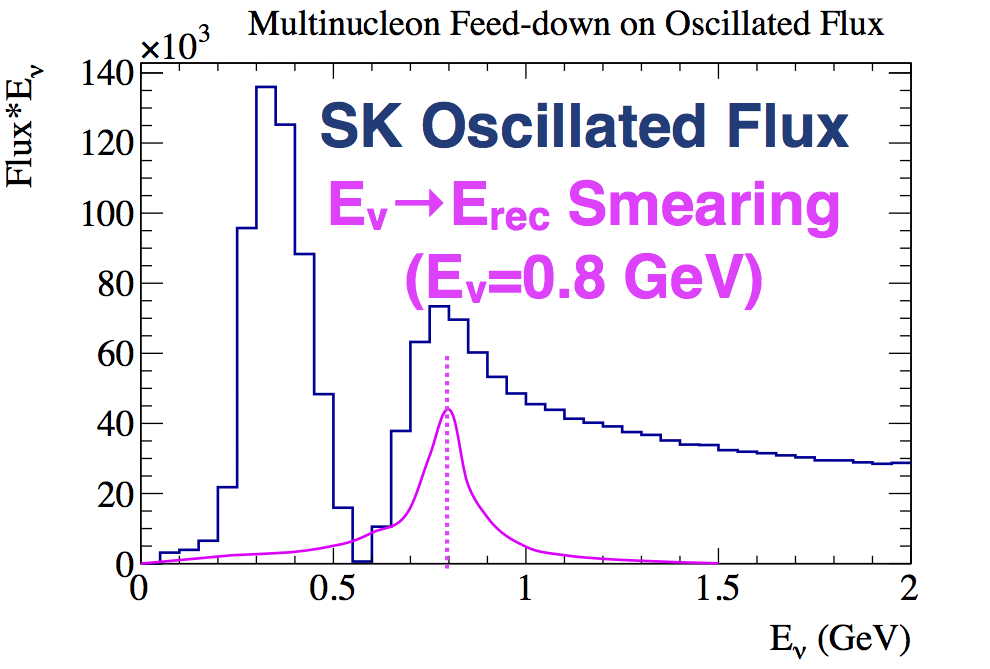}
       \caption{A cartoon of the effect of energy reconstruction biases is shown for both the T2K near detector (top) and the far detector (bottom). At the far detector, these biases directly impact the measurement of the oscillation dip, but the biases are largely unconstrained at the near detector due to the large unoscillated sample of unbiased CCQE events.}
       \label{fig:badnd}
     \end{center}
\end{figure}

Figure~\ref{fig:badnd} illustrates the challenge associated with using near detector data to constrain the interaction model that predicts far detector event rates. The detectors measure the convolution of the neutrino spectrum with the interaction model.  Since the near and far detector spectra are different due to neutrino oscillations, the measurement of this convolution in the near detector does not directly constrain the event rate in the far detector, and neutrino energies that represent a small fraction of the event rate in the near detector can be a significantly impact the measurement of oscillation parameters in the far detector.
%Figure~\ref{fig:badnd} illustrates the difficulties associated with using near detectors to constrain the effects of neutrino interaction modeling on the predictions for a far detector.

In addition to multinucleon effects, other effects such as long range correlations and final state interactions within the target nucleus can also produce distortions to the neutrino energy spectrum that can be difficult to model. In order to perform precision oscillation measurements with uncertainties at the level of the few percent statistical errors expected for $7.8\times10^{21}$ POT, it will be necessary to provide a data-driven constraint on these neutrino interaction model uncertainties.

\subsection{ND280 Capabilities and Limitations}

T2K oscillation analyses rely on precise constraints of flux and cross section model parameters from ND280. While a 3.2\% uncertainty on the predicted number of electron neutrinos at the far detector has been achieved for the combination of flux and cross section parameters that are currently constrained by the near detector, there remains a 4.7\% uncertainty on the far detector event sample due to additional cross section parameters that remain unconstrained. This unconstrained uncertainty is dominated by uncertainties in the modeling of the target oxygen nucleus, and largely depends on the theoretical model used to extrapolate measurements on carbon to oxygen.

The full capabilities of the T2K near detector have not yet been exploited. For example, the near detector analyses have thus far used interactions in the most-upstream Fine-Grained Detector (FGD1), which is composed entirely of alternating layers of horizontally- and vertically-oriented scintillator bars. Since the FGD scintillator layers are predominantly composed of carbon and hydrogen, FGD1 measurements cannot directly probe interactions on oxygen. An additional FGD (FGD2) contains layers of water interspersed within its scintillator layers. A simultaneous fit of the interactions in both FGDs can provide a constraint on nuclear uncertainties in oxygen, and may potentially reduce the corresponding nuclear model uncertainties.
%The ultimate event samples in both FGDs are shown in Table~\ref{tab:futurerates}.
%The statistical precision of a subtraction of interactions on scintillator from interactions on water is better than 1\%, which is more precise current detector systematic uncertainties ($\sim$3\%).

Another expected improvement to ND280 is the extension of the measured phase space of the outgoing lepton kinematics from a charged-current neutrino interaction. In the currently available analyses, muons are required to be produced in an FGD and traverse a minimum distance through the downstream TPC in order to make a measurement of both muon momentum and particle identification. This limits the muon acceptance to forward angles. Improvements to detector timing calibration and to track matching to the Electromagnetic Calorimeters (ECALs) and Side Muon Range Detectors (SMRDs) surrounding the FGDs and TPCs will allow for the reconstruction of charged-current events with backward-going and sideways-going muons. These additional samples will add less than 20\% to the total even sample with a degraded energy resolution relative to events that enter a TPC, however they may be able to improve constraints on the cross section modeling in previously inaccessible and potentially important new regions of phase space.

An additional sample of events that has not yet been incorporated into the oscillation analysis are charged-current interactions in the pi-zero detector (P0D). The P0D is capable of operating with and without water targets dispersed throughout its active volume, and by measuring the event rates separately in these two configurations, it is possible to extract constraints on interactions in water. The requirement to match a track in the TPC limits the angular acceptance for muons produced in the P0D, however the larger fiducial volume of the P0D produces a higher event sample.

In order for any of these new samples to reduce systematic uncertainties, it is necessary to choose a neutrino interaction model that can characterize all possible variations of the neutrino cross sections as a function of both neutrino energy and the final state particle kinematics. In other words, model dependent choices will have to be made that will directly impact the strength of the constraint that can be extracted. Given the difficulties in understanding neutrino-nucleus interactions, it may not be possible to justify reductions in cross section systematic uncertainties beyond their current level without a direct experimental constraint. In addition, the aforementioned uncertainties due to multinucleon effects have not yet been incorporated into T2K oscillation analyses. Preliminary studies within T2K indicate that these effects would be difficult to constrain using only lepton kinematics from ND280 at the level required for the full-statistics T2K sensitivity, and may be as large as the current dominant systematic uncertainties. The use of additional hadronic information is being explored, but any such constraint would be subject to even further model dependence.

\subsection{Detector Overview}

The \nuprismlite detector uses the same water Cherenkov detection technology as Super-K with a cylindrical water volume that is taller than Super-K (50-100~m vs 41~m) but with a much smaller diameter (10-12~m vs 39~m). The key requirements are that the detector span the necessary off-axis range (1$^\circ$-4$^\circ$) and that the diameter is large enough to contain the maximum required muon momentum. The baseline design considers a detector location that is 1~km downstream of the neutrino interaction target with a maximum contained muon momentum of 1~GeV/c. This corresponds to a 50~m tall tank with a 6~m diameter inner detector (ID) and a 10~m diameter outer detector (OD). A larger, 8~m ID is also being considered at the expense of some OD volume at the downstream end of the tank. As the \nuprismlite analysis studies mature, the exact detector dimensions will be refined to ensure sufficient muon momentum, \nue statistics and purity, etc.

The instrumented portion of the tank is a subset of the full height of the water volume, currently assumed to be 10~m for the ID and 14~m for the OD. The novel feature of this detector is the ability to raise and lower the instrumented section of the tank in order to span the full off-axis range in 6 steps. The inner detector will be instrumented with either 5-inch or 8-inch PMTs to ensure sufficient measurement granularity for the shorter light propagation distances relative to Super-K. Also under consideration is to replace the OD reflectors with large scintillator panels, such as those used in the T2K Side Muon Range Detector (SMRD), although this has not yet been integrated into the overall detector design. More details regarding the detector hardware can be found in Section~\ref{sec:detector}

\section{Physics Capabilities}\label{sec:physics}

The physics goals of \nuprismlite include reducing systematic uncertainties on the T2K oscillation analyses, using electron-like events to search for sterile neutrino oscillations and constrain electron neutrino cross sections, and making the first ever energy dependent neutral current (NC) and charged current (CC) cross section measurements that do not rely on neutrino generators to provide the incident neutrino energy. 

\subsection{Off-Axis Fluxes}

\begin{figure}[htpb]
    \begin{center}
      \includegraphics[width=6cm] {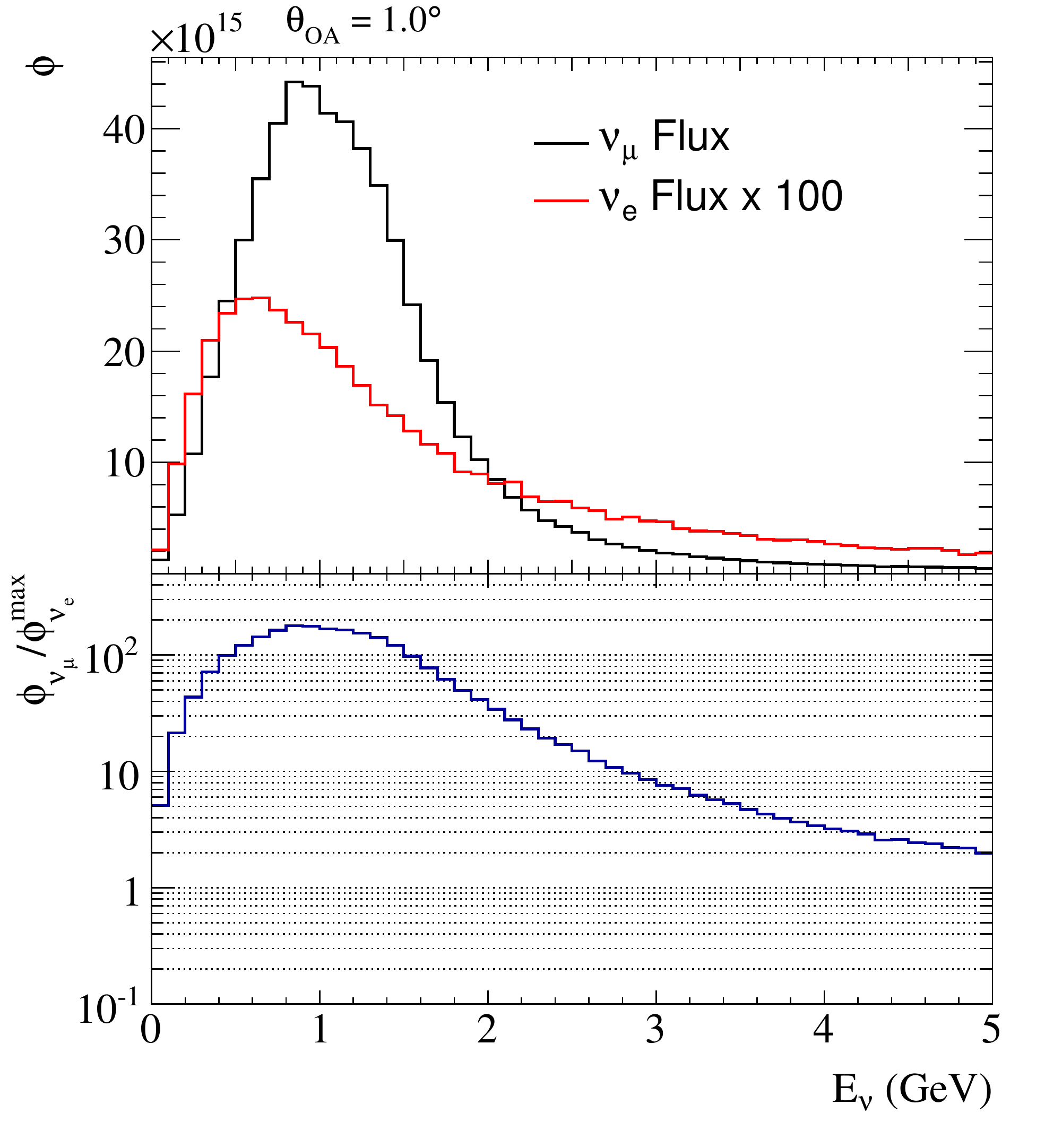}
      \includegraphics[width=6cm] {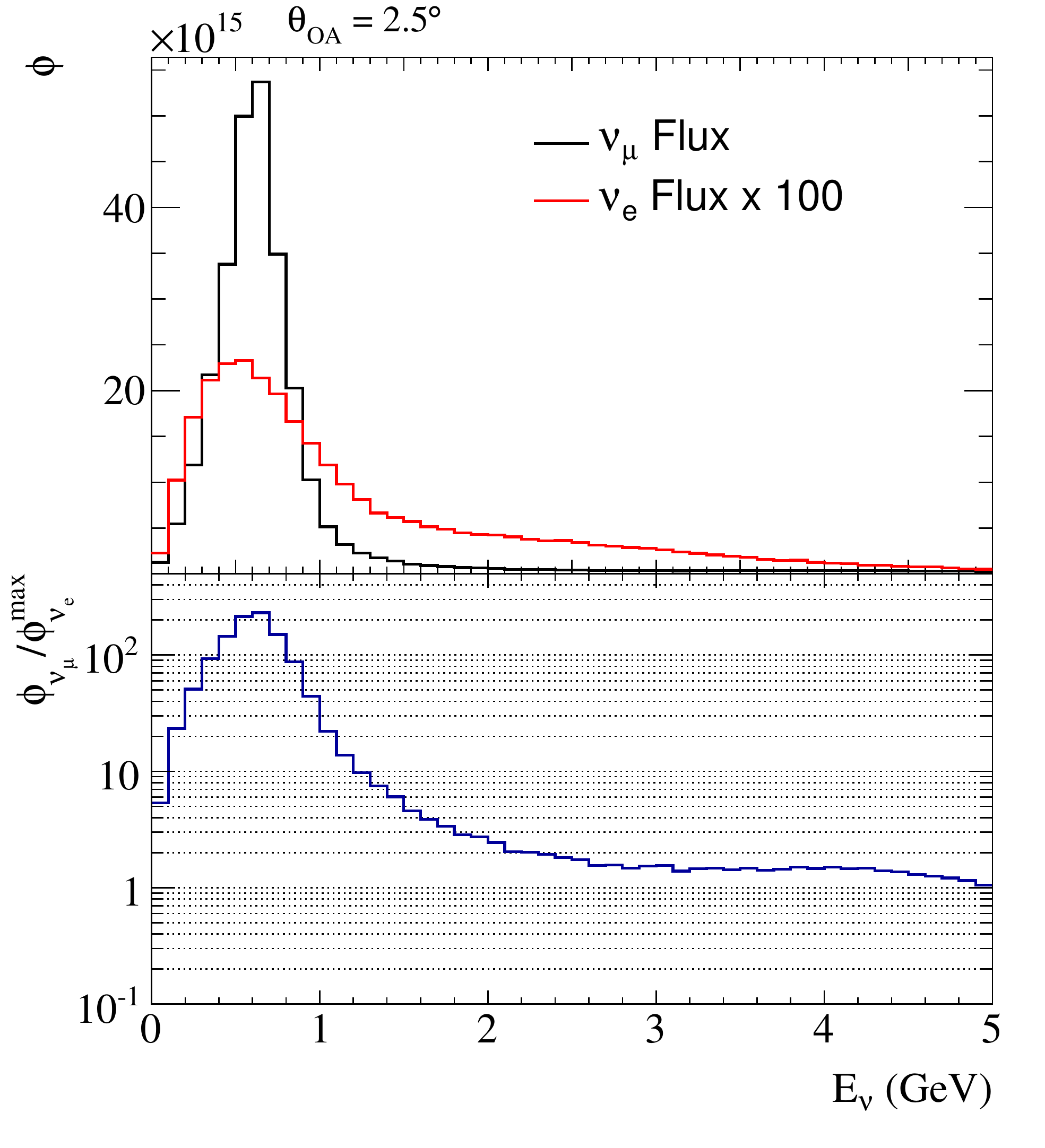}
      \includegraphics[width=6cm] {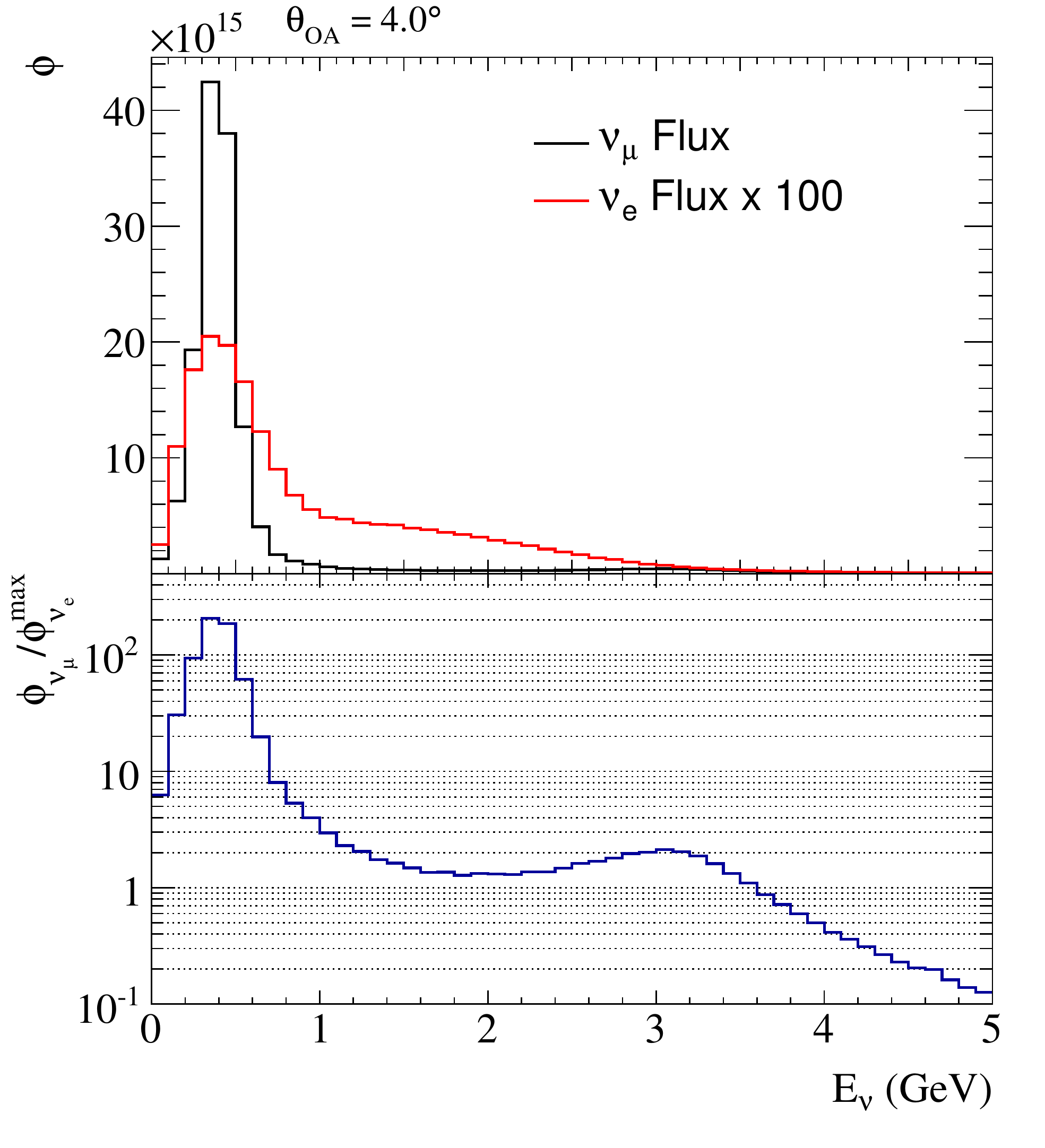}
    \end{center}
\caption{The neutrino energy spectra for $\nu_{\mu}$ and $\nu_{e}$ fluxes in the T2K beam operating in neutrino mode are shown for off-axis angles of 1$^{\circ}$, 2.5$^{\circ}$, and 4$^{\circ}$. The $\nu_{\mu}$ flux normalized by the maximum $\nu_e$ flux is shown at the bottom of each plot, demonstrating that feed-down from high energy NC backgrounds to $\nu_{e}$ candidates can be reduced by going further off-axis. }
\label{fig:offaxisfluxes}
\end{figure}

The \nuprism detector concept exploits the fact that as a neutrino detector is moved to larger off-axis angles relative to the beam direction, the peak energy of the neutrino energy spectrum is lowered and the size of the high-energy tail is reduced. This effect can be seen in Figure~\ref{fig:offaxisfluxes}, which shows the neutrino energy spectra at several different off-axis angles in the T2K beam line.  Since the off-axis angle for a single neutrino interaction can be determined from the reconstructed vertex position, this extra dimension of incident neutrino energy dependence can be used to constrain the interaction rates and final state particles in a largely model independent way.

A typical \nuprism detector for the T2K beam line would span a continuous range of off-axis angles from 1$^\circ$ to 4$^\circ$. For T2K, the best choice of technology is a water Cherenkov detector in order to use the same nuclear target as Super-K, and to best reproduce the Super-K detector efficiencies.

\subsection{Monochromatic Beams}

\begin{figure}[htpb]
\includegraphics[width=0.45\textwidth]{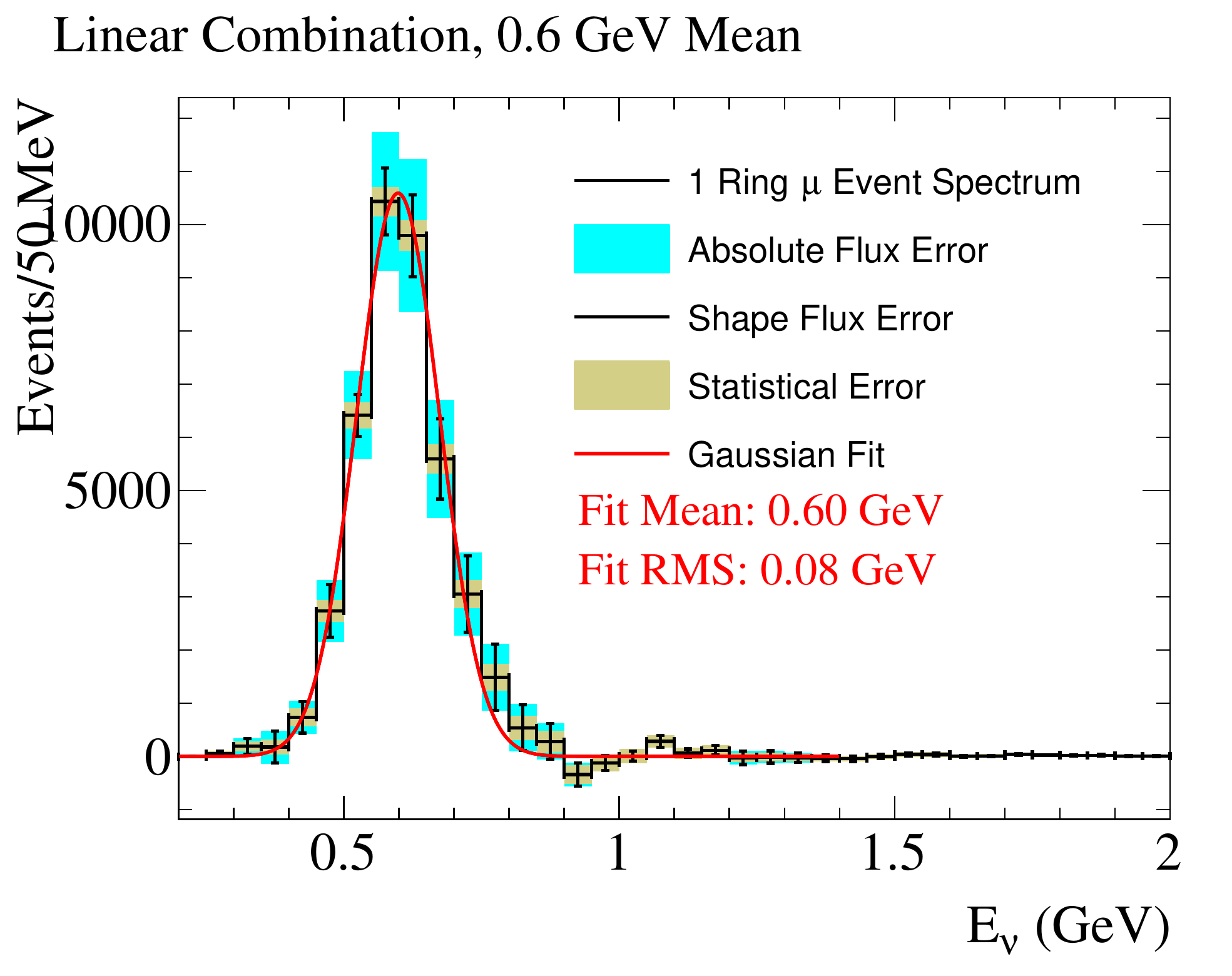} \\
\includegraphics[width=0.45\textwidth]{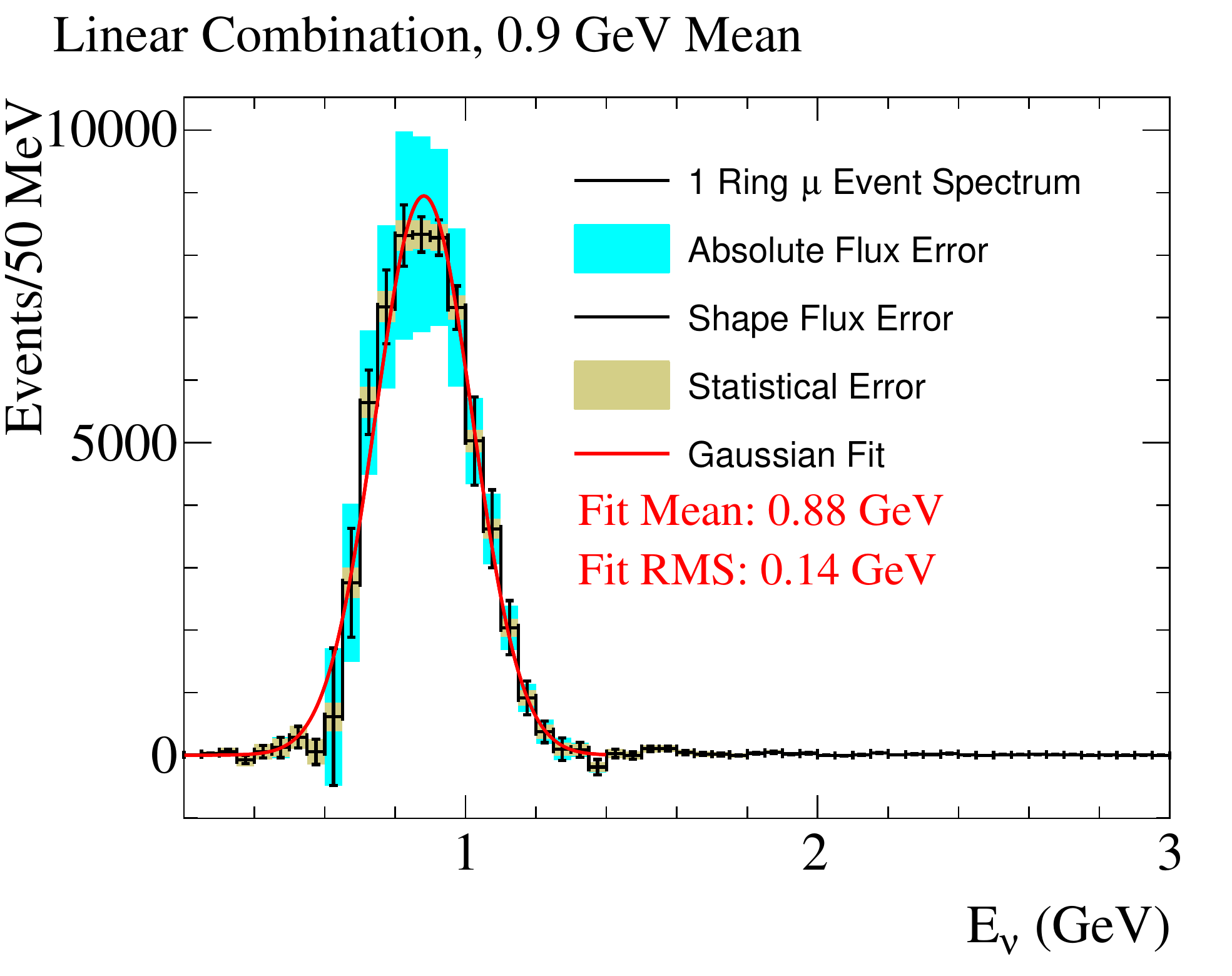} \\
\includegraphics[width=0.45\textwidth]{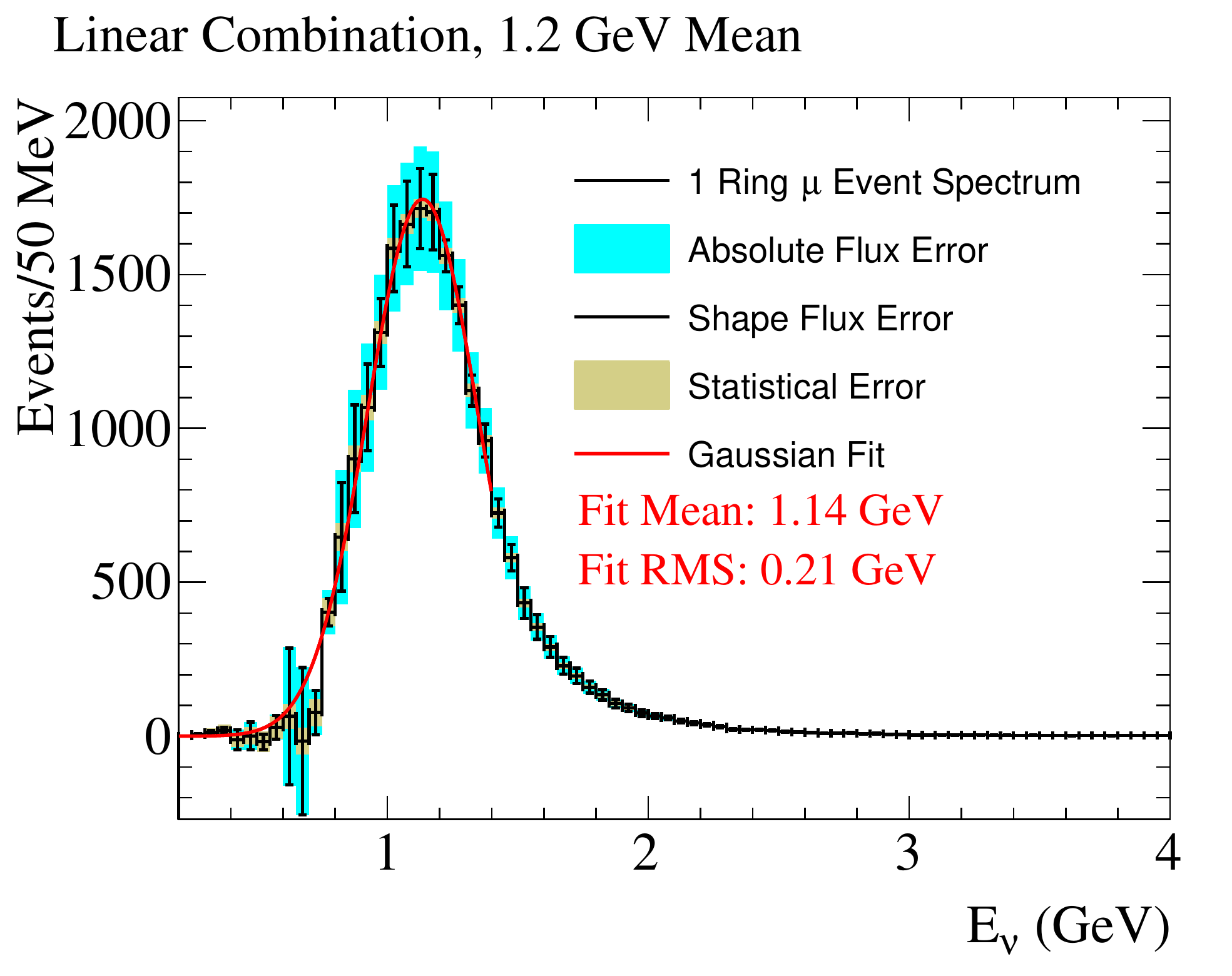} 
\caption{Three ``pseudo-monochromatic" spectra centered at 0.6 (top), 0.9 (middle) and 1.2 (bottom) GeV.  The aqua error bars show
the 1 $\sigma$ uncertainty for flux systematic variations, while the black error bars show the flux systematic variation after the
overall normalization uncertainty is removed.  The tan error bars show the statistical uncertainty for samples corresponding to 
 $4.5\times10^{20}$ protons on target.  }
\label{fig:mono_beam}
\end{figure}

The detector can be logically divided into slices of off-axis angle based on the reconstructed vertex of each
event. In each slice, the muon momentum and angle relative to the mean neutrino direction can be measured. By
taking linear combinations of the measurements in each slice, it is possible to produce an effective muon 
momentum and angle distribution for a Gaussian-like beam at energies between 0.4 and 1.2 GeV. Qualitatively, any
desired peak energy can be chosen by selecting the appropriate off-axis angle, as shown in Figure~\ref{fig:mono_beam}, and then
the further on-axis measurements are used to subtract the high energy tail, while the further off-axis 
measurements are used to subtract the low energy tail. Figure~\ref{fig:mono_beam} shows three such 
``pseudo-monochromatic" neutrino energy spectra constructed in this manner. These spectra are for selected 
1-ring muon candidates and systematic errors from the flux model are applied using the T2K flux systematic 
error model.  The statistical errors for an exposure of $4.5\times10^{20}$ protons on target are also shown. 
In all cases the high energy and low energy tails are mostly canceled over the full energy range and the 
monochromatic nature of the spectrum is stable under the flux systematic and statistical variations.

\begin{figure}[htpb]
\includegraphics[width=0.45\textwidth]{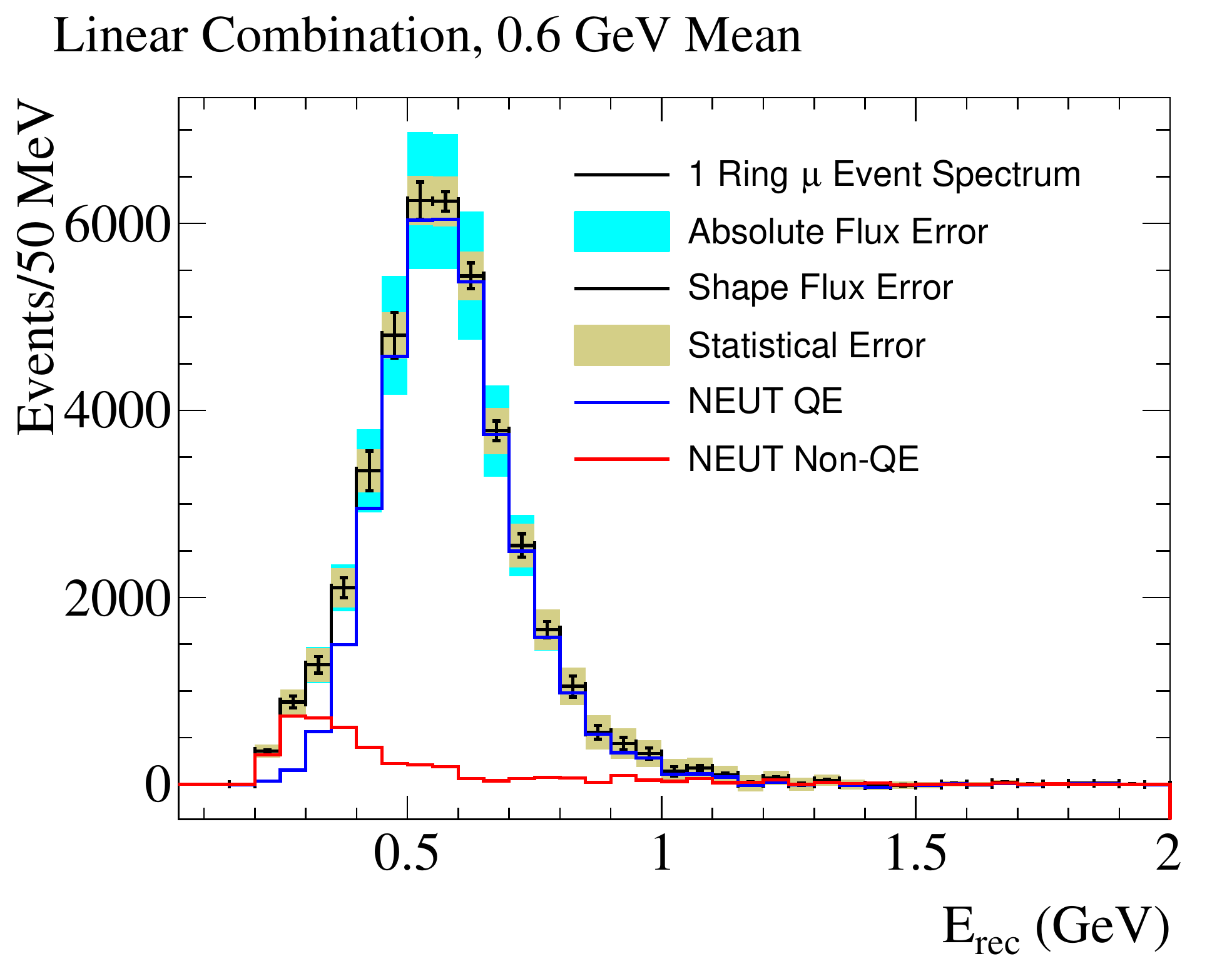} \\
\includegraphics[width=0.45\textwidth]{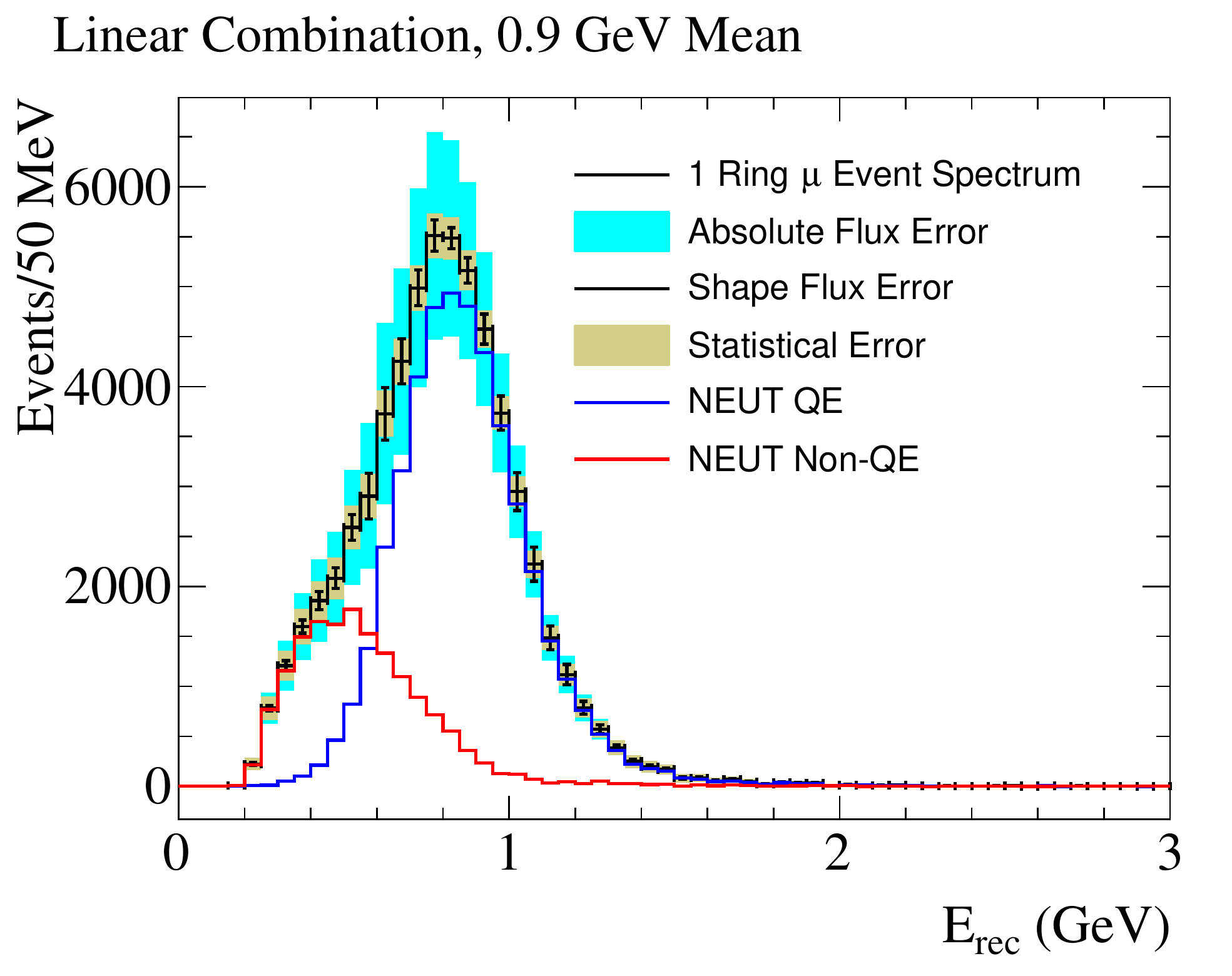} \\
\includegraphics[width=0.45\textwidth]{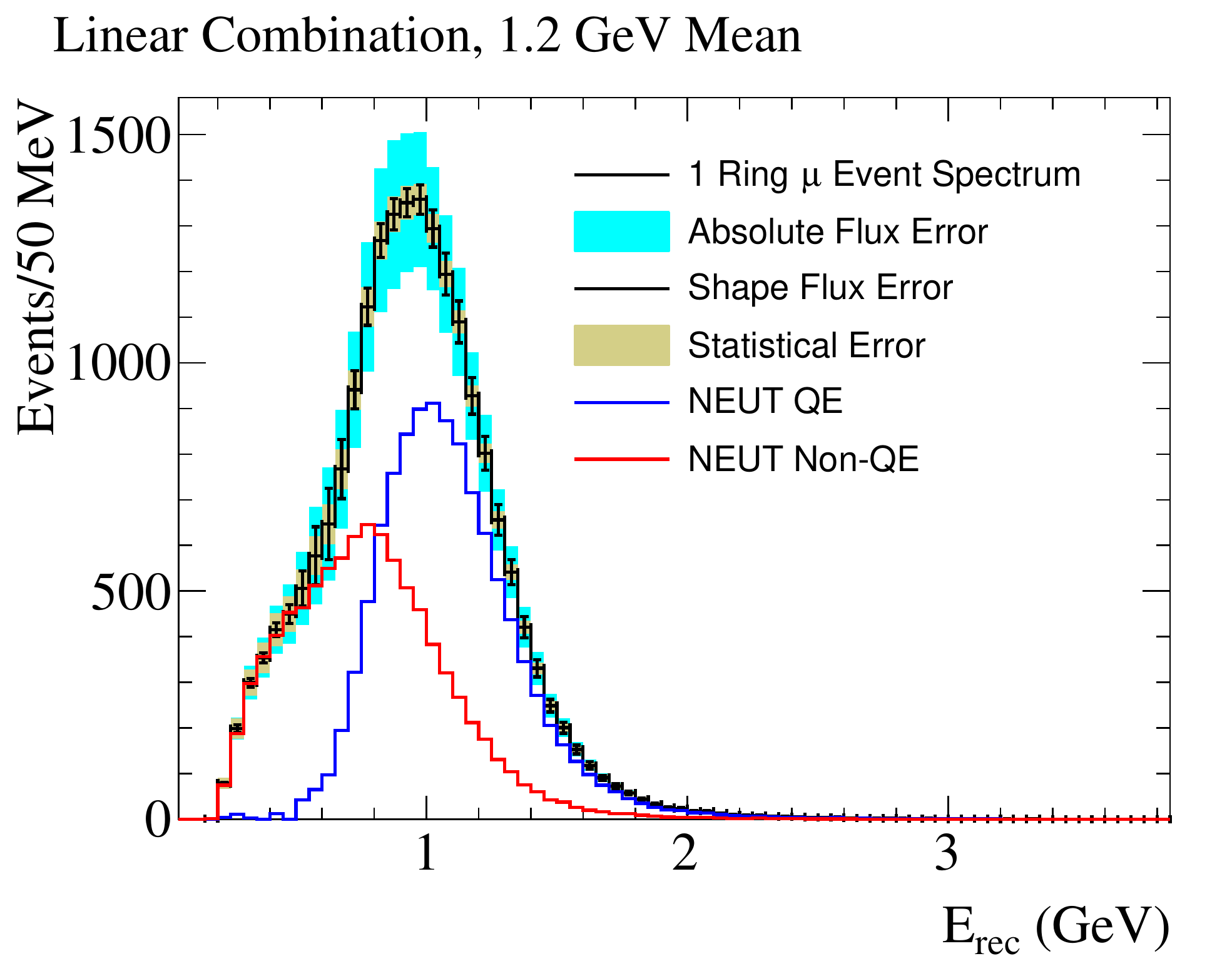} 
\caption{The reconstructed energy distributions for 1-ring muon candidate events produced using 
``pseudo-monochromatic" spectra centered at 0.6 (top), 0.9 (middle) and 1.2 (bottom) GeV.  The aqua error bars show
the 1 $\sigma$ uncertainty for flux systematic variations, while the black error bars show the flux systematic variation after the
overall normalization uncertainty is removed.  The tan error bars show the statistical uncertainty for samples corresponding to 
 $4.5\times10^{20}$ protons on target.  The red and blue histograms show the contributions from non-quasi-elastic and quasi-elastic
scatters respectively. }
\label{fig:mono_beam_erec}
\end{figure}

Figure~\ref{fig:mono_beam_erec} shows the reconstructed energy distributions for 1-ring muon candidates 
observed with the pseudo-monochromatic beams shown in Figure~\ref{fig:mono_beam}. The candidate events
are divided into quasi-elastic scatters and non-quasi-elastic scatters, which include contributions from
processes related to nuclear effects such as multinucleon interactions or pion absorption in final state
interactions.  With these pseudo-monochromatic beams, one sees a strong separation between the quasi-elastic 
scatters and the non-quasi-elastic scatters with significant energy reconstruction bias, especially in the
0.8 to 1.2 GeV neutrino energy range.  These measurements can be used to directly predict the effect of
non-quasi-elastic scatters in oscillation measurements and can also provide a unique constraint on nuclear 
models of these processes.

The \nuprism technique can be expanded beyond these pseudo-monochromatic beams. This linear combination method can be used to reproduce a wide variety of flux shapes between 0.4 and 1.0 GeV. In particular, as described later in this section, it is possible to reproduce all possible oscillated Super-K spectra with a linear combination of \nuprism measurements, which significantly reduces many of the uncertainties associated with neutrino/nucleus interaction modeling.

\subsection{Simulation Inputs \label{sec:nuprism_sim}}

To perform \nuprismlite sensitivity analyses, the official T2K flux production and associated flux uncertainties have been extended to cover a continuous range of off-axis angles, and the standard T2K package used to generate vertices in ND280 has also been modified to handle flux vectors with varying energy spectra across the detector. However, for the analysis presented in this note, full detector simulation and reconstruction of events were not available. Instead, selection efficiencies and reconstruction resolutions for vertex, direction, and visible energy were tabulated using the results of fiTQun run on Super-K events. The efficiency for electrons (muons) was defined as events passing the following cuts: OD veto, 1-ring, e-like ($\mu$-like), 0 (1) decay electrons, and the T2K fiTQun $\pi^0$ rejection (no $\pi^0$ cut). The efficiency tabulation was performed in bins of the true neutrino energy, the visible energy and distance along the track direction to the wall of the most energetic ring, and separate tables were produced for charged current events with various pion final states (CC0$\pi$, CC1$\pi^\pm$0$\pi^0$, CC0$\pi^\pm$1$\pi^0$, CCN$\pi^\pm$0$\pi^0$, and CCother) for both \nue and \numu events, as well as a set of neutral current final states, also characterized by pion content (NC0$\pi$, NC1$\pi^\pm$0$\pi^0$, NC0$\pi^\pm$1$\pi^0$, NCN$\pi^\pm$0$\pi^0$, and NCother). To determine the smearing of true quantities due to event reconstruction, vertex, direction, and visible energy resolution functions were also produced for the 1-ring e-like and $\mu$-like samples in bins of visible energy and distance along the track direction to the wall of the most energetic ring.

\begin {figure}[htp]
  \begin{center}
    \includegraphics[width=0.45\textwidth]{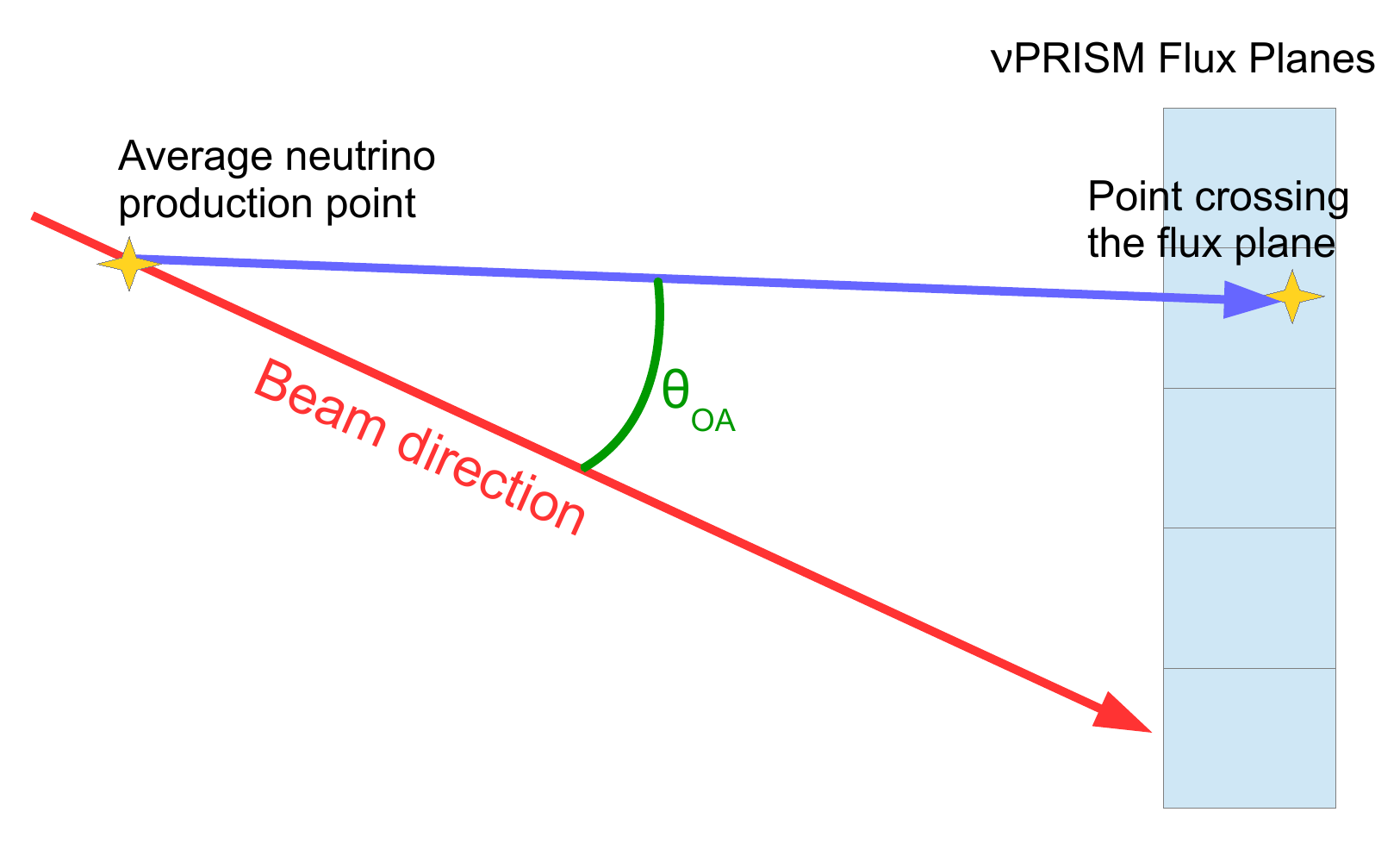}
    \caption{The definition of the off-axis angle for individual neutrinos.}
    \label{fig:oa_def}
  \end{center}
\end {figure}

\begin {figure*}[htp]
  \begin{center}
    \includegraphics[width=0.9\textwidth]{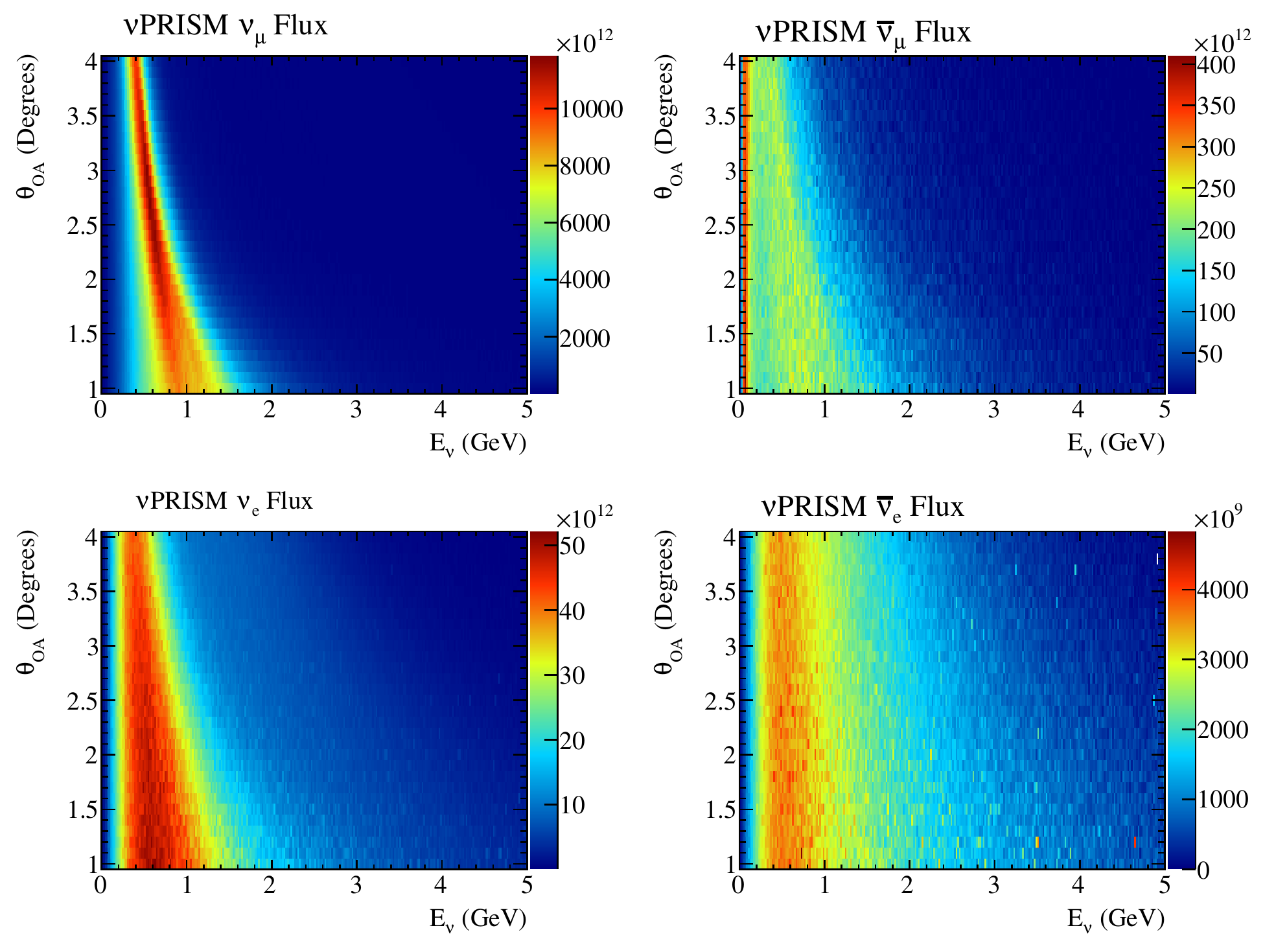}
    \caption{The neutrino flux (arbitrary normalization) as a function of off-axis angle and energy
for each neutrino flavor with the horn in neutrino-mode operation.}
    \label{fig:sim_oa_enu}
  \end{center}
\end {figure*}

The neutrinos in \nuprism are simulated with the T2K flux simulation tool called JNUBEAM.  The version of JNUBEAM used is consistent with what is currently used by T2K and it includes the modeling of hadronic interactions based on data from the NA61/SHINE experiment.  We define the off-axis angle for a particular neutrino as the angle between 
the beam axis and the vector from the average neutrino production point along the beam axis to the point
at which the neutrino crosses the flux plane, as illustrated in Fig.~\ref{fig:oa_def}.  The off-axis angle
is defined in terms of the average neutrino production point so that an off-axis angle observable can be
constructed based on the location of the interaction vertex in \nuprism.  The off-axis angle and energy 
dependence for each neutrino flavor is shown in Fig.~\ref{fig:sim_oa_enu}.  The neutrino flux files are produced
for both neutrino mode (focussing positively charged hadrons) and antineutrino mode (focussing negatively 
charged hadrons), although only the neutrino mode flux is used for the analysis presented in this note. 

%The neutrino interactions in the \nuprism water volume are modeled using NEUT 5.1.4.2.  The neutgeom package
%used for ND280 neutrino vector production has been adapted to handle arbitrary geometries. By adapting this package,
%we are able to generate vectors in the detector volume while accounting for the flux dependence on off-axis angle.
%A simple water column
%geometry was produced for \nuprism and interactions were generated using the previously described flux.  Interactions
%for the equivalent of 3.6e21 POT have been generated for neutrino mode operation, while a smaller sample equivalent to
%6.5e20 POT has been generated for antineutrino mode operation.  Interactions are also generated with NEUT 5.3.2 and the
%multinucleon events are skimmed and used for studies.  

\begin {figure}[htp]
  \begin{center}
    \includegraphics[width=9cm]{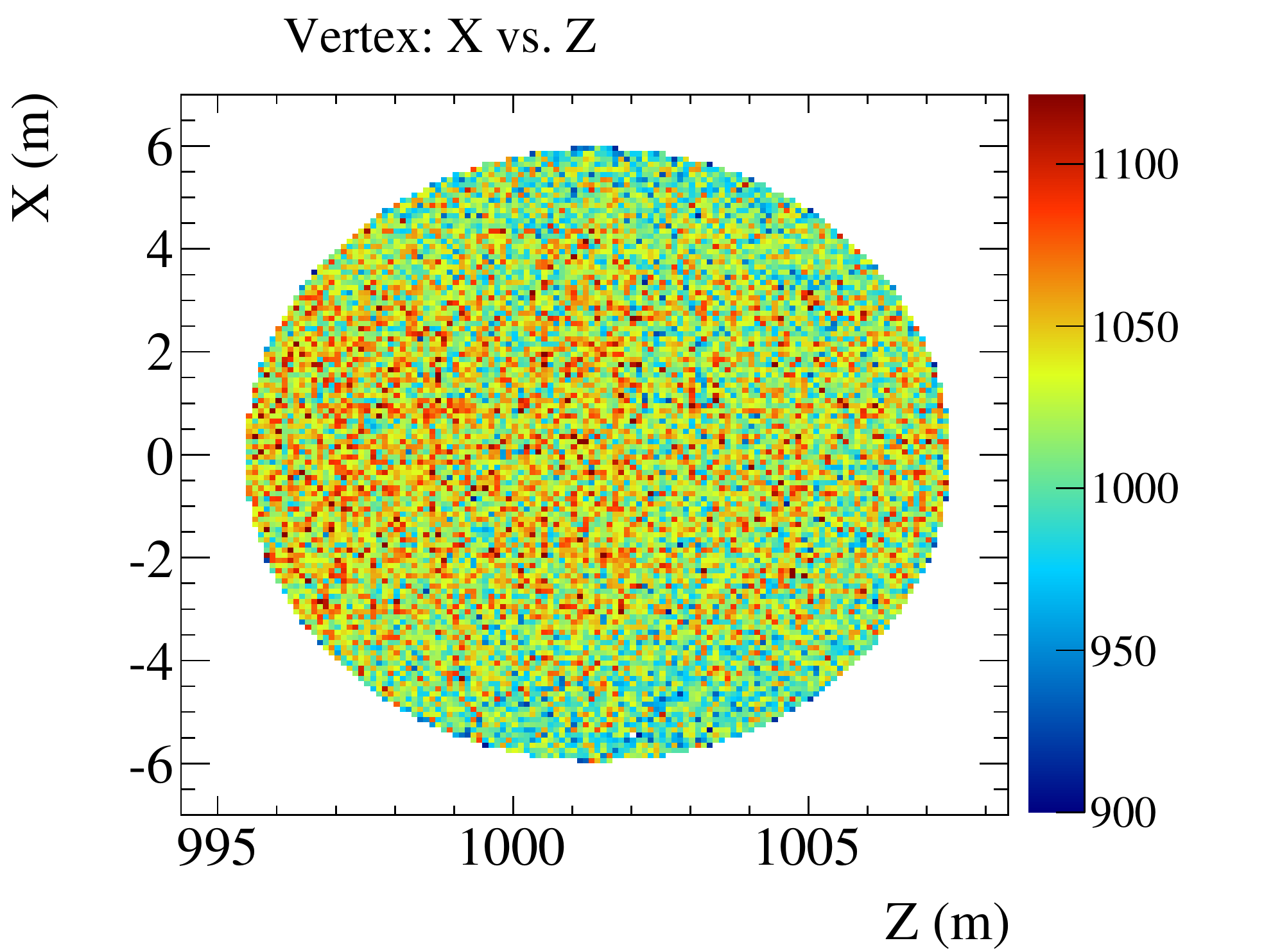}
    \includegraphics[width=9cm]{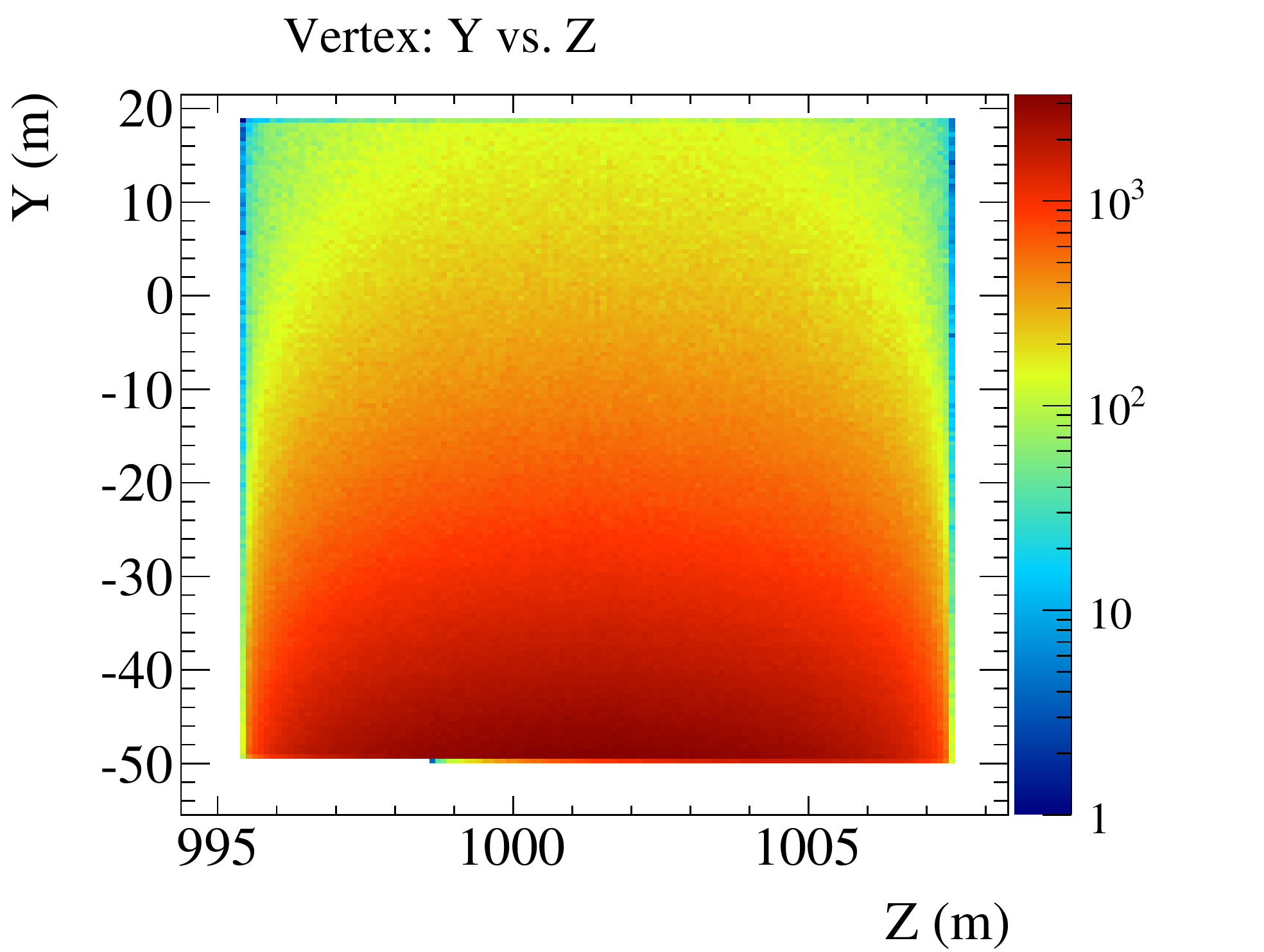}
    \includegraphics[width=9cm]{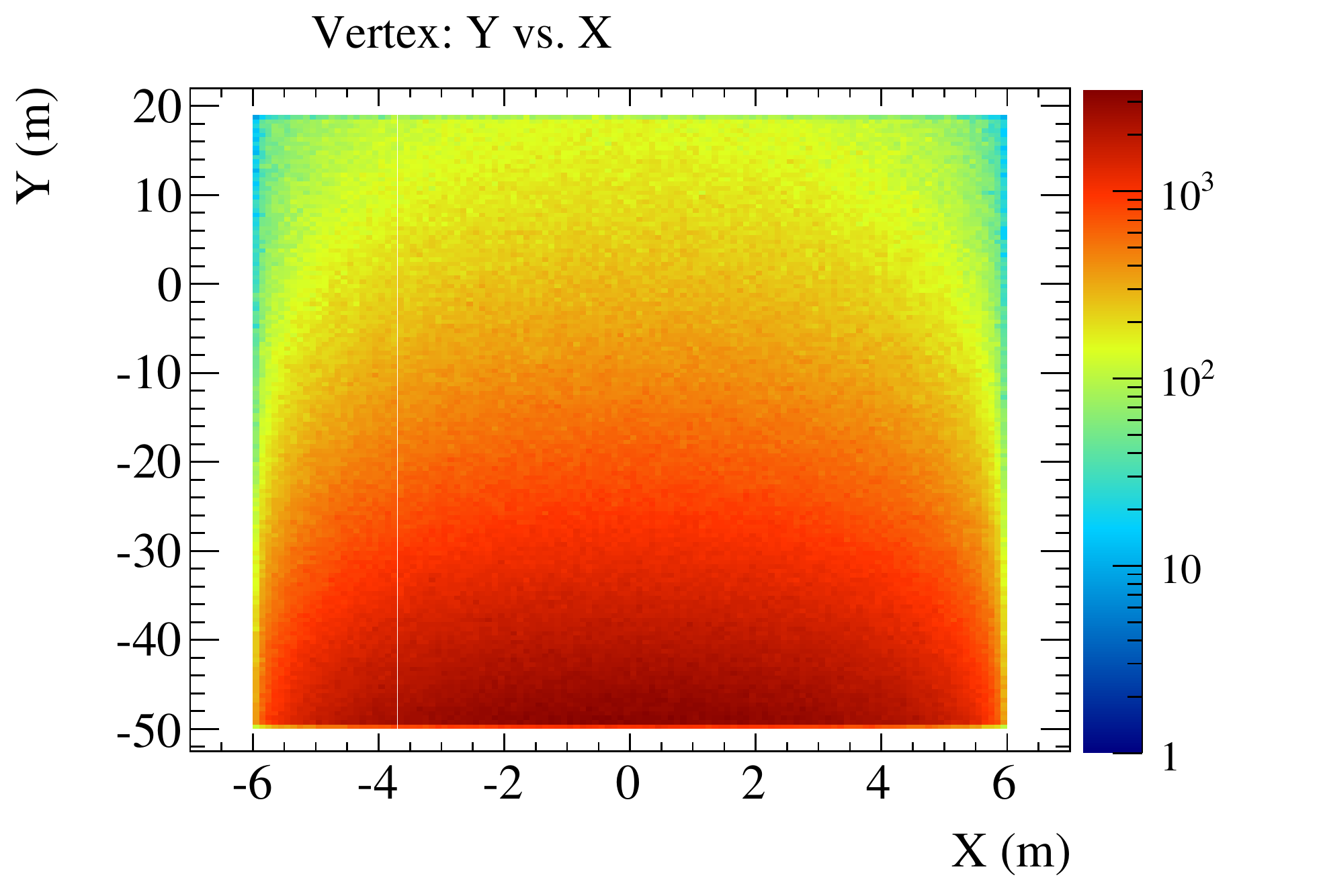}
    \caption{The distribution of simulated vertices shown in projections to the x-z (top), y-z (middle) and x-y (bottom) planes.  Here
    x is defined as the horizontal axis perpendicular to the beam, z is the horizontal axis in the beam direction and y is the vertical axis. }
    \label{fig:nu_vertex}
  \end{center}
\end {figure}

The positions of the neutrino interaction vertices in the \nuprism water volume are shown in Fig.~\ref{fig:nu_vertex}.
The rate of simulated interactions has been cross checked against the observed INGRID rates% in Section~\ref{sec:nuprism_pileup}
and found to be consistent.

%A full detector simulation for \nuprism is not yet ready, so we model the efficiency and resolution for the detection of 
%single electron or muon rings using the known performance of the SK detector and the new reconstruction software developed for T2K, called fiTQun.  Using the SK Monte Carlo,
%efficiency tables are generated for each true event topology, and these tables are binned by the true quantities: distance to the wall of the most energetic particle, $ToWall$; the visible energy of the most energetic particle, $E_{vis}$; and the true neutrino energy, $E_{true}$.  Smearing of the reconstruced $E_{vis}$, ring direction and vertex are also applied.  The resolution histograms are constructed from the SK Monte Carlo for muon and electron ring hypotheses separately and evaluated for bins in true $ToWall$ and $E_{vis}$. 
%
%For each neutrino interaction generated for \nuprism, the true topology, most energetic ring, $E_{vis}$ and $ToWall$ are evaluated.  Then the
%corresponding efficiencies is found from the SK/fiTQun tables and random rejection is applied according to the efficiency.  If the event passes the efficiency step, the resolution histograms are found and the observed quantities are evaluated by smearing the true quantities.

\subsection{Event Pileup } \label{sec:nuprism_pileup} % I assume this will be sand background as well? TL
\begin {figure}[htbp]
  \begin{center}
    \includegraphics[width=9cm]{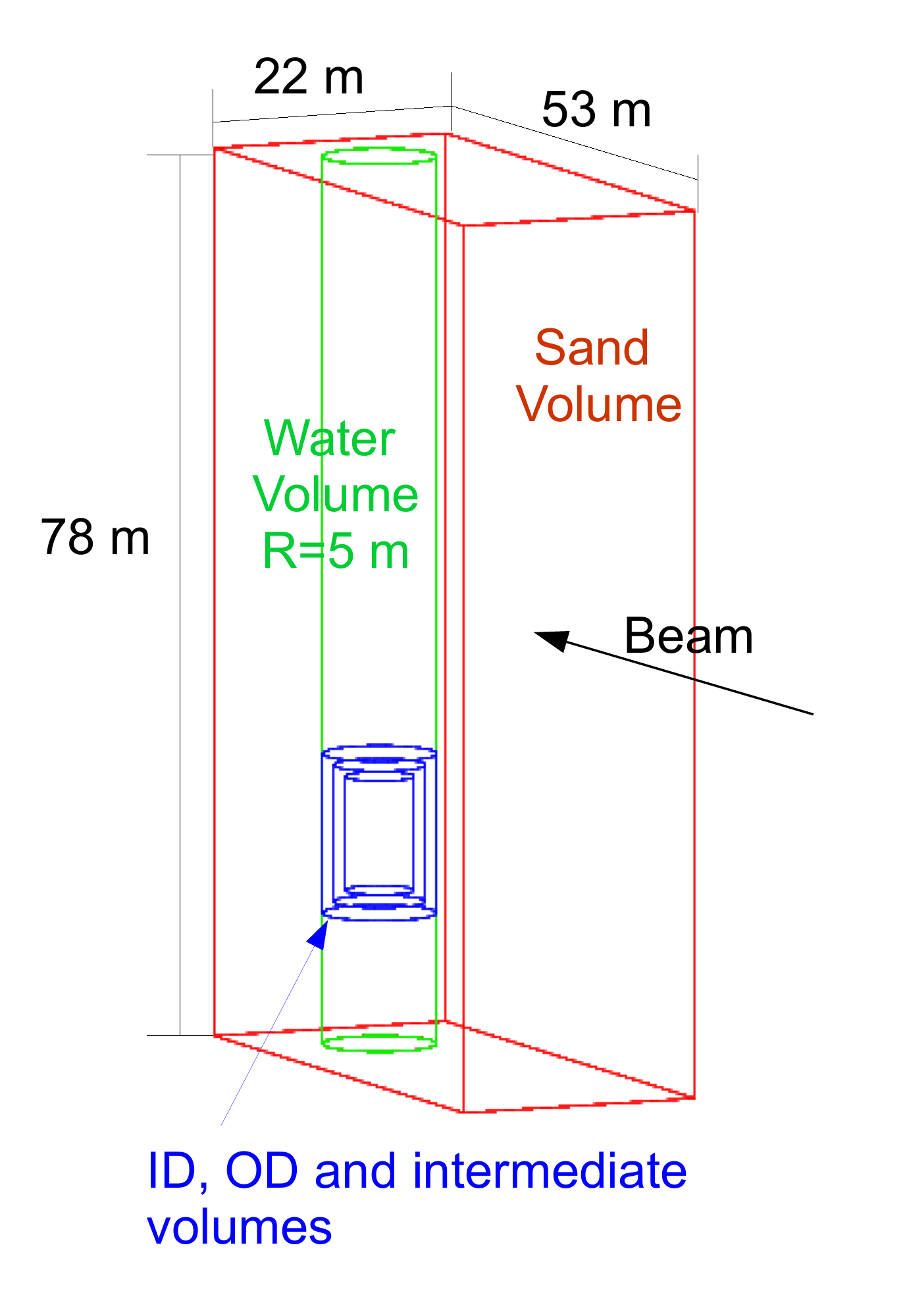}
    \caption{The GEANT4 geometry used in the pile-up simulation.}
    \label{fig:sand_geom}
  \end{center}
\end {figure}

The baseline design of \nuprismlite is an outer detector (OD) volume with radius of 5 m and 
height of 14 m, and an inner detector (ID) volume with a radius of 3 m and height of 10 m, located
1 km from the T2K target.  We have carried out a simulation of events in the 
\nuprismlite ID and OD volumes, as well as the surrounding earth to study the event pile-up
in \nuprismlite.  The simulation is carried out for the earth+\nuprismlite 
geometry shown in Fig.~\ref{fig:sand_geom}.  The flux at the upstream end of the volume is simulated
using the JNUBEAM package with horn currents set to 320 kA.  Interactions in the earth and detector 
volumes are generated using the same tools from the NEUT package used for ND280 neutrino vector
generation.  The earth volume is filled with SiO$_{2}$ with a density of 1.5 g/cm$^{3}$.  The water volume
has three detector sub-volumes: the ID detector, the OD detector and an intermediate volume.
The vertical position of the detector volumes in the water column can be adjusted to study the
event pile-up at different off-axis angles.  A GEANT4 simulation of the particles from the neutrino
vectors is carried out and all particles with visible energy greater than 10 MeV are recorded if they 
originate in any of the detector volumes or cross any of the detector volume boundaries.

We break up the visible events into five categories for the pile-up studies:
\begin{enumerate}
\item Events originating outside of the ID and entering the ID.
\item Events originating inside the ID with visible particles escaping the ID.  These are called partially contained (PC) ID events.
\item Events originating inside the ID with no visible particles escaping the ID.  These are called fully contained (FC) ID events.
\item Events originating in the OD with no visible particles entering the ID.  
\item Events originating outside the OD with visible particles entering the OD, but not the ID.
\end{enumerate}
The first three categories represent the event rate in the ID, while all but the second category represent the event rate
in the OD.  Table~\ref{tab:pileup} shows the simulated event rates per $2.5\times10^{13}$ protons on target, the assumed protons per bunch for full
750 kW operation.  Rates are shown for the \nuprismlite configurations where the ID covers off-axis angle ranges of 0.0-0.6, 1.0-1.6,
2.0-2.6 or 3.0-3.6 degrees.  While the current design does not include a pit that extends to on-axis, the 0.0-0.6 degree position is used
to make comparisons to the INGRID event rates.

\begin{table*}
\begin{center}
\caption{The event rates per 2e13 POT for \nuprismlite with horn currents at 320 kA.}
\label{tab:pileup}
\begin{tabular}{l|c|c|c|c|c}
\hline
Off-axis Angle ($^{\circ}$) & Entering ID & PC ID  & FC ID  & OD Contained  & Entering OD    \\ \hline
0.0-0.6                     & 0.4179      & 0.2446 & 0.3075 & 1.2904        & 0.7076 \\
1.0-1.6                     & 0.1005      & 0.0550 & 0.0741 & 0.3410        & 0.1939 \\
2.0-2.6                     & 0.0350      & 0.0198 & 0.0230 & 0.1234        & 0.0635 \\
3.0-3.6                     & 0.0146      & 0.0092 & 0.0156 & 0.0564        & 0.0291 \\ \hline
\end{tabular}
\end{center}
\end{table*}

For the off-axis angle 1.0-1.6 degree position, the total rate of ID+OD visible events in a spill (8 bunches) is 6.12.  If a bunch contains 
an event, the probability that the next bunch contains at least one visible event is 53\%.  This suggests that \nuprismlite should employ
deadtime-less electronics that can record events in neighboring bunches and that the after-pulsing of PMTs
should be carefully considered. 
The rate of ID events per bunch is 0.230 and the probability of two or more visible ID
events in a single bunch with at least one visible event is 20\%.  Hence, most bunches will not require the reconstruction of multiple interactions
in the ID volume.  
However, the probability of 2 or more ID events per spill is 84\%, so the reconstruction of out of time events such
as decay electrons needs to be carefully studied.  Decay electrons in a spill may potentially be matched to their parent
 interactions using both spatial and timing information.  For interactions inside the ID, a spatial likelihood matching
 the decay electron to the primary vertex may be constructed based on the reconstructed decay electron vertex position and 
the reconstructed primary vertex or reconstructed stopping point of the candidate muons or charged pions in the event. For decay electrons
 originating from muons produced outside of the ID, a similar spatial likelihood may be constructed using OD light, ID light, and hits from 
scintillator panels (if they are installed between the OD and ID) from the entering particle.   Since the muon mean lifetime (2.2 $\mu$s) is 
shorter than the spill length (~5 $\mu$s), there will also be statistical power to match decay electrons to their primary vertex based on 
the time separation of the decay electron vertex and primary vertex.  On the other hand, the muon lifetime may provide a cross-check for the 
spatial matching of primary and decay electron vertices since significant mismatching would tend to smear the time separation distribution 
beyond the muon lifetime.  Studying the matching of decay electrons to primary interactions is a high priority and work is underway to address 
this issue with a full simulation of \nuprism and the surrounding rock.

The rate of events producing light in the OD is 0.690 per bunch.  Hence, the 
probability that an FC ID event will have OD activity in the same bunch is 50\%.  Neglecting out of time events,
the rejection rate of FC ID events would be 50\% if a veto on any OD activity in the bunch is applied.  This rejection rate falls
to 21\% and 10\% in the 2.0-2.6 and 3.0-3.6 degree off-axis positions respectively.  Of the OD events, about 30\% are entering
from the surrounding earth, and most of those are muons.  The scintillator panels may be used to relax the veto on these types of 
pile-up events by providing additional spatial and timing separation between the OD and ID activity in the same bunch.
If the veto can be removed for all events entering the OD from the earth,
then rejection rates due to OD pile-up drop to 39\%, 16\% and 8\% for the 1.0-1.6, 2.0-2.6 and 3.0-3.6 degree
off-axis angle positions respectively.

We can cross-check the estimated \nuprism event rates by extrapolating from the event rates observed by INGRID.  We assume that
the rate of interactions inside the detector will scale with the detector mass, and the rate of entering events from the earth
will scale with the cross-sectional area of the detector.  The rates should also scale with $1/d^{2}$, were $d$ is the
distance from the average neutrino production point to the detector, about 240~m for INGRID and 960~m for \nuprism. 
INGRID observes 1.74 neutrino events per $1\times10^{14}$ POT in 14 INGRID modules with a total mass of $5.7\times10^4$ kg.  For an OD mass
of $8.2\times10^5$~kg, we extrapolate the INGRID rate, assuming 60\% detection efficiency in INGRID, to obtain 0.66 interactions 
in the OD for each $2.5\times10^{13}$ POT bunch.  The simulated rate of visible OD interactions in \nuprismlite is 1.50 and 0.39 for
the 0.0-0.6 and 1.0-1.6 degree positions respectively.  Since INGRID covers an angular range of about $\pm1$ degree, it is
reasonable that the extrapolated value from INGRID falls between the simulated \nuprismlite values at these two positions.

INGRID also observes a event rate from earth interactions of 4.53 events per $1\times10^{14}$ POT in 14 modules with a cross-sectional area
of 21.5~m$^{2}$.  These earth interaction candidates are INGRID events failing the upstream veto and fiducial volume cuts.  The selection
of entering earth-interaction events is $>99$\% efficient and 85.6\% pure.  Scaling to the OD cross-sectional area and
distance while
correcting for the efficiency and purity gives a rate of 0.31 events entering the OD per bunch.  The rate from
the \nuprism pile-up simulation is 0.903 or 0.239 for the 0.0-0.6 and 1.0-1.6 degree positions respectively.  Once again, the extrapolated
INGRID rate falls between the simulated rates for these two \nuprismlite positions.

In summary, the event pile-up rates for \nuprism appear manageable.  Even for the most on-axis position and high power beam, most bunches
with interactions will only have a single interaction with visible light in the ID.  The OD veto rate from pile-up can be as large as 50\%,
hence careful studies of the OD veto are needed.
The OD veto rate may be reduced and better understood with the inclusion of scintillator panels at the outer edge of the OD or at the OD/ID 
boundary.  The electronics for \nuprismlite should be deadtime-less to handle multiple events per spill.  

Further studies of the event rates will be carried out.  These will include the study of entering neutral particles to be used
in the optimization of the OD and fiducial volume sizes, more realistic studies of how the scintillator panels may be used to
optimize the OD veto cut, and updates to the earth density to better reflect the surveyed density of the rock strata at potential \nuprism sites.

%Section goals:
%\begin{itemize}
%\item MC study of pileup including rates of muons and neutrons as a function of off-axis angle
%\end{itemize}

\subsection{Event Selection for Sensitivity Studies}
We select samples of single ring muon and electron candidates for the long and short baseline sensitivity studies described in the following sections.  
As described in Section~\ref{sec:nuprism_sim}, the efficiencies for single ring electron or muon selections are applied using tables calculated from the SK MC.  
The efficiency tables are calculated with the following requirements for muon and electron candidates:
\begin{itemize}
\item Muon candidate requirements: fully contained, a single muon-like ring, 1 or fewer decay electrons
\item Electron candidate requirements: fully contained, a single electron-like ring, no decay electrons, passes the fiTQun $\pi^{0}$ cut
\end{itemize}
Additional cuts are applied on the smeared $\nu$PRISM MC.  For the muon candidates the cuts are similar to the SK selection for the T2K disappearance analysis:
\begin{itemize}
\item Muon candidate cuts: $dWall>100$ cm, $toWall>200$ cm, $E_{vis}>30$ MeV, $p_{\mu}>200$ MeV/c
\end{itemize}
where $dWall$ is the distance from the event vertex to the nearest wall, and $toWall$ is the distance from the vertex to the wall along the direction of the particle.

For the single ring electron candidates, the cuts on $toWall$ and $E_{vis}$ were reoptimized since the separation between electrons and muons or electrons
and $\pi^{0}$s degrades closer to the wall.  The cut on $dWall$ is set to 200 cm to avoid entering backgrounds.  The 
cuts are:
\begin{itemize}
\item Electron candidate cuts: $dWall>200$ cm, $toWall>320$ cm, $E_{vis}>200$ MeV
\end{itemize}
The tight fiducial cuts for the electrons candidates are needed to produce a relatively pure sample, but there is a significant impact to the electron candidate statistics.
A simulation with finer PMT granularity may allow for the $toWall$ cut to be relaxed, increasing the statistics without degrading the purity.

\subsection{T2K \numu Disappearance Sensitivities}
\label{sec:disap}

The most straightforward application of the \nuprism concept to T2K is in the \numu disappearance measurement. A full \numu analysis has been performed in which \nuprismlite completely replaces ND280. In the future, it will be useful to incorporate ND280 into \nuprismlite analyses, particularly the sterile neutrino searches, but for simplicity this has not yet been done.

The main goal of this \numu disappearance analysis is to demonstrate that \nuprismlite measurements will remove most of the neutrino cross section systematic uncertainties from measurements of the oscillation parameters. This is achieved by directly measuring the muon momentum vs angle distribution that will be seen at Super-K for any choice of $\theta_{23}$ and $\Delta m^2_{32}$.

\begin{figure}[htpb]
\begin{center}
  \begin{minipage}[t]{.45\textwidth}
    \begin{center}
      \includegraphics[width=\textwidth] {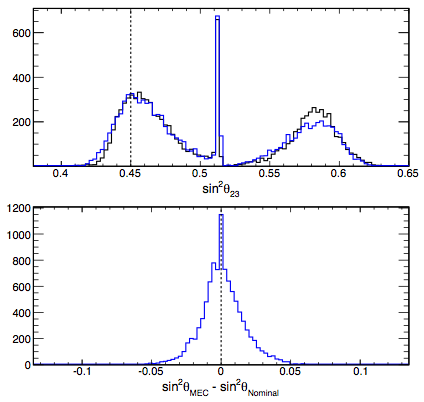}
    \end{center}
  \end{minipage}
  \begin{minipage}[t]{.45\textwidth}
    \begin{center}
      \includegraphics[width=\textwidth] {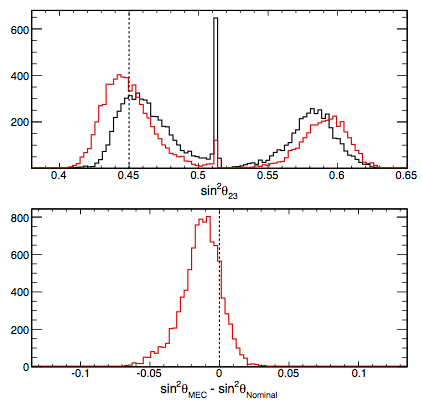}
    \end{center}
  \end{minipage}
\end{center}
\caption{The results of fitting fake data with and without multinucleon effects are shown. The measured differences in $\sin^2\theta_{23}$ when comparing the Nieves model (blue) to default neut (black) and the Martini model (red) to default neut give RMS values of 3.6\% and 3.2\%, respectively, and biases of 0.3\% and -2.9\%, respectively.}
\label{fig:nievesmartini}
\end{figure}

To clearly compare the \nuprismlite \numu analysis with the standard T2K approach, the full T2K analysis is reproduced using \nuprismlite in place of ND280. This is done by generating fake data samples produced from throws of the flux and cross section systematic parameters and fitting these samples using the standard oscillation analysis framework. In each flux, cross section and statistical throw, three fake data samples using different cross section models were produced at both ND280 and Super-K: default NEUT with pionless delta decay, NEUT with the Nieves multinucleon model replacing pionless delta decays, and NEUT with an ad-hoc multinucleon model that uses the final state kinematics of the Nieves model and the cross section from Martini {\it et al}.
For each throw, all three fake data samples were fit to derive estimates of the oscillation parameters.  The differences between the fitted values of  $\sin^2\theta_{23}$ for the NEUT nominal and NEUT+Nieves or NEUT+Martini fake data fits are shown in Figure~\ref{fig:nievesmartini}.
The systematic uncertainty associated with assuming the default NEUT model rather than the model of Martini or Nieves is given by the quadrature sum of the RMS and mean (i.e. bias) of these distributions. For the ND280 analysis, there is a 3.6\% uncertainty when comparing with the Nieves model, and a 4.3\% uncertainty in the measured value of $\sin^2\theta_{23}$ when comparing with the Martini model. These uncertainties would be among the largest for the current T2K \numu disappearance analysis, and yet they are based solely on model comparisons with no data-driven constraint.

%The first stage of the \nuprismlite \numu analysis is to separate the 1-4 degree off-axis range of the detector into 30 0.1 degree slices in off-axis angle. The neutrino energy spectrum in each off-axis bin is predicted by the T2K neutrino flux simulation.  For each set of oscillation parameters to be tested, the Super-K neutrino energy spectrum is also predicted using the T2K flux simulation. For each set of oscillation parameters to be included in the final oscillation fit, the oscillated Super-K energy spectrum is also extracted from the flux Monte Carlo simulation. A linear combination of the 30 off-axis fluxes is then taken to reproduce each of the Super-K oscillated spectra,
%\begin{equation}
%\Phi^{SK} \left(E_\nu;\theta_{23},\Delta m^2_{32}\right)E_\nu =\sum_{i=1}^{30}c_i\left(\theta_{23},\Delta m^2_{32}\right)E_\nu\Phi^{\nu P}_i(E_\nu),
%\end{equation}
%where $c_i\left(\theta_{23},\Delta m^2_{32}\right)$ is the weight of each off-axis slice, $i$. The extra factors of $E_\nu$ are inserted to approximate the effect of cross section weighting. An example linear combination of nuPRISM off-axis fluxes that reproduces the SK flux flux is shown in Figure~\ref{fig:fluxfit}. These fits can successfully reproduce Super-K oscillated spectra, except at neutrino energies below $\sim 400$~MeV. The maximum off-axis angle is 4$^\circ$, which peaks at 380~MeV, so at lower energies it is difficult to reproduce an arbitrary flux shape. This could be improved by extending the detector further off-axis.

As was discussed in Section~\ref{sec:enu_determine} the limitation of using ND280 data to predict observed particle distributions at Super-K is that the neutrino flux at these two detectors is different due to oscillations.  Therefore, any extrapolation has significant and difficult to characterize cross section model dependent uncertainties.  In the \nuprismlite based analysis, this limitation is resolved by deriving linear combinations of the fluxes at different off-axis angles to produce a flux that closely matches the predicted oscillated flux at Super-K.  The observed particle distributions measured by \nuprismlite are then combined with the same linear weights to predict the particle distribution at Super-K.  In this way, the analysis relies on the flux model to determine the weights that reproduce the oscillated flux while minimizing cross section model dependence in the extrapolation.

\begin{figure}[htpb]
\begin{center}
      \includegraphics[width=8cm]{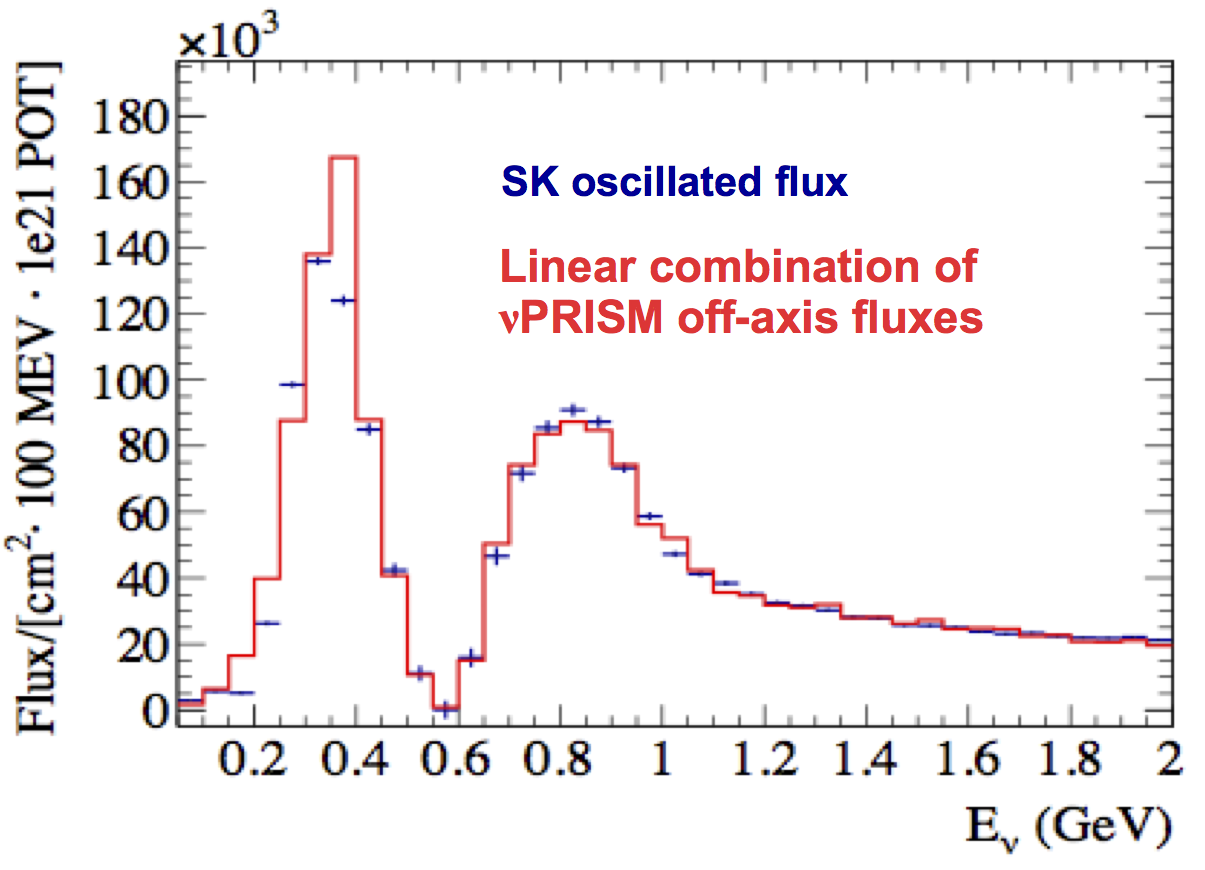}
\end{center}
\caption{A sample fit of the flux in 30 \nuprismlite fluxes to an oscillated Super-K flux is shown. Good agreement can be achieved, except at low energies due to the 4$^\circ$ maximum off-axis angle seen by \nuprismlite.}
\label{fig:fluxfit}
\end{figure}

The first stage of the \nuprismlite \numu analysis is to separate the 1-4 degree off-axis range of the detector into 30 0.1 degree or 60 0.05 degree bins in off-axis angle. The neutrino energy spectrum in each off-axis bin is predicted by the T2K neutrino flux simulation.  For each hypothesis of oscillation parameter values that will be tested in the final oscillation fit, the oscillated Super-K energy spectrum is also predicted by the T2K neutrino flux simulation. A linear combination of the 30 (60) off-axis fluxes is then taken to reproduce each of the Super-K oscillated spectra,
\begin{equation}
\Phi^{SK} \left(E_\nu;\theta_{23},\Delta m^2_{32}\right)E_\nu =\sum_{i=1}^{30}c_i\left(\theta_{23},\Delta m^2_{32}\right)E_\nu\Phi^{\nu P}_i(E_\nu),
\end{equation}
where $c_i\left(\theta_{23},\Delta m^2_{32}\right)$ is the weight of each off-axis bin, $i$. The extra factors of $E_\nu$ are inserted to approximate the effect of cross section weighting. The $c_i\left(\theta_{23},\Delta m^2_{32}\right)$ are determined by a fitting routine that seeks agreement between the Super-K flux and the linear combination over a specified range of energy. An example linear combination of nuPRISM off-axis fluxes that reproduces the SK flux is shown in Figure~\ref{fig:fluxfit}. These fits can successfully reproduce Super-K oscillated spectra, except at neutrino energies below $\sim 400$~MeV. The maximum off-axis angle is 4$^\circ$, which peaks at 380~MeV, so at lower energies it is difficult to reproduce an arbitrary flux shape. This could be improved by extending the detector further off-axis.

\begin{figure}[htpb]
    \begin{center}
      \includegraphics[width=8cm] {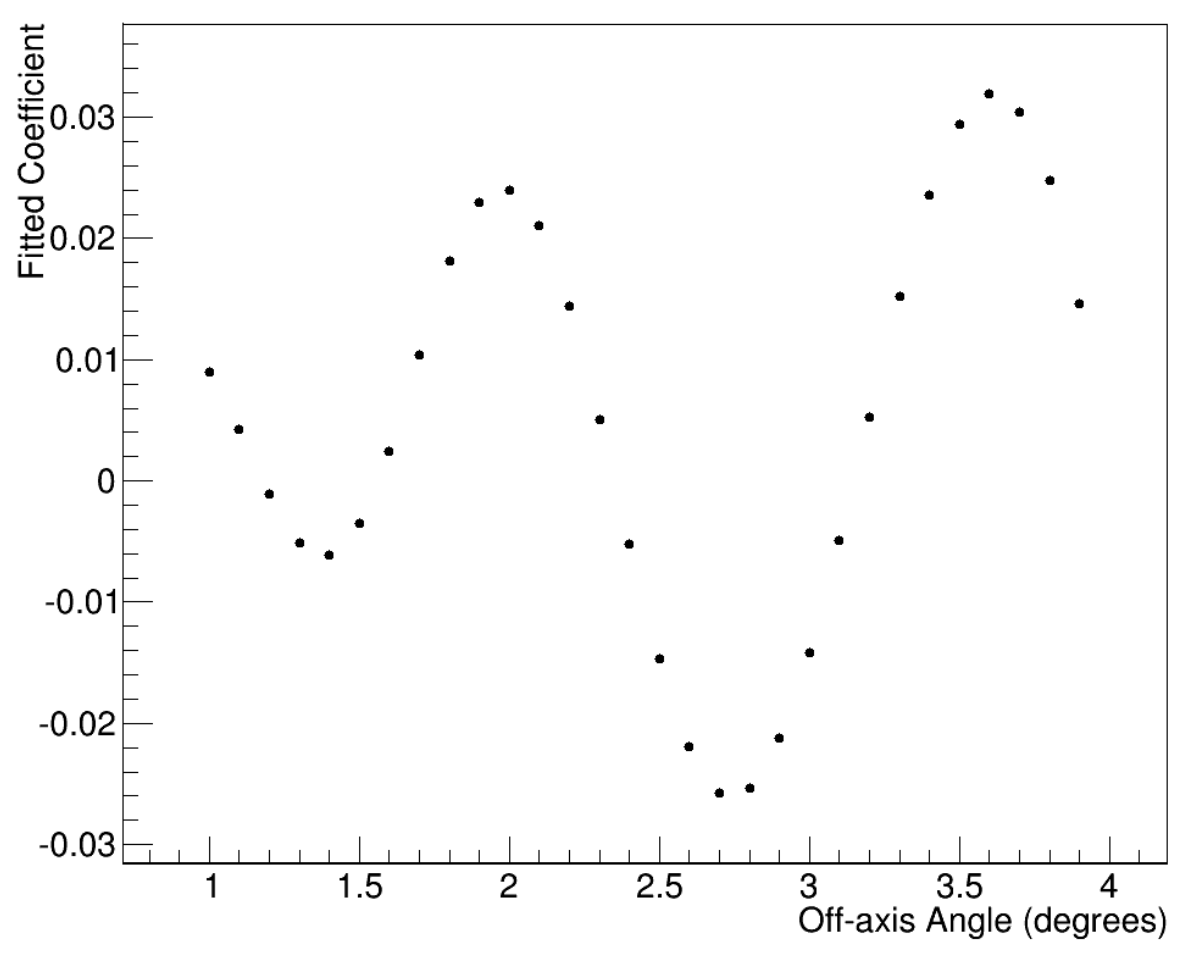}
    \end{center}
    \begin{center}
      \includegraphics[width=8cm] {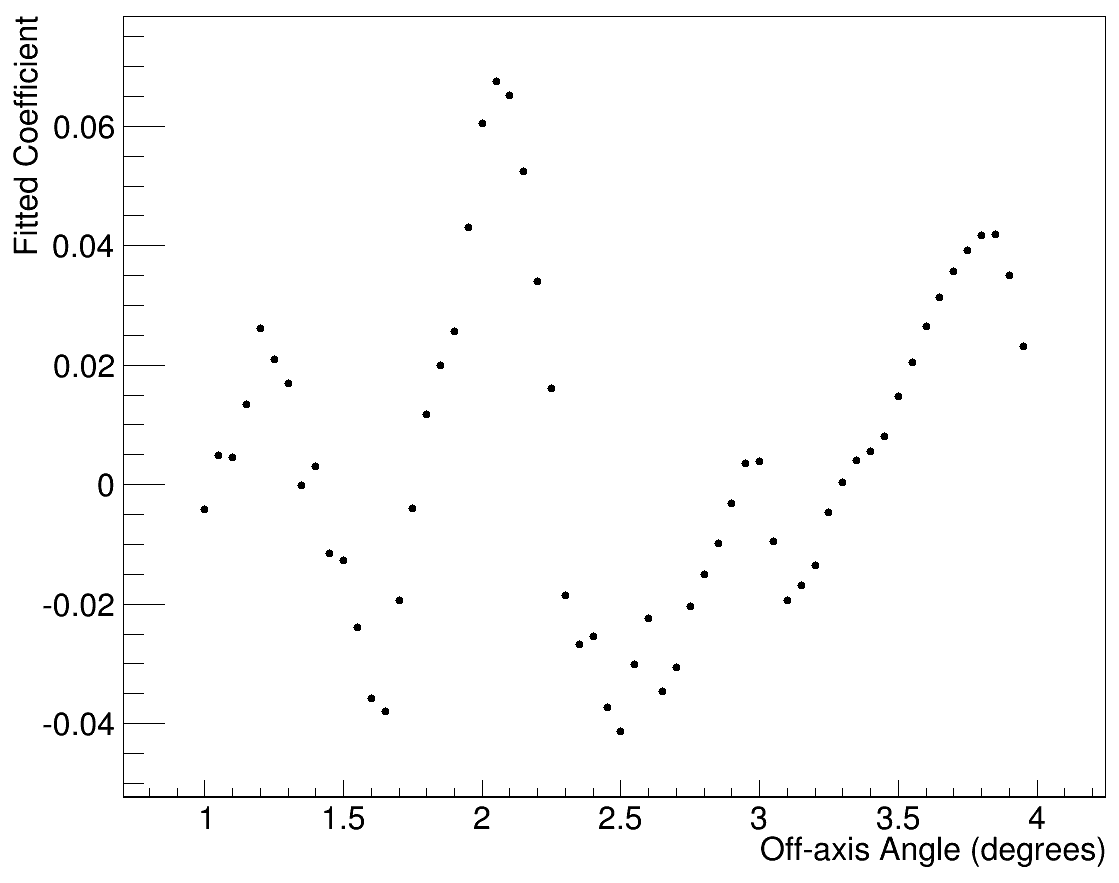}
    \end{center}
\caption{The weights for each off-axis bin produced in the \nuprismlite flux fits are shown after requiring that neighboring bins have similar values (top; as in Figure~\ref{fig:fluxesevents_smooth} left column) and with neighboring bins allowed to vary more freely relative to each other (bottom; as in Figure~\ref{fig:fluxesevents_smooth} right column).}
%without any constraint (top; as in Figure~\ref{fig:fluxfit}) and after requiring that adjacent slices have similar weights (bottom; as in Figure~\ref{fig:fluxesevents}).
%The variance of the constrained weights are smaller by an order of magnitude, which significantly reduces the statistical uncertainty in each measured bin.}
\label{fig:weightvariance}
\end{figure}

\begin{figure*}
\begin{minipage}[t]{1.0\textwidth}
\begin{center}
\includegraphics[width=0.45\textwidth] {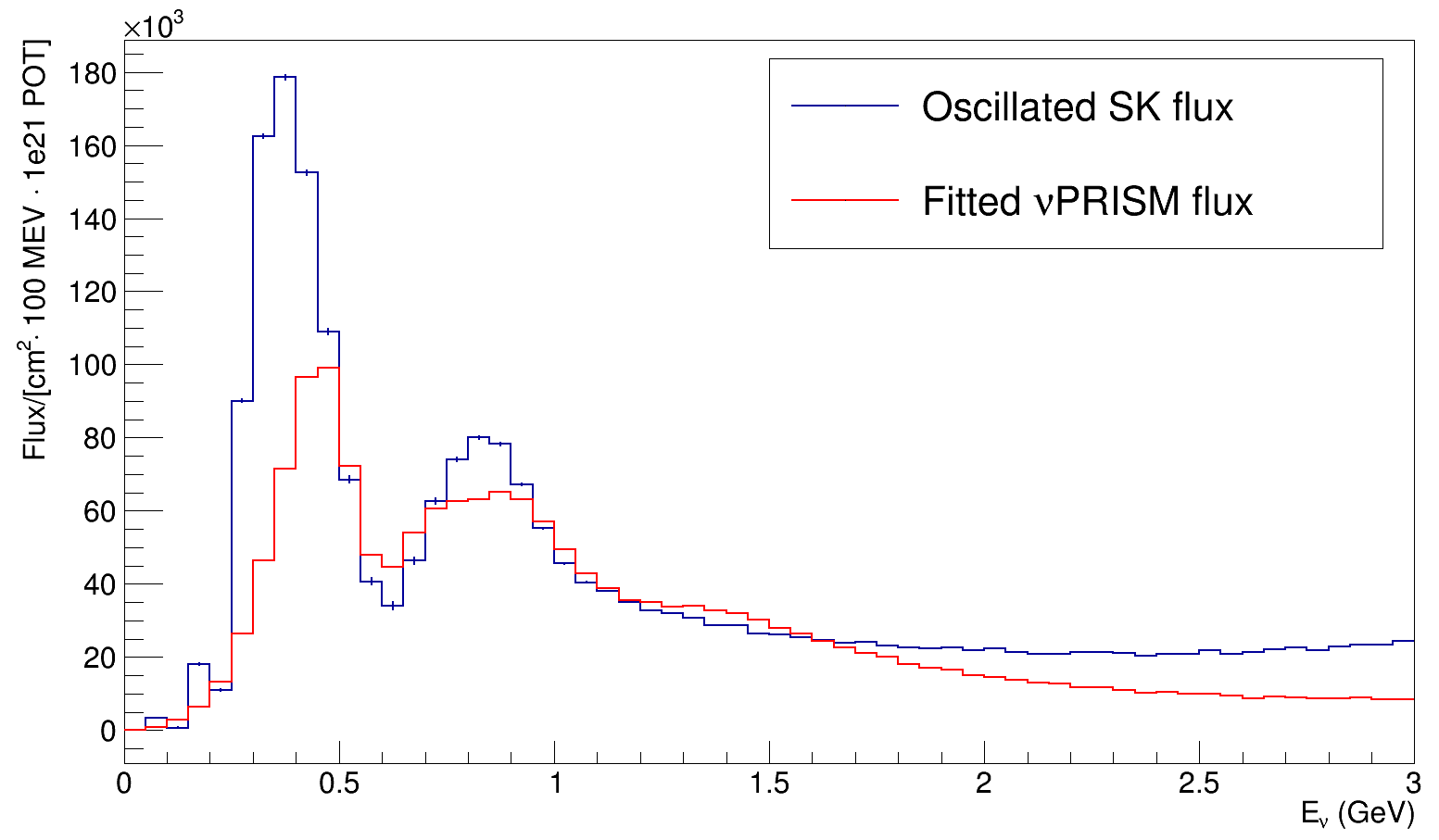}
\includegraphics[width=0.46\textwidth] {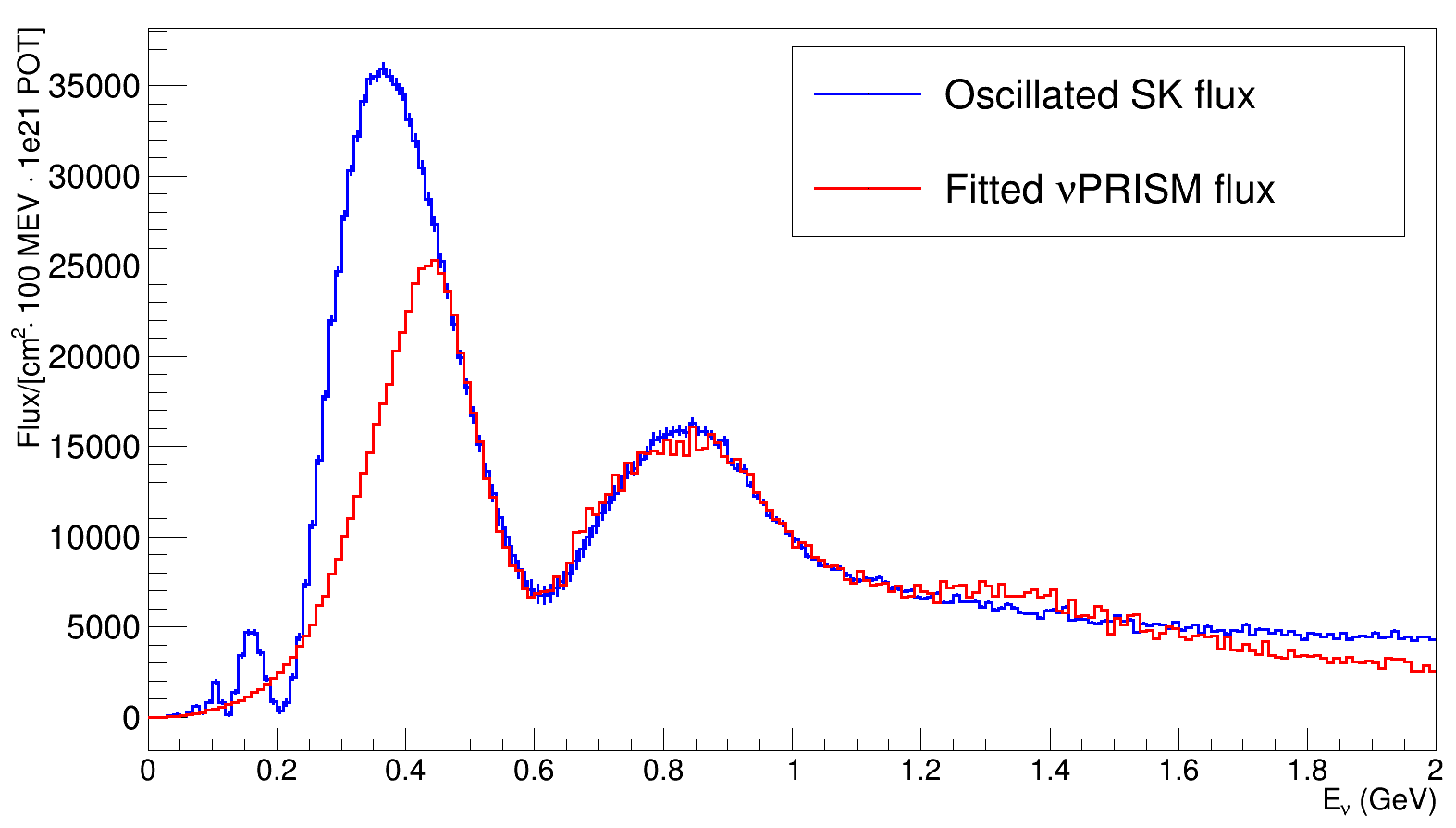}
\includegraphics[width=0.45\textwidth] {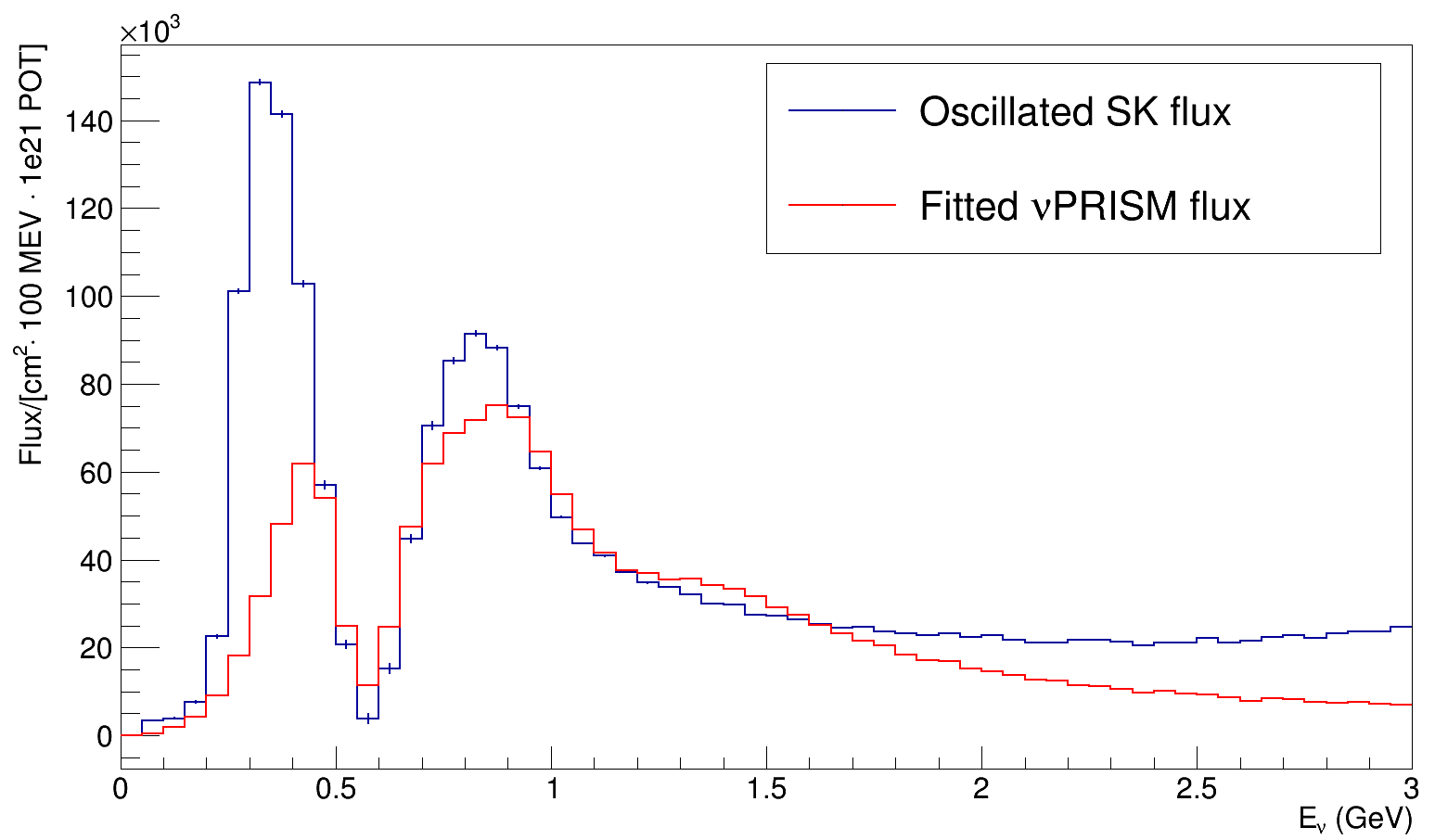}
\includegraphics[width=0.46\textwidth] {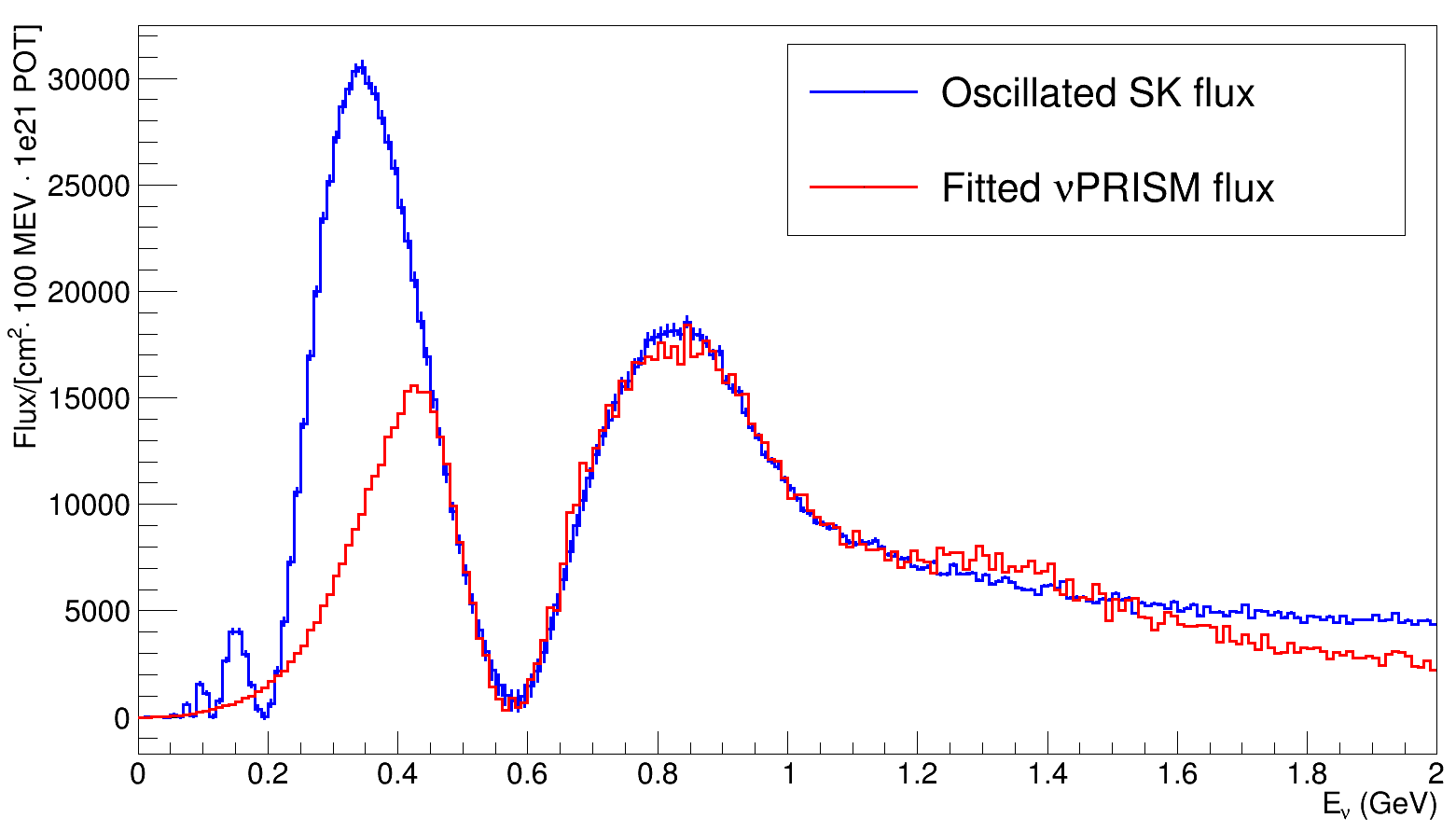}
\includegraphics[width=0.45\textwidth] {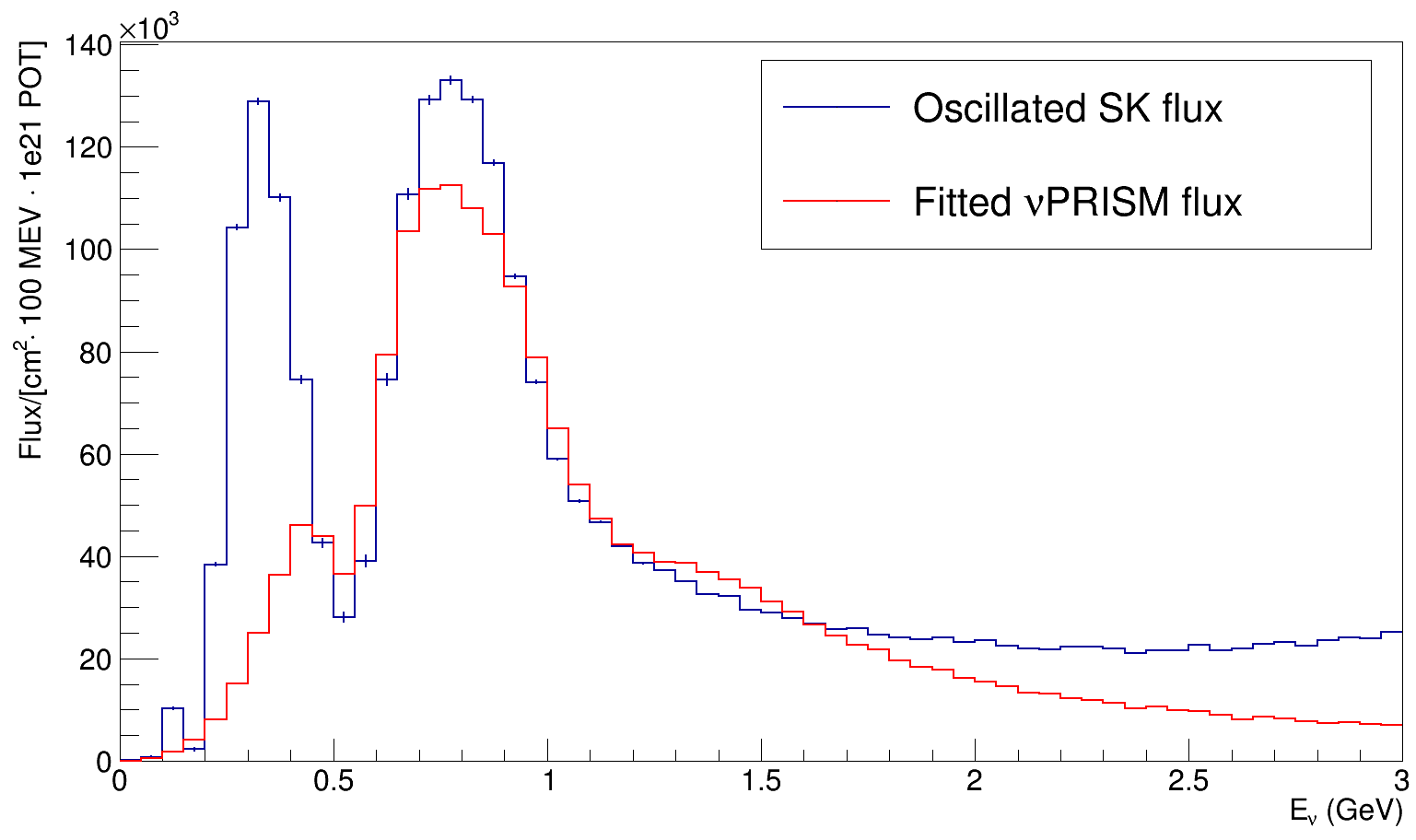}
\includegraphics[width=0.46\textwidth] {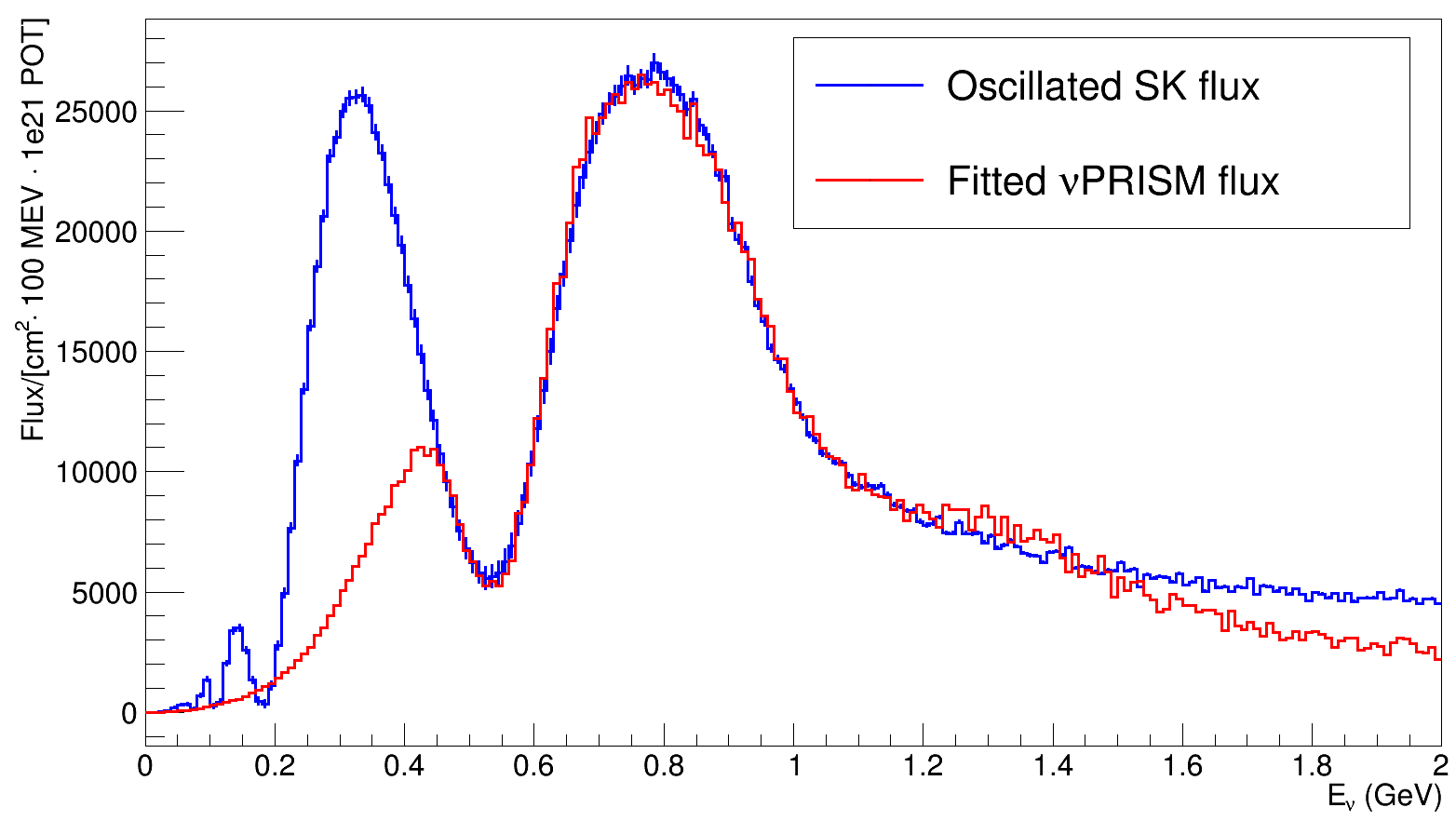}
\end{center}
\end{minipage}
\caption{Fits of the \nuprismlite flux bins to oscillated Super-K fluxes are shown for three different sets of $\left(\theta_{23},\Delta m^2_{32}\right)$: top - (0.61, $2.56*10^{-3}$), middle - (0.48, $2.41*10^{-3}$), and bottom - (0.41, $2.26*10^{-3}$). In the left column, the weights for the off-axis bins are forced to vary smoothly with off-axis angle, while in the right column they are allowed to vary more freely.}
\label{fig:fluxesevents_smooth}
\end{figure*}

The determination of the $c_i\left(\theta_{23},\Delta m^2_{32}\right)$ weights to reproduce the oscillated flux is subject to some optimization. Figure~\ref{fig:weightvariance} shows two sets of weights for a particular oscillation hypothesis.  In the first case a smoothness constrain was applied to the weights so that they vary smoothly between neighboring off-axis angle bins. In the second case the weights are allowed to vary more freely relative to their neighbors.  Figure~\ref{fig:fluxesevents_smooth} shows the comparisons of the \nuprismlite flux linear combinations with the Super-K oscillated flux for a few oscillation hypotheses in the smoothed and free weight scenarios.  The oscillated flux in the maximum oscillation region is nearly perfectly reproduced when the weights are allowed to vary more freely.  When they are constrained to vary smoothly, the agreement is less perfect, although still significantly better than the agreement between ND280 and Super-K fluxes.  An analysis using the free weights is less dependent on the cross section model assumptions in the extrapolation to Super-K since the Super-K flux is more closely matched.  On the other hand, the analysis with the smoothed weights is less sensitive to uncertainties on the flux model and \nuprismlite detector model that have an off-axis angle dependence since neighboring bins have similar weight values.  The statistical errors are also smaller for the smoothed weight case since the sum in quadrature of the weights in a given neutrino energy bin is smaller when there are less fluctuations in weight values.  In the analysis presented here, the smoothed weights are used, although the optimization of the level of smoothness is an area where the analysis will be improved in the future.

The \nuprism candidate events are events with a single observed muon ring and no-other observed particles, matching the selection applied at Super-K.  After the $c_i\left(\theta_{23},\Delta m^2_{32}\right)$  coefficients are derived, they are used to make linear combination of observed candidate event distributions from each \nuprism off-axis bin.  In this case the observables are the momentum and polar angle of the scattered muon candidate, and hence the expected Super-K distribution of these observables is predicted by the linear combination of observed \nuprism events.

In order to use these \nuprismlite measurements to make an accurate prediction of Super-K muon kinematics, a series of corrections are required. First, non-signal events from either neutral current events or charged current events with another final state particle above Cherenkov threshold, must be subtracted from each near detector slice. This is particularly important for neutral current events, which depend on the total flux rather than the oscillated flux at Super-K, but depend on the oscillated flux in the \nuprismlite linear combination. This background subtraction is model dependent, and is a source of systematic uncertainty, although neutral current interactions can be well constrained by in situ measurements at \nuprismlite. The differences in detector efficiency and resolution must also be corrected. The efficiency differences are due to differences in detector geometry and are largely independent of cross section modeling. Detector resolutions must be well determined from calibration data, but this effect is somewhat mitigated due to the fact that the near and far detector share the same detector technology. Finally, for the present analysis, the two dimensional muon momentum vs angle distribution is collapsed into a one dimensional $E_{rec}$ distribution using a transfer matrix, $M_{i,p,\theta}\left(E_{rec}\right)$. This is an arbitrary choice that does not introduce model dependence into the final result, and has only been used for consistency with existing T2K $\nu_\mu$ disappearance results. Future analyses can be conducted entirely in muon momentum and angle variables.

\begin{figure}[htpb]
\begin{center}
      \includegraphics[width=8cm] {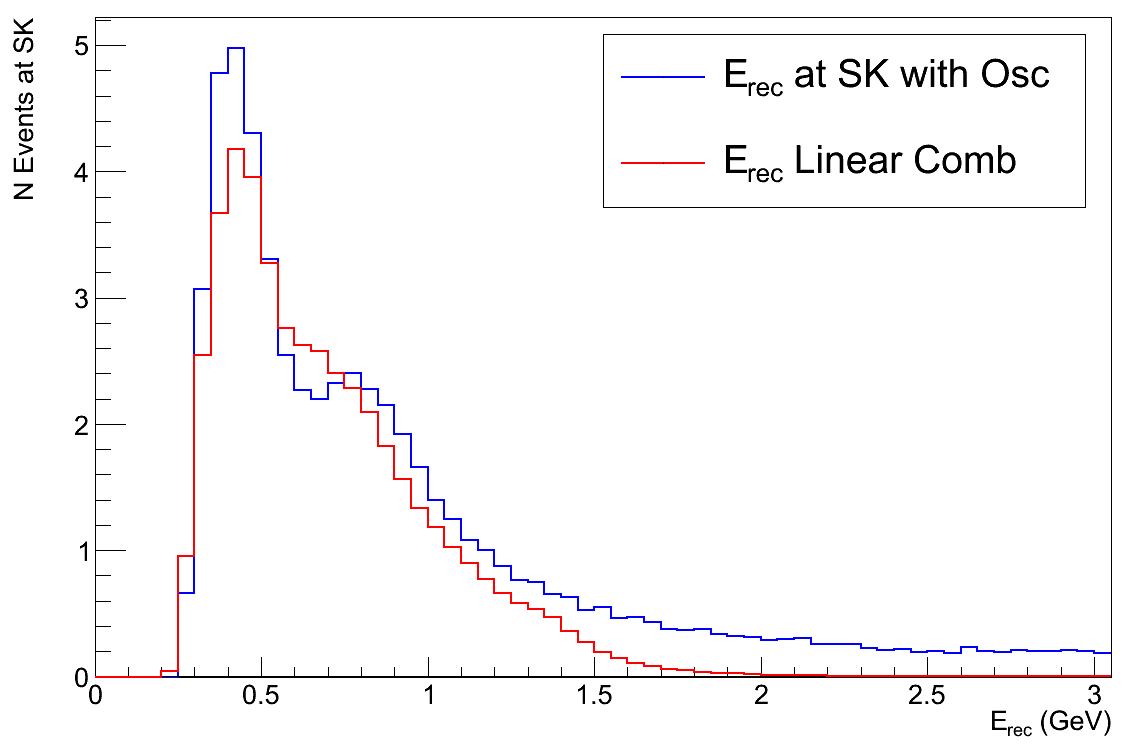}
      \includegraphics[width=8cm] {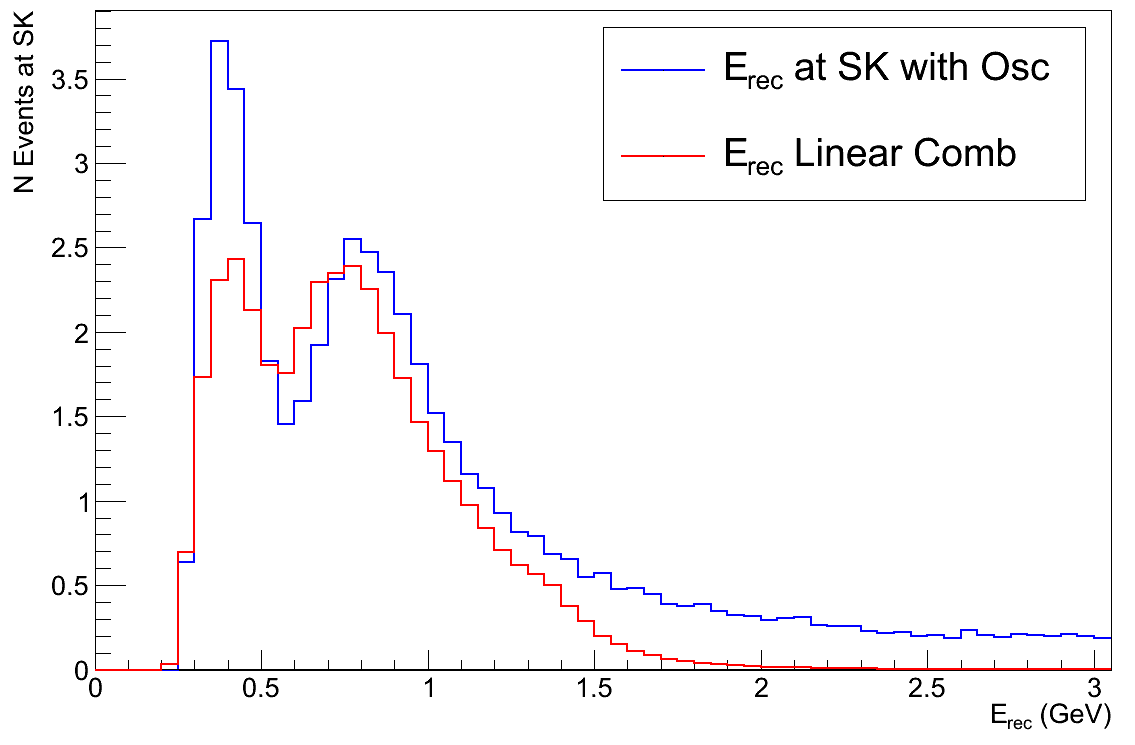}
      \includegraphics[width=8cm] {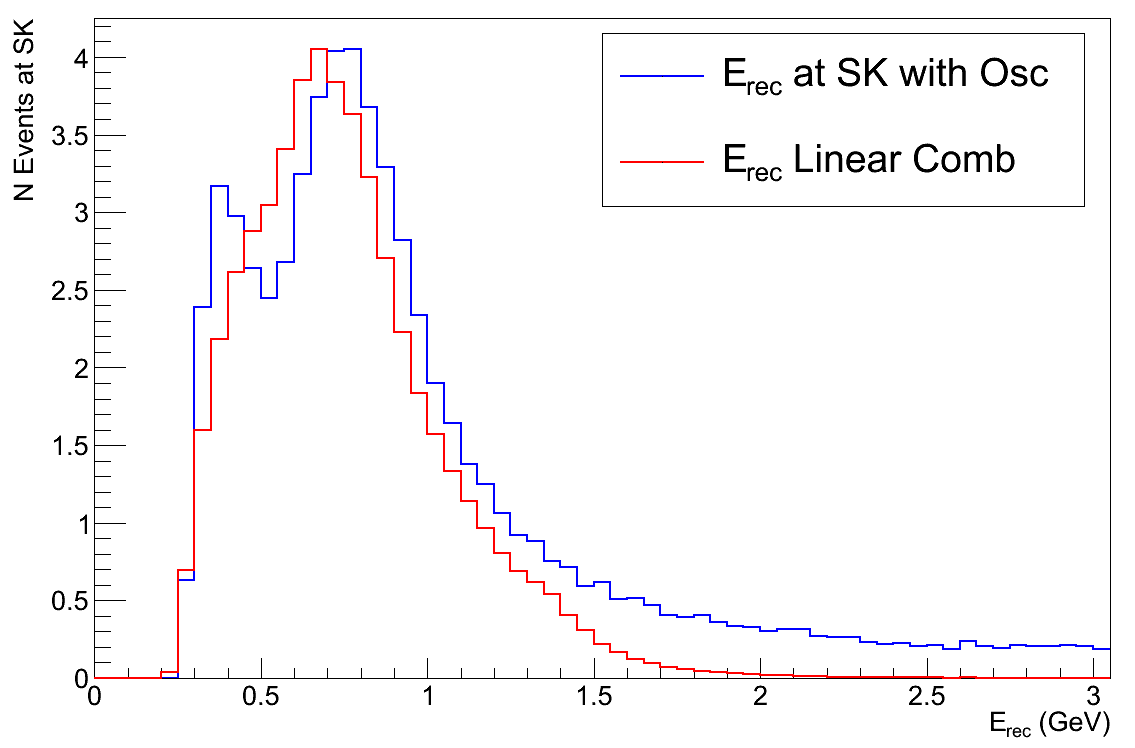}
\end{center}
\caption{The Super-K $E_{rec}$ distributions and \nuprismlite $E_{rec}$ predictions corresponding to the flux fits in Figure~\ref{fig:fluxesevents_smooth} (left column) are shown prior to applying the $\delta\left(E_{rec}\right)$ correction factor.}
\label{fig:fluxesevents2}
\end{figure}

The final expression for the \nuprismlite prediction for the Super-K event rate is then
\begin{equation}
\begin{split}
N&^{SK} \left(E_{rec};\theta_{23},\Delta m^2_{32}\right)= \delta\left(E_{rec}\right)+B^{SK}\left(E_{rec};\theta_{23},\Delta m^2_{32}\right) \\
& +\sum_{i=1}^{30}\sum_{p,\theta}c_i\left(\theta_{23},\Delta m^2_{32}\right)\left(N^{\nu P}_{i,p,\theta}-B^{\nu P}_{i,p,\theta}\right) \\
& \times\frac{\epsilon^{SK}_{p,\theta}}{\epsilon^{\nu P}_{i,p,\theta}}M_{i,p,\theta}\left(E_{rec}\right),
\end{split}
\end{equation}
where $N^{SK}\left(E_{rec}\right)$ and $N^{\nu P}_{i,p,\theta}$ are the number of expected events in Super-K $E_{rec}$ bins and \nuprismlite off-axis angle, muon momentum, and muon angle bins, respectively, $B^{SK}\left(E_{rec}\right)$ and $B^{\nu P}_{i,p,\theta}$ are the corresponding number of background events in these samples, and $\epsilon^{SK}_{p,\theta}$ and $\epsilon^{\nu P}_{i,p,\theta}$ are the efficiencies in each detector. The final correction factor, $\delta\left(E_{rec}\right)$, accounts for any residual differences between the \nuprismlite prediction and the Super-K event rate predicted by the Monte Carlo simulation. These are mostly due to the previously described imperfect flux fitting, and the fact that \nuprismlite is not sensitive to neutrino energies above $\sim 1.5$~GeV since most muons at that energy are not contained within the inner detector.
Comparisons of the Super-K event rate and the \nuprismlite prediction for Super-K prior to applying the $\delta\left(E_{rec}\right)$ correction factor are given in Figure~\ref{fig:fluxesevents2}.

%\begin{figure}[h!]
%\begin{center}
%      \includegraphics[width=8cm] {figures/FittedFlux_dm2_2_56_theta23_0_61.png}
%      \includegraphics[width=8cm] {figures/FittedFlux_dm2_2_41_theta23_0_48.png}
%      \includegraphics[width=8cm] {figures/FittedFlux_dm2_2_26_theta23_0_41.png}
%\end{center}
%\caption{Fits of the \nuprismlite flux slices to oscillated Super-K fluxes are shown for three different sets of $\left(\theta_{23},\Delta m^2_{32}\right)$: top - (0.61, $2.56*10^{-3}$), middle - (0.48, $2.41*10^{-3}$), and bottom - (0.41, $2.26*10^{-3}$).}
%\label{fig:fluxesevents}
%\end{figure}

\begin{figure}[htpb]
\begin{center}
  \includegraphics[width=0.45\textwidth] {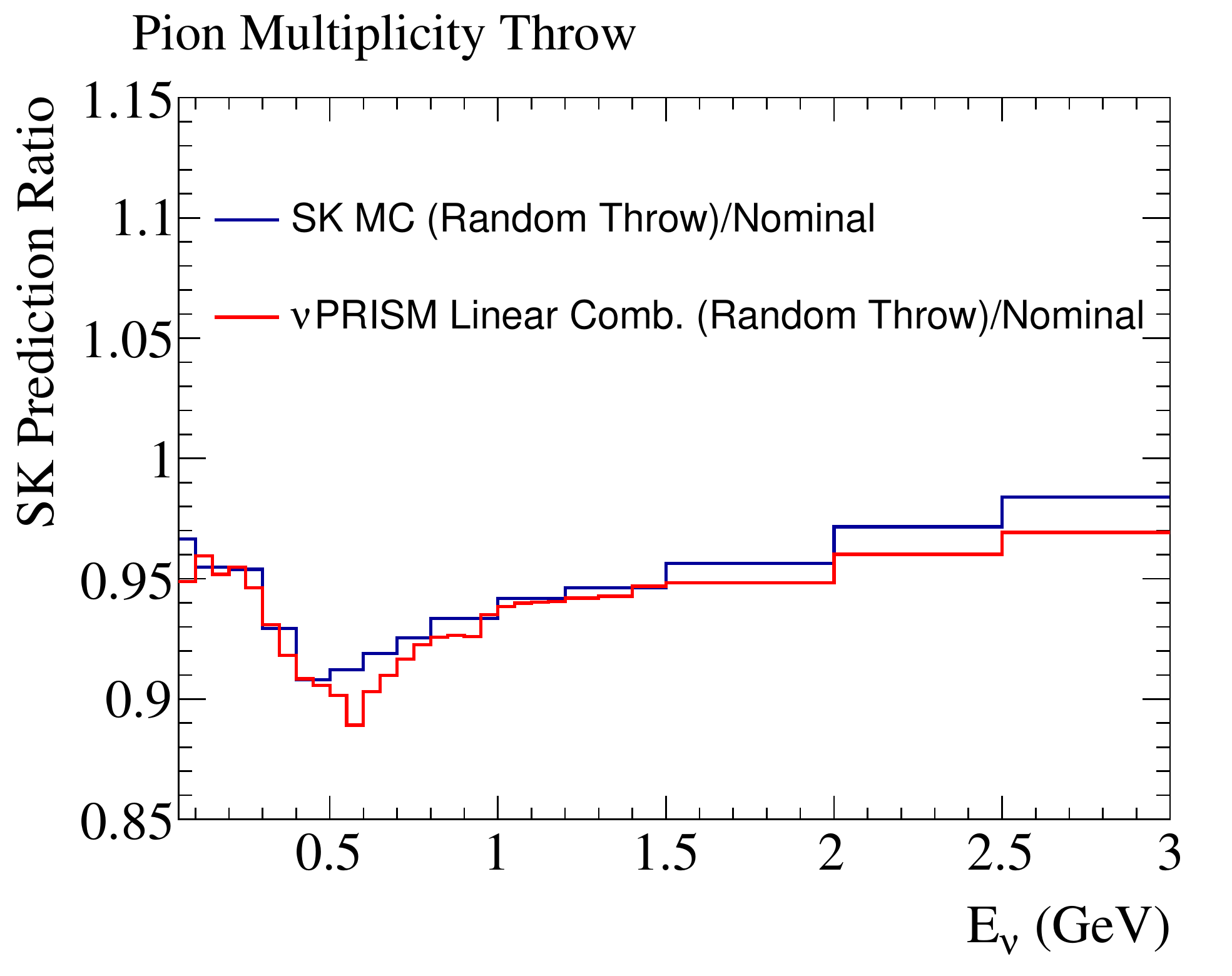}
  \includegraphics[width=0.45\textwidth] {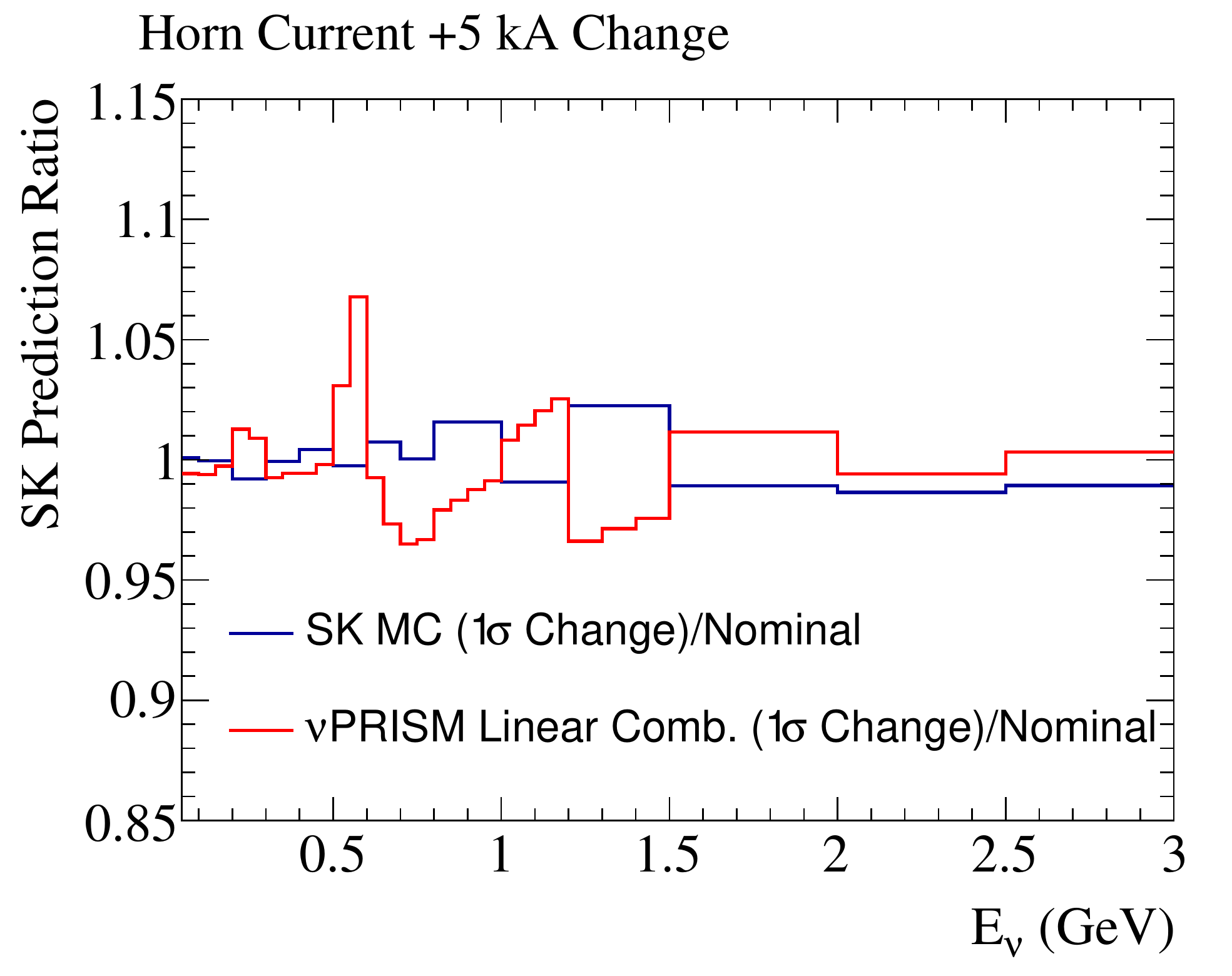}
  \includegraphics[width=0.45\textwidth] {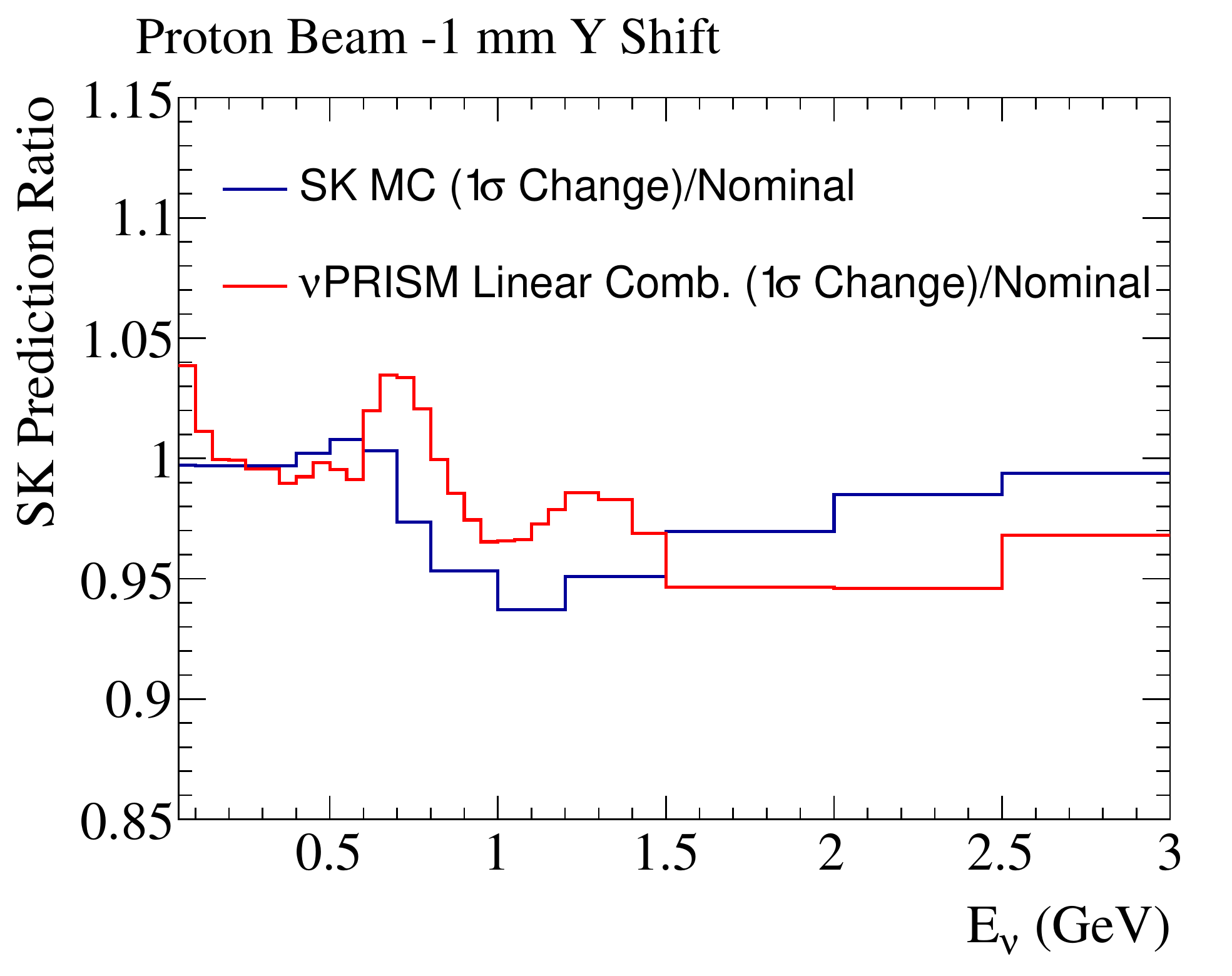}
\end{center}
\caption{Systematic uncertainties on the neutrino flux prediction due to pion production (top), horn current (middle), and proton beam y-position (bottom) are shown.}
\label{fig:fluxerrors}
\end{figure}

\begin{figure}[htpb]
\begin{center}
  \begin{minipage}[t]{.45\textwidth}
    \begin{center}
      \includegraphics[width=\textwidth] {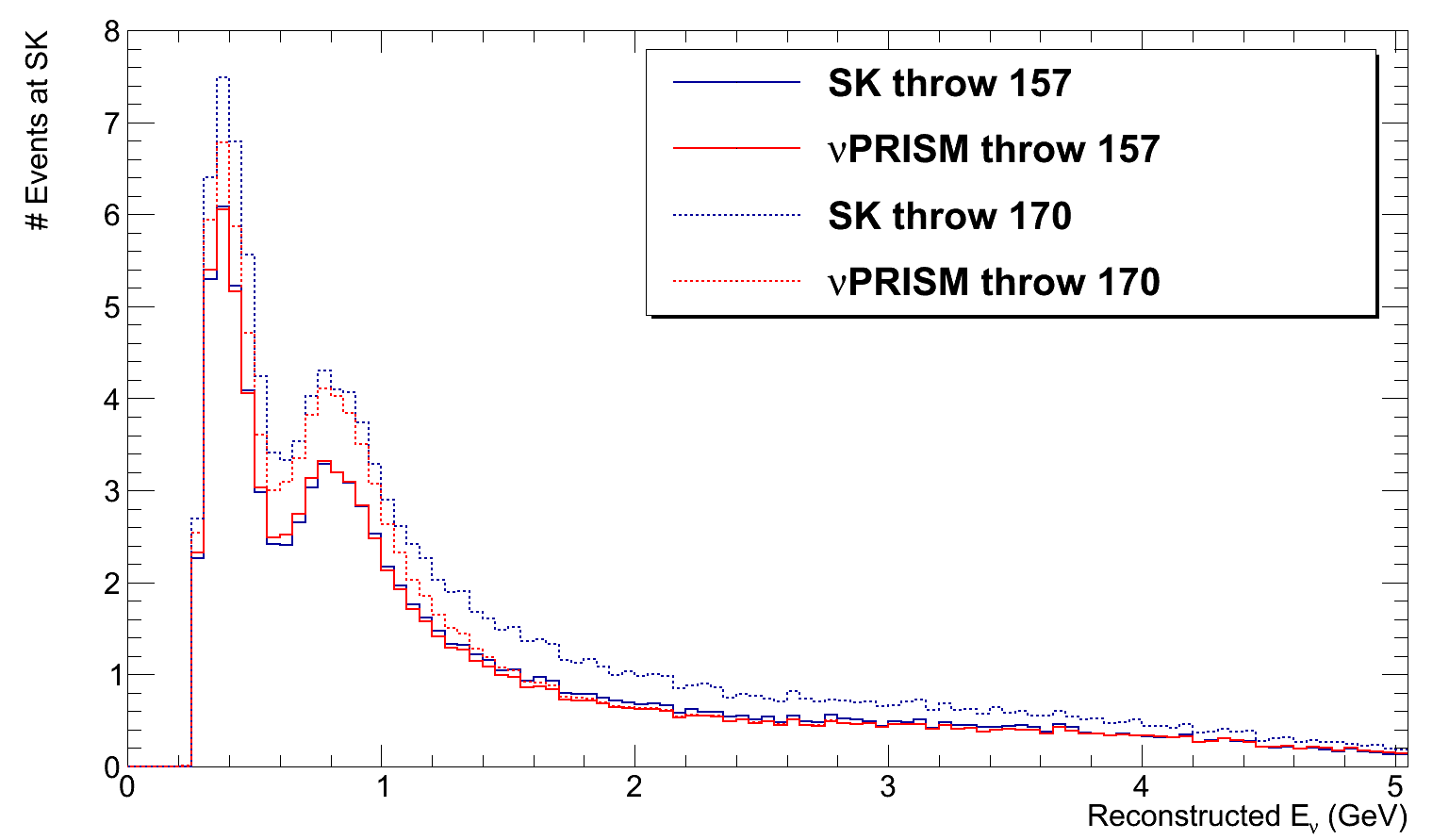}
    \end{center}
  \end{minipage}
  \begin{minipage}[t]{.45\textwidth}
    \begin{center}
      \includegraphics[width=\textwidth] {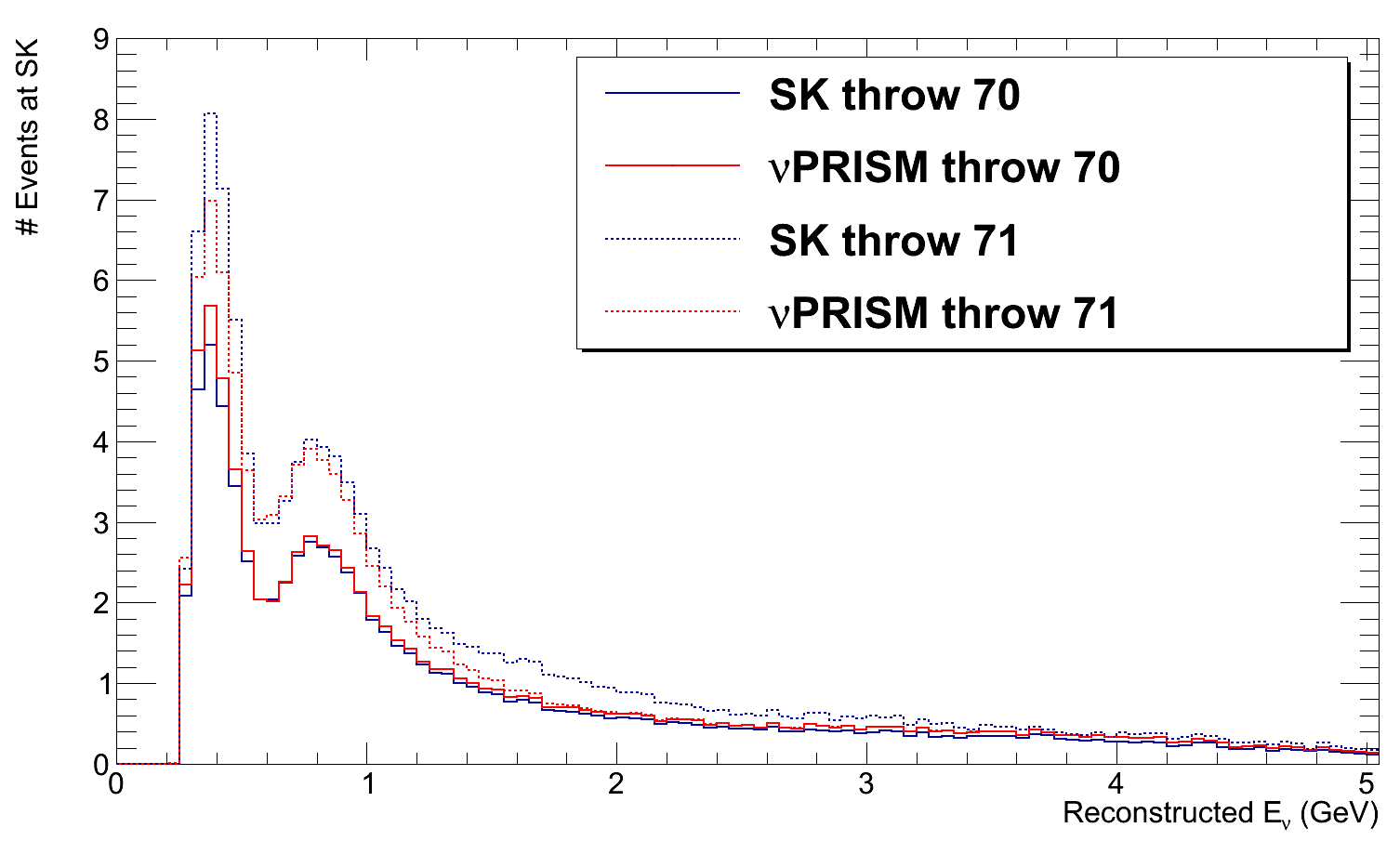}
    \end{center}
  \end{minipage}
\end{center}
\caption{Variations in the Super-K $E_{rec}$ spectrum and the corresponding \nuprismlite prediction are shown for 4 throws of all the flux and cross section parameters. Significant correlations exist between the the near and far detector, which help to reduce the systematic uncertainty.}
\label{fig:erecerrors}
\end{figure}

The \nuprism technique effectively shifts uncertainties in neutrino cross section modeling into flux prediction systematic uncertainties. This is quite helpful in oscillation experiments since many flux systematic uncertainties cancel, and the important physical processes in the flux prediction, the hadronic scattering, can be directly measured by dedicated experiments using well characterized proton and pion beams. Figure~\ref{fig:fluxerrors} shows the effect of a few selected flux uncertainties on the Super-K energy spectrum and the \nuprismlite linear combination. The largest flux uncertainty is due to pion production in proton-carbon interactions, but this uncertainty mostly cancels when applied at both the near and far detector. The more problematic uncertainties are those that affect the off-axis angle, such as horn current and proton beam positioning, since these effects will impact Super-K and the \nuprismlite linear combinations differently. Figure~\ref{fig:erecerrors} shows four examples of how the Super-K $E_{rec}$ distribution and the corresponding \nuprismlite predicted distribution vary for different throws of all the flux and cross section systematic uncertainties. The predicted spectra from the nuPRISM linear combination closely tracks the true spectrum at SK, indicating a correlated effect from most systematic parameters on the nuPRISM linear combination and SK event rates.

\begin{figure}[htpb]
\begin{center}
  \begin{minipage}[t]{.45\textwidth}
    \begin{center}
      \includegraphics[width=\textwidth] {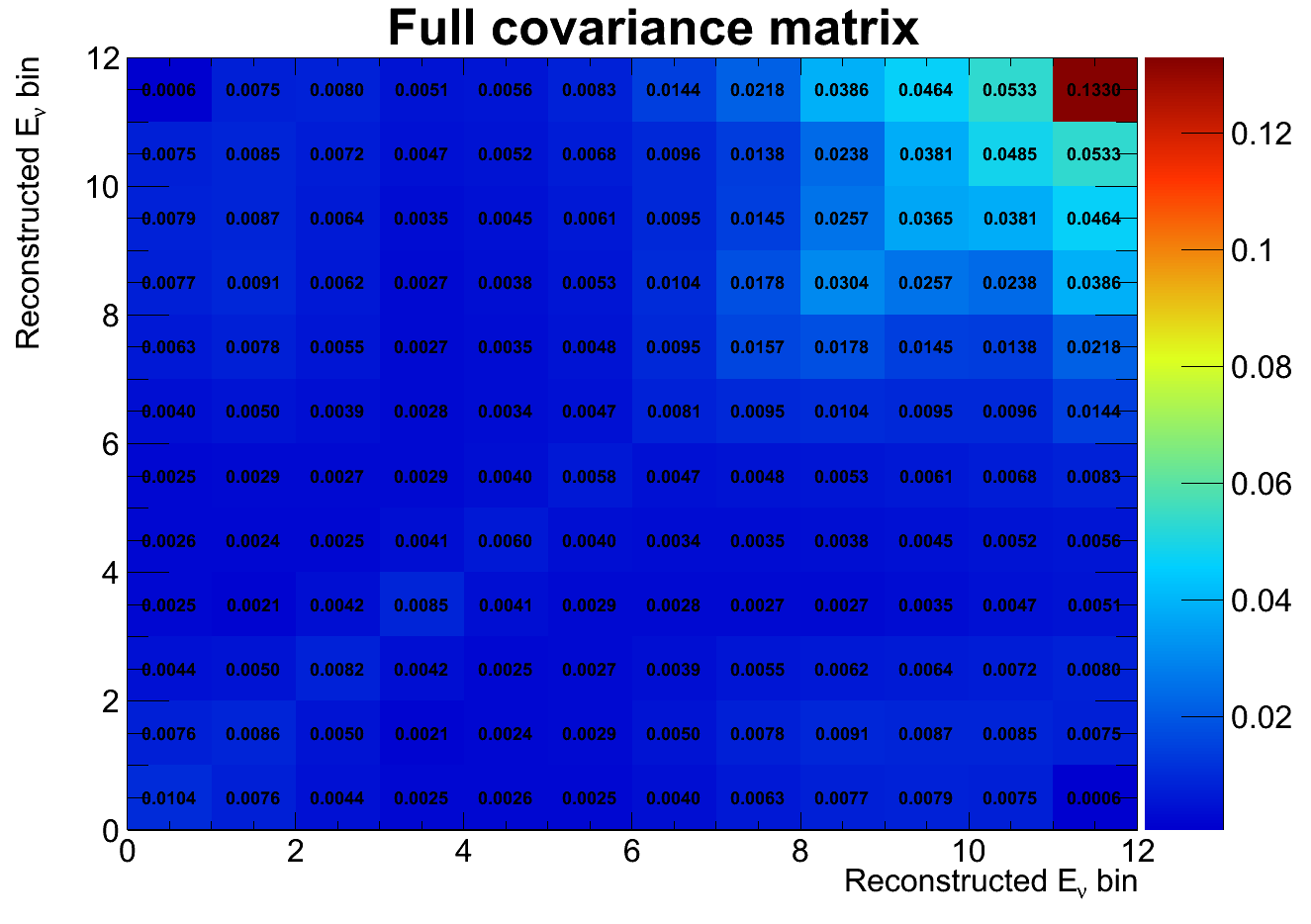}
    \end{center}
  \end{minipage}
  \begin{minipage}[t]{.45\textwidth}
    \begin{center}
      \includegraphics[width=\textwidth] {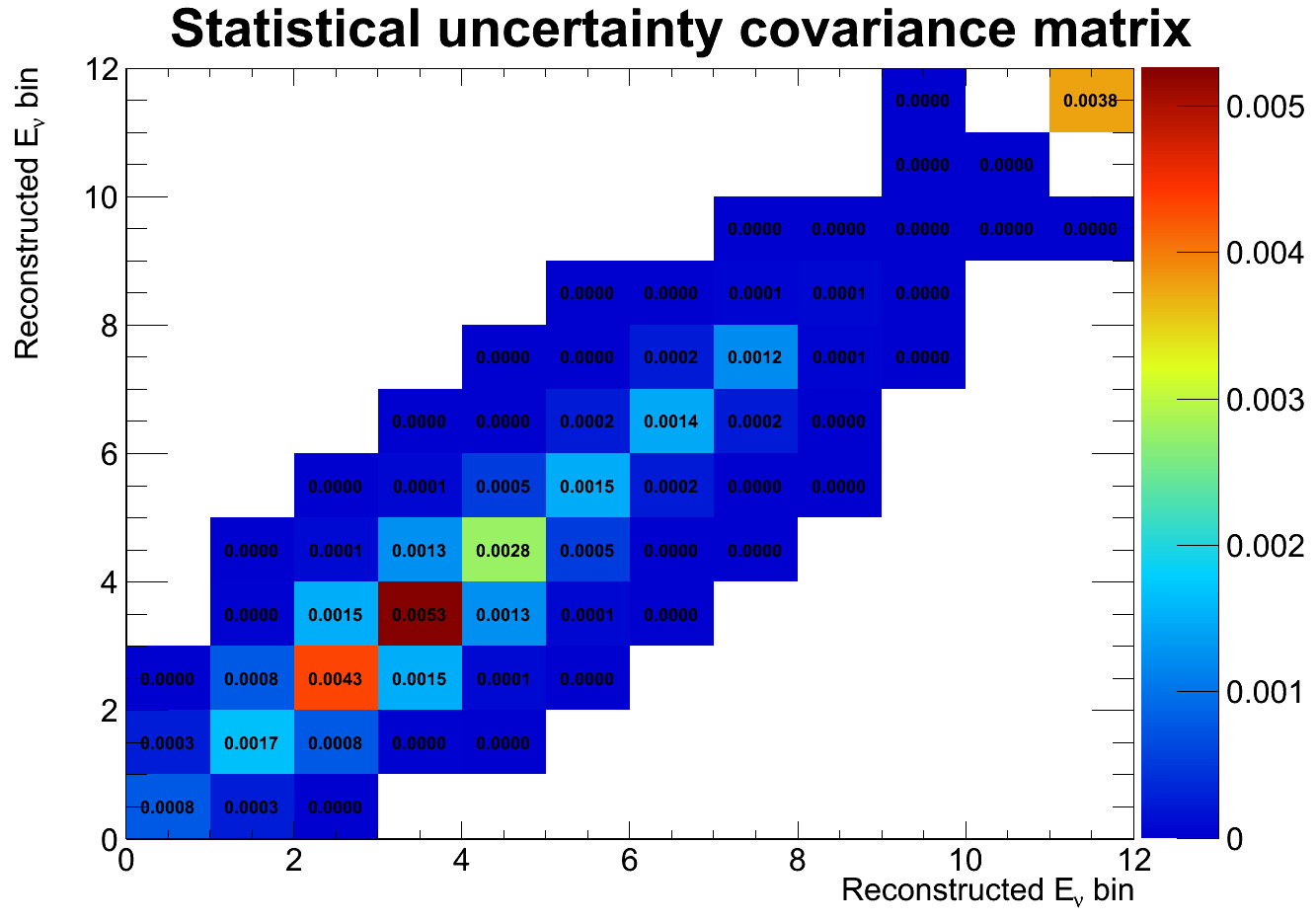}
    \end{center}
  \end{minipage}
  \begin{minipage}[t]{.45\textwidth}
    \begin{center}
      \includegraphics[width=\textwidth] {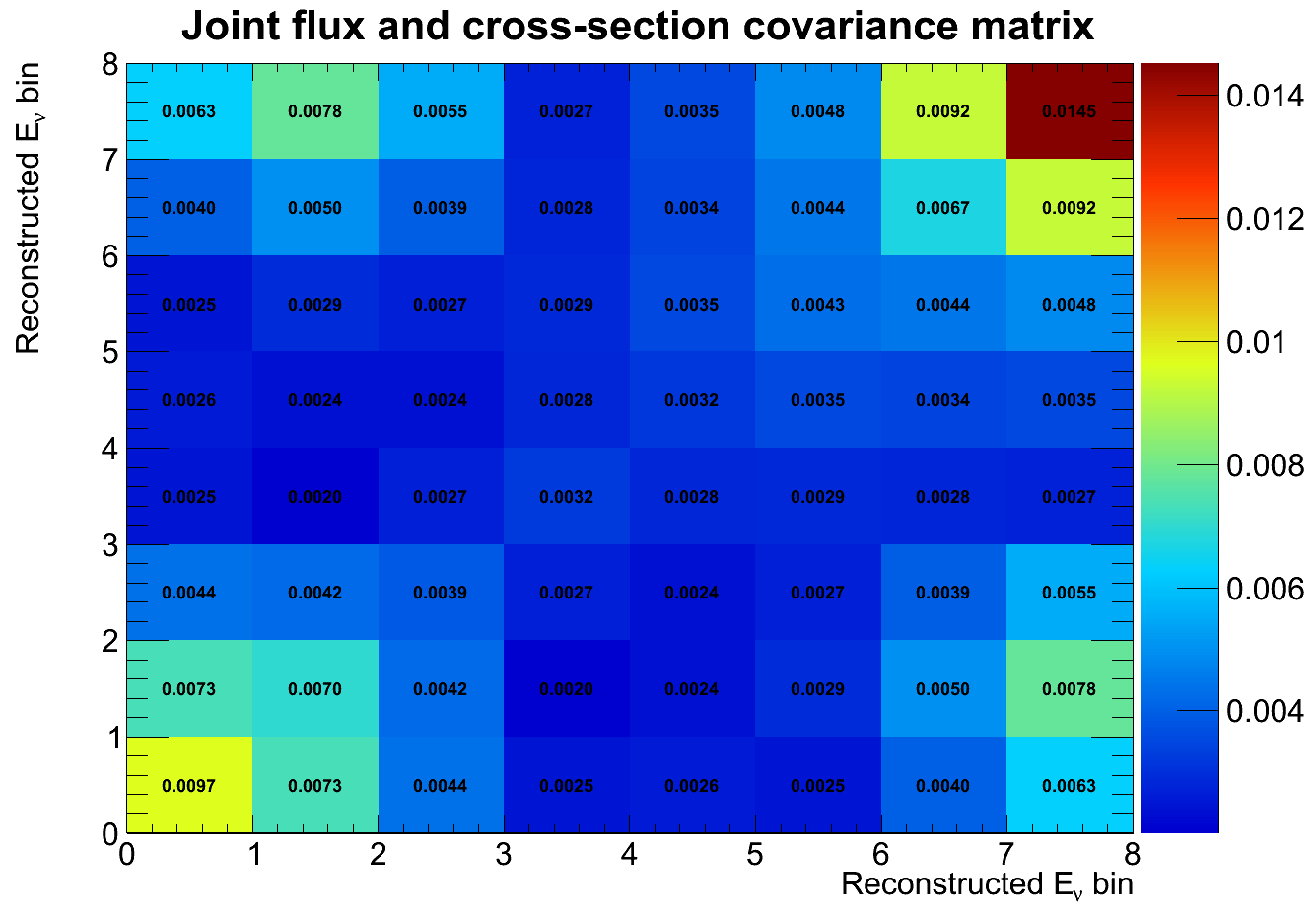}
    \end{center}
  \end{minipage}
  \begin{minipage}[t]{.45\textwidth}
    \begin{center}
      \includegraphics[width=\textwidth] {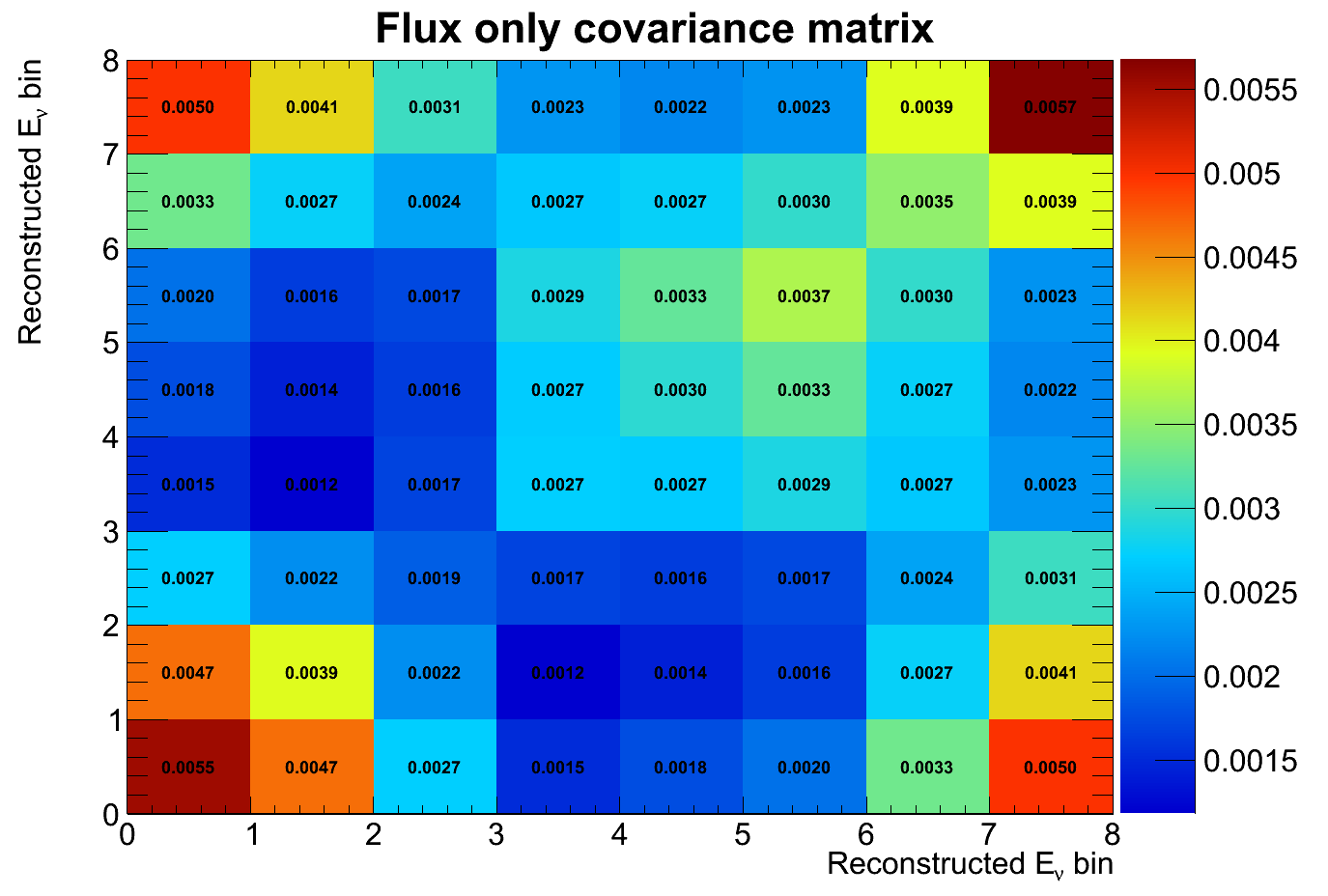}
    \end{center}
  \end{minipage}
\end{center}
\caption{Covariance matrices are shown (from top to bottom) for the total, statistical, systematic, and flux only uncertainties. The bin definitions (in GeV) are 0: (0.0,0.4), 1: (0.4,0.5), 2: (0.5,0.6), 3: (0.6,0.7), 4: (0.7,0.8), 5: (0.8,1.0), 6: (1.0,1.25), 7: (1.25,1.5), 8: (1.5,3.5), 9: (3.5,6.0), 10: (6.0,10.0), 11: (10.0,30.0)}
\label{fig:covmat}
\end{figure}

The final covariance matrices are shown in Figure~\ref{fig:covmat}. The largest errors are at high energies where no \nuprismlite events are present due to the smaller diameter of the detector relative to Super-K. In this region, the Super-K prediction is subject to the full flux and cross section uncertainties with no cancelation at the near detector. Similarly, at energies below 400 MeV the errors get larger since the current 4$^\circ$ upper bound in off-axis angle prohibits the \nuprismlite flux fit from matching the Super-K spectrum at low energies.

\begin{figure}[htpb]
\begin{center}
  \begin{minipage}[t]{.45\textwidth}
    \begin{center}
      \includegraphics[width=\textwidth] {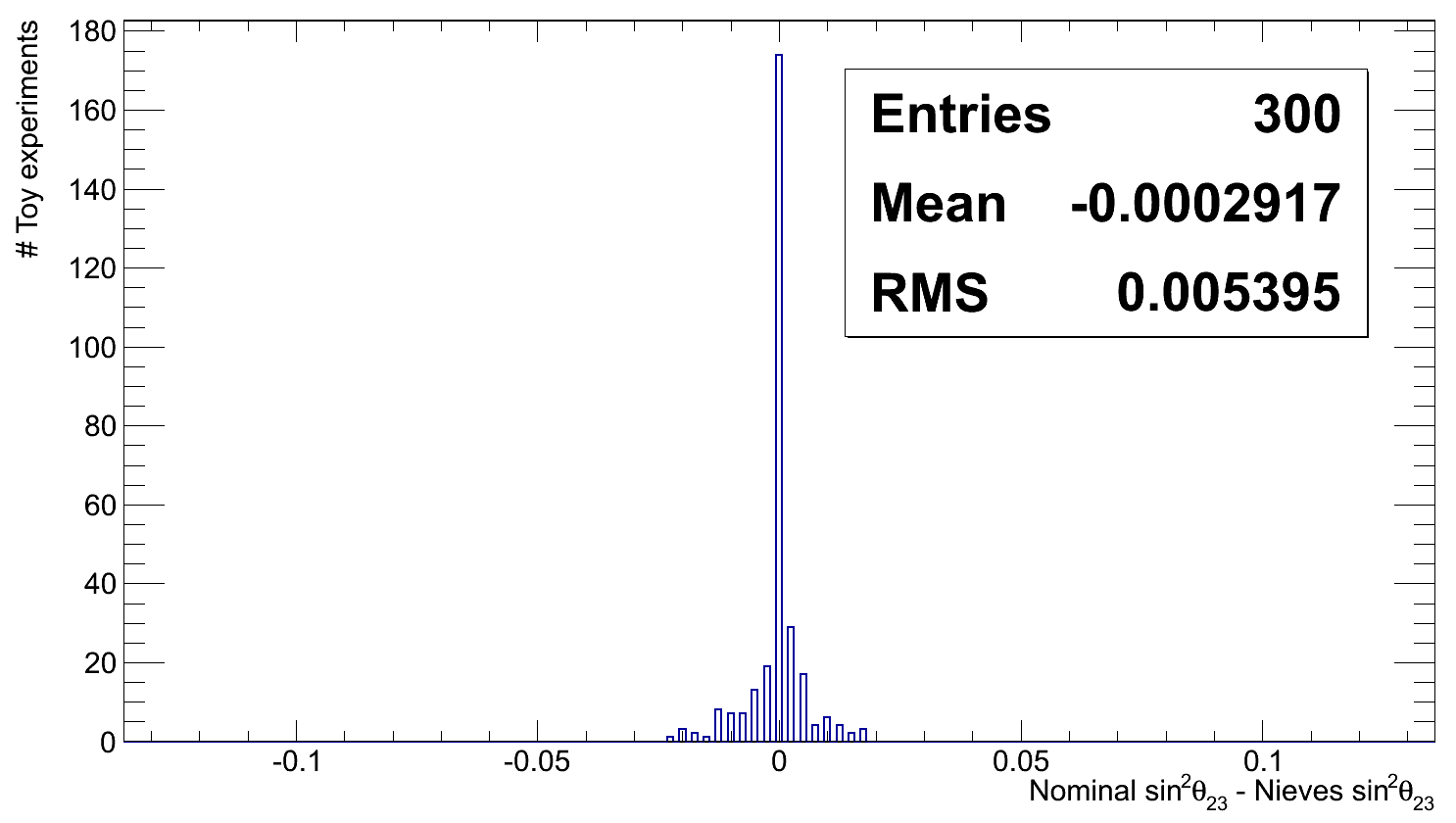}
    \end{center}
  \end{minipage}
  \begin{minipage}[t]{.45\textwidth}
    \begin{center}
      \includegraphics[width=\textwidth] {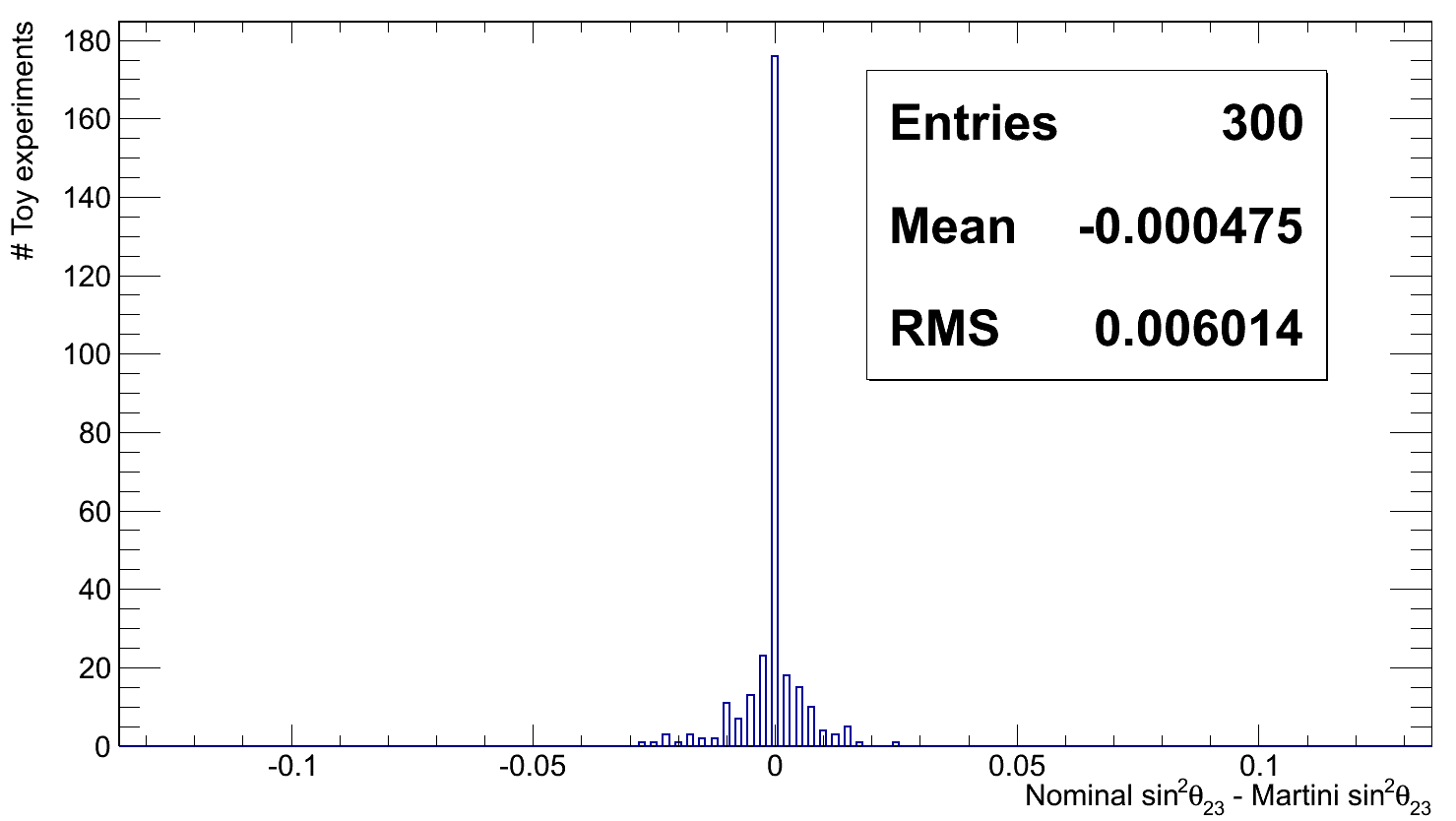}
    \end{center}
  \end{minipage}
\end{center}
\caption{The variation in the measured $\sin^2\theta_{23}$ due to multinucleon effects in the \nuprismlite \numu analysis are shown. For the Nieves and Martini fake datasets, the RMS produces 1.0\% and 1.2\% uncertainties, respectively, with no measurable bias. This is a large improvement over the standard T2K results shown in Figure~\ref{fig:nievesmartini}}
\label{fig:numuresults}
\end{figure}

Using the \nuprismlite covariance matrices shown in Figure~\ref{fig:covmat} in place of those produced by ND280, the standard T2K \numu disappearance oscillation analysis is repeated. The results are shown in Figure~\ref{fig:numuresults}. As expected, the \nuprismlite analysis is largely insensitive to cross section modeling. Replacing the default new model with the Nieves multinucleon model now produces a 1.0\% uncertainty in $\sin^2\theta_{23}$, and the corresponding Martini uncertainty is 1.2\%. More importantly, this uncertainty is now constrained by data rather than a pure model comparison. These uncertainties are expected to be further reduced as the flux fits are improved, and \nuprismlite constraints on NC backgrounds and information from ND280 are incorporated into the analysis.

\clearpage

\subsection{\nuprismlite 1-Ring e-like Ring Measurements}

Single ring e-like events in \nuprismlite at an off-axis angle of 2.5$^\circ$ in principle provide a reliable estimate of the $\nu_e$ appearance background at SK, since the near-to-far extrapolation correction is small. This includes beam $\nu_e$, NC$\pi^0$, and NC single $\gamma$ (NC$\gamma$) backgrounds with production cross section and detection efficiency in water folded in. For a $\nu_e$ background study with better than $\sim$10\% precision, more careful studies are required: for example, the $\gamma$ background from outside the detector scales differently between the near and far detectors due to the differences in surface to volume ratio. Contributions from CC backgrounds, e.g. CC$\pi^0$ events created outside the detector, would also be different between near and far detector due to oscillation. Careful identification of each type of single ring e-like event is required. As described below, the \nuprismlite capability of covering wide off-axis ranges makes such a study possible. It also enables relative cross section measurements between $\nu_e$ and $\nu_\mu$, which are likely to be limiting systematic uncertainties for measuring CP violation.

The \nuprismlite detector will also provide a unique and sensitive search for sterile neutrinos in the $\nu_\mu\rightarrow\nu_e$ channel, and eventually the $\nu_\mu\rightarrow\nu_\mu$ channel, particularly when ND280 is incorporated into the analysis. The 1km location of nuPRISM for the off-axis peak energies of 0.5-1.0GeV matches the oscillation maximum for the sterile neutrinos hinted by LSND and MiniBooNE. The presence or absence of an excess of $\nu_e$ events as a function of off-axis angle will provide a unique constraint to rule out many currently proposed explanations of the MiniBooNE excess, such as feed-down in neutrino energy due to nuclear effects. The off-axis information also allows for a detailed understanding of the backgrounds, since they have a different dependence on off-axis angle than the oscillated signal events.

%The backgrounds inND280 tracker $\nu_e$ analysis is currently dominated by the background coming from outside. The 2 meters of outer veto and 2m of fiducial cut from the wall as in SK analysis would greatly reduce these main backgrounds.
\begin{figure}[htpb]
\centering\includegraphics[width=8cm,angle=0]{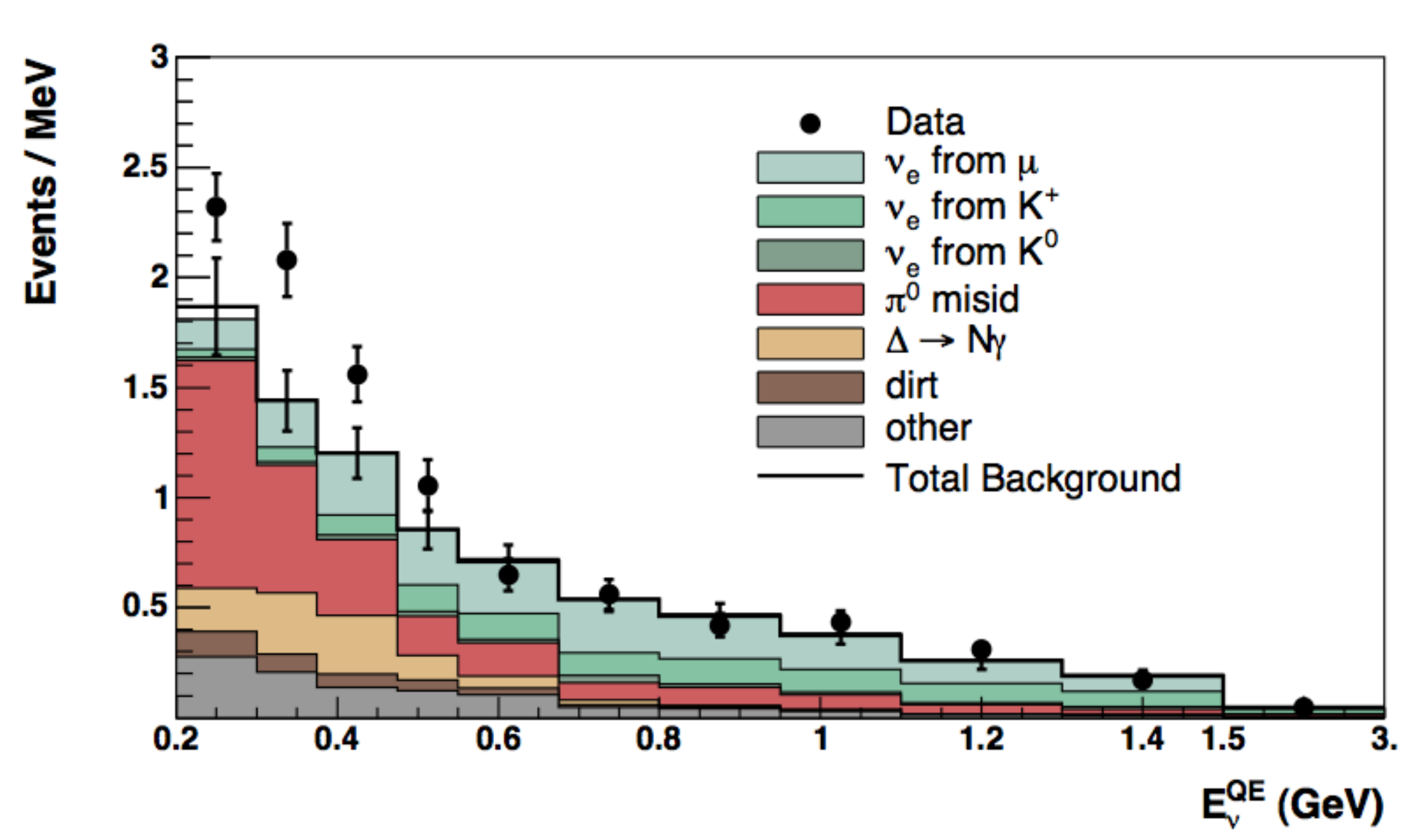}
\caption{Reconstructed neutrino energy distribution for the $\nu_e$ appearance analysis of MiniBooNE
\cite{miniboone-nue}.}
% MiniBooNE collaboration, Phys.Rev.Lett.102(2009)101802
\label{fig:miniboone}
\end{figure}
Figure~\ref{fig:miniboone} shows the single ring e-like events observed by MiniBooNE. There are several sources of events:
\begin{itemize}
  \item Beam $\nu_e$ from muons and kaons
  \item NC$\pi^0$ with one of the photons missed
  \item NC$\gamma$ ($\Delta \rightarrow N \gamma$) 
  \item "Dirt" events: background $\gamma$ coming from outside
  \item Others, such as CC events with $\mu$ misidentified as electron
  \item Possible sterile neutrino contribution causing $\nu_\mu\rightarrow\nu_e$ oscillation
\end{itemize}
There is a significant discrepancy between data and the Monte Carlo prediction. For precision $\nu_e$ appearance studies, such as CP violation, it is essential to understand the origin of this discrepancy.

\subsubsection{Beam $\nu_e$ and $\nu_e$ cross section study}\label{sec:nuexsec}
The beam $\nu_e$ represents only 1\% of the total neutrino flux and about 0.5\% at the off-axis peak energy at $E_\nu$=600MeV. Thanks to the excellent $\mu$/e particle identification and $\pi^0$ suppression in water Cherenkov detectors when using fiTQun, the $\nu_\mu$ background is expected to be suppressed, similar to the suppression seen at Super-K. Since the beam $\nu_e$'s originate from three body decays of muons and kaons, their off-axis dependence is more mild than the dependence seen in the $\nu_\mu$ flux. By taking advantage of the steep off-axis angle dependence of the $\nu_\mu$ flux, it is possible to study background contamination in detail. For example, the $\nu_\mu$ backgrounds are largely suppressed compared to beam $\nu_e$ at an off-axis angle larger than 3 degrees. The beam $\nu_e$ events at \nuprismlite provide an opportunity to precisely study $\nu_e$ cross sections, for which there is currently very little data available. The cross section difference between $\nu_e$ and $\nu_\mu$, which does not cancel in the near to far detector extrapolation in $\nu_\mu \rightarrow \nu_e$ appearance, is considered to be an eventual limitation of the CP violation sensitivity~\cite{nuSTORM}. 
%nuSTORM - Neutrinos from STORed Muons: Proposal to the Fermilab PAC, arXiv:1308.6822
The differences in the $\nu_e$ and $\nu_\mu$ cross sections come from kinematical phase space differences due to the difference in mass between electron and muons, radiative corrections, possible second class currents, which also depend on lepton mass, and nuclear effects~\cite{nue-numu-cross-section}.
% 	Melanie Day, Kevin S. McFarland, Differences in Quasi-Elastic Cross-Sections of Muon and Electron Neutrinos Phys.Rev. D86 (2012) 053003 
%The nuSTORM project proposes to study the $\nu_e$/$\nu_\mu$ cross section ratio down to 1-2\% level using neutrinos from a muon storage ring. The advantage of nuSTORM is that the flux ratio between $\nu_e$ and $\nu_\mu$ is well known from muon decays in the ring. 
%In T2K, the current systematic uncertainty on the beam $\nu_e$ rate at Super-K, constrained by the ND280 $\nu_\mu$ measurement, is 4.8\%, without including the cross section difference. Considering the farther distance ($\sim$1km) with less near-to-far extrapolation error and the measurement of different off-axis $\nu_\mu$ rate for \nuprismlite, we would expect significantly better than 4.8\% in flux systematics for the relative cross section measurement between $\nu_e$ and $\nu_\mu$. 
%In the next several months, we will study how much more we can reduce this relative $\nu_e$ and $\nu_\mu$ flux uncertainties.

\nuprismlite provides a unique method for canceling the flux differences between $\nu_e$ and $\nu_\mu$. Using a technique similar to that used in the \nuprismlite \numu disappearance analysis, it is possible to use linear combinations of \numu measurements at different off-axis angles to reproduce the shape of the intrinsic \nue flux in the large off-axis angle section of \nuprism:
\begin{equation}
\Phi_{\nu_e}(E_\nu) = \Sigma c_i \Phi^i_{\nu_\mu}(E_\nu),
\end{equation}
where $\Phi_{\nu_e}(E_\nu)$ is the \nuprism $\nu_e$ flux of interest, $\Phi^i_{\nu_\mu}(E_\nu) $ is the $\nu_\mu$ flux at the $i^{th}$ off-axis position and $c_i$ is the weight factor for the $i^{th}$ off-axis position. Using this combination, the ratio of the \nue and \numu double differential cross sections in momentum and angle can be directly measured, averaged over the $\nu_e$ flux spectrum.

\begin{figure}[htpb]
\begin{center}
  \begin{minipage}[t]{.47\textwidth}
    \begin{center}
      \includegraphics[width=\textwidth] {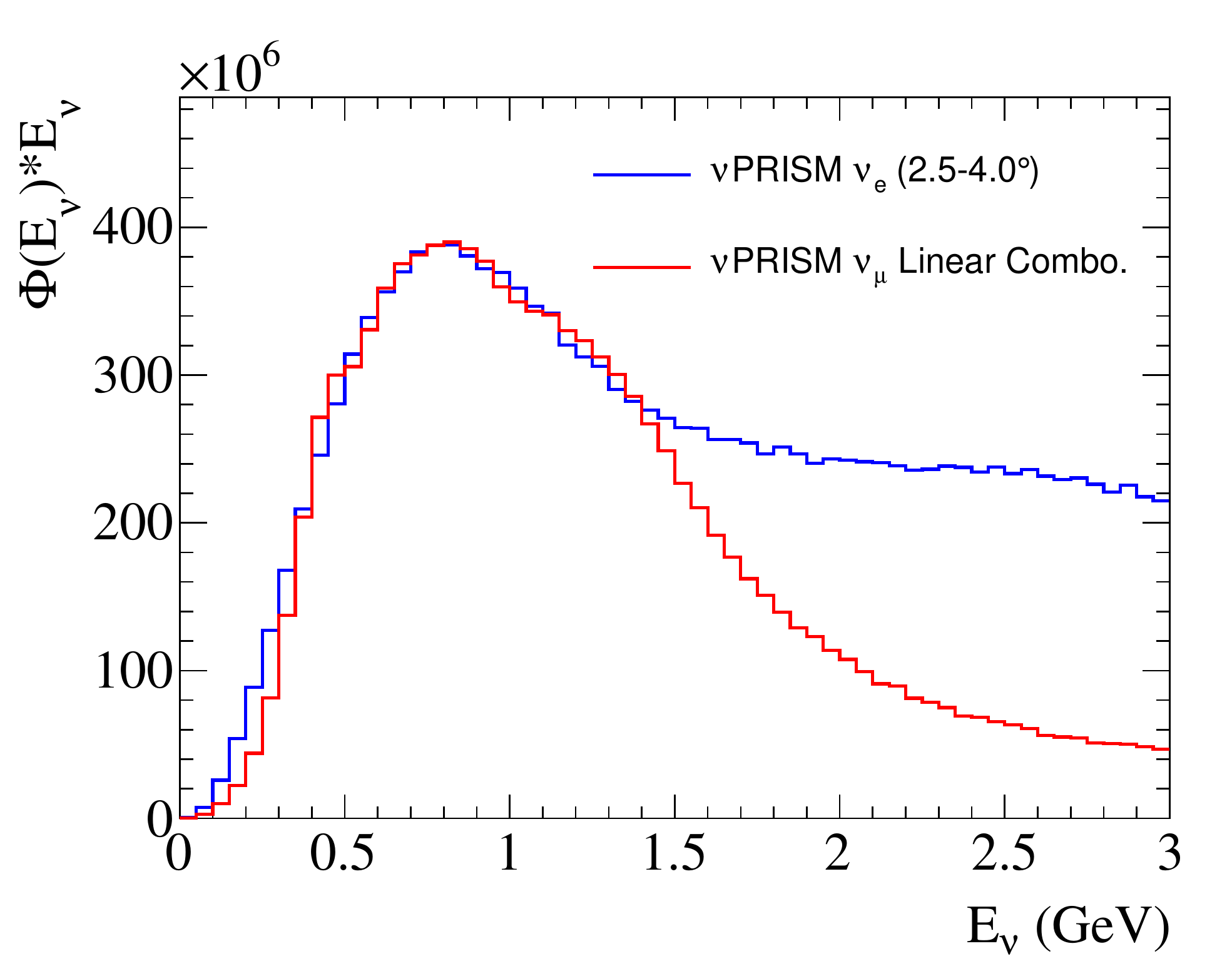}
    \end{center}
  \end{minipage}
  \begin{minipage}[t]{.47\textwidth}
    \begin{center}
      \includegraphics[width=\textwidth] {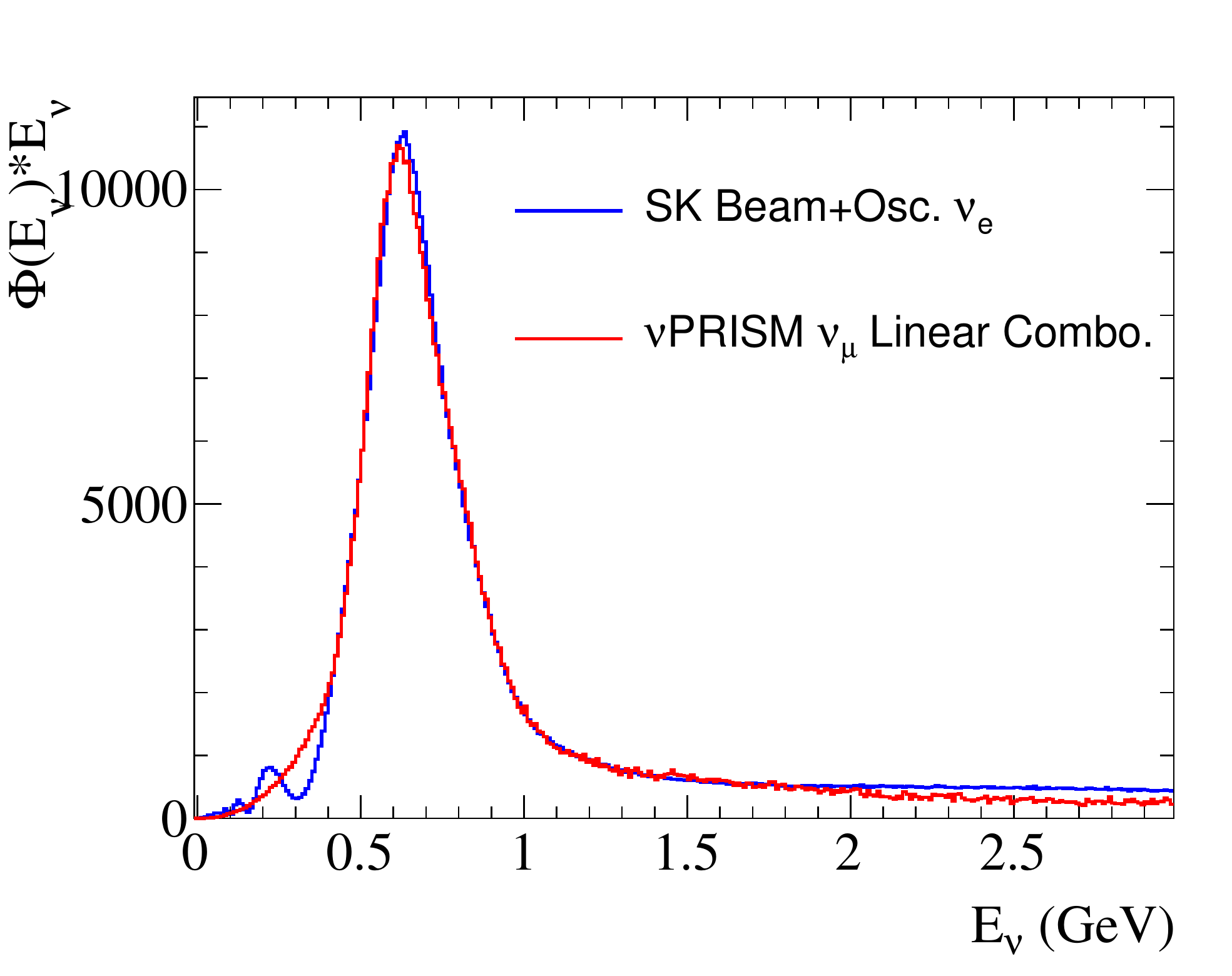}
    \end{center}
  \end{minipage}
\end{center}
\caption{Fits of the off-axis \nuprism $\nu_{\mu}$ fluxes to the \nuprism $2.5^{\circ}-4.0^{\circ}$ off-axis $\nu_e$ flux (top) and the oscillated+intrisic beam $\nu_e$ at SK (bottom) assuming sin$^{2}2\theta_{13}$=0.094, $\delta_{cp}$=0, $\Delta m^{2}_{32}=2.4\times10^{-3}$eV$^{2}$ and sin$^{2}\theta_{23}$=0.5. }
\label{fig:nuefluxfits}
\end{figure}

Fig.~\ref{fig:nuefluxfits} shows that the \nuprism $2.5^{\circ}-4.0^{\circ}$ off-axis $\nu_e$ flux can be reproduced by the linear combination of $\nu_{\mu}$ fluxes for the 0.3-1.5 $GeV$ energy range.  Above 1.5 $GeV$ the $\nu_{e}$ flux cannot be produced since the fall-off of the $\nu_{\mu}$ fluxes is steeper.  However, this region will have little impact for the ratio measurement for a couple of reasons.  First, Fig.~\ref{fig:nuefluxfits} shows the flux multiplied by the energy to approximate the effect of the cross section, but the cross section for CC interactions producing no detectable pions is growing more slowly than this linear dependence and the rate from the high energy flux will be lower than it appears in the figure.  Second, the analysis will be applied in the limited lepton kinematic range where the \nuprism muon acceptance is non-zero, cutting out forward produced high momentum leptons.  This will also suppress the contribution from the high energy part of the flux.  

\subsubsection{Predicting oscillated $\nu_e$ for the appearance measurement} 

As discussed in the previous section, the cross section ratio of $\sigma_{\nu_{e}}/\sigma_{\nu_{\mu}}$ can be measured using beam $\nu_{e}$ and $\nu_{\mu}$ interaction candidates in \nuprism.  The measured cross section ratio can be used to apply the \nuprism extrapolation method to predict the $\nu_{e}$ candidates at SK for the appearance measurement.  Following the procedure used for the disappearance analysis, the oscillated+intrinsic beam $\nu_{e}$ flux is described by a linear combination of the \nuprism off-axis $\nu_{\mu}$ fluxes:
\begin{equation}
\begin{split}
\Phi^{SK}_{\nu_{\mu}}(E_{\nu})P_{\nu_{\mu}\rightarrow\nu_{e}}(E_{\nu}|\theta_{13},\delta_{cp},...)+\Phi^{SK}_{\nu_{e}}(E_{\nu})  \\
 = \sum c_i(\theta_{13},\delta_{cp},...) \Phi^i_{\nu_\mu}(E_\nu).
\end{split}
\end{equation}

$\Phi^{SK}_{\nu_{\mu}}(E_{\nu})$ and $\Phi^{SK}_{\nu_{e}}(E_{\nu})$ are the predicted $\nu_{\mu}$ and $\nu_{e}$ fluxes at SK in the absence of oscillations. $P_{\nu_{\mu}\rightarrow\nu_{e}}$ is the $\nu_{\mu}$ to $\nu_{e}$ oscillation probability.  $\Phi^i_{\nu_\mu}(E_\nu)$ is the $i^{th}$ off-axis $\nu_{\mu}$ flux in \nuprism and the $c_i$ are the derived coefficients that depend on the oscillation hypothesis being tested.  Fig.~\ref{fig:nuefluxfits} shows the level of agreement that can be achieved between the linear combination of \nuprism fluxes and the predicted SK $\nu_{e}$ flux for a particular oscillation hypothesis.  The agreement is excellent between 0.4 and 2.0 GeV.  Below 0.4 GeV, the second oscillation maximum is not reproduced, but the rate from this part of the flux is small.

Using the derived $c_i$ coefficients, the measured muon $p,\theta$ distributions from \nuprism are used to predict the SK $p,\theta$ distribution for the $\nu_e$ flux.  An additional correction must be applied to correct from the predicted muon distribution for $\nu_{\mu}$ interactions to the predicted electron distribution for $\nu_e$ interactions.  This correction is derived from the cross section models which are constrained by the ratio measurement described in the previous section.

\subsubsection{Backgrounds from $\nu_\mu$'s}
The backgrounds from $\nu_\mu$ comes from NC$\pi^0$ events with one $\gamma$ missed, NC$\gamma$ events ($\Delta \rightarrow N \gamma$), CC events with e/$\mu$ mis-ID, $\gamma$'s coming from $\nu$ (mainly $\nu_\mu$) interaction outside the detector (dirt or sand events). Because the $\nu_\mu$ energy spectrum changes dramatically as a function of vertex positions (= off-axis angles) in nuPRISM, these background processes can be studied and verified by comparing their vertex distributions. 

The NC$\pi^0$ rate can be measured by detecting two $\gamma$'s in nuPRISM. By using the hybrid $\pi^0$ technique used in T2K-SK analysis, the $\pi^0$ backgrounds with a missing $\gamma$ can be estimated using the beam $\nu_e$ and Michel electrons as electron samples combined with a Monte Carlo $\gamma$ event. The NC$\pi^0$ rate can also be used to estimate the NC$\gamma$ rate. As mentioned above, dirt/sand background is suppressed by having fully active outer veto detector and the fiducial volume cut. The vertex distribution of the $\nu_e$ events as a function of the distance from the (upstream) wall provides an excellent confirmation of the suppression of the background, as is done in the T2K-SK analysis.

%\subsubsection{Sterile neutrino search}
%The 1km location of nuPRISM for the off-axis peak energies of 0.5-1.0GeV matches the oscillation maximum of the sterile neutrinos hinted by LSND and miniBooNE. One can choose the off-axis position which maximizes the oscillation for given oscillation parameter, and thus enhance the signal. By comparing the different off-axis angles, we can positively identify the signal by its enhancement at the oscillation maximum off-axis position. The off-axis information also helps detailed understanding of the backgrounds, which have different off-axis dependence, in particular for the main beam $\nu_e$ events.
%For 700kW beam for one year ($3\times 10^{20}$ POT), T2K predicts 200 $\nu_\mu$ events at SK. For a 20 tonne fiducial (2m diameter and 6m height)  at 1km, this corresponds to 56,000$\nu_\mu$ events. Even for the $\sin^22\theta_{\mu e}=10^{-3}$, we would expect 56 $\nu_e$ appearance events. The beam $\nu_e$ events at this peak energy is 0.5\% of $\nu_\mu$ or 280 events, and thus this corresponds to 56/$\sqrt{280}$=3.3$\sigma$ significance. More detailed sensitivity studies is presented in the following section.

\subsubsection{Sterile Neutrino Sensitivity}

\def\stme       {\ensuremath{\textrm{sin}^2(2\theta_{\mu e})}\xspace}
\def\dmsqfo     {\ensuremath{\Delta m^{2}_{41}\xspace}}
\def\nue        {\ensuremath{\nu_e}\xspace}
\def\nueb       {\ensuremath{\nub_e}\xspace}
\def\nuenueb    {\ensuremath{\nue\nueb}\xspace}
\def\numu        {\ensuremath{\nu_\mu}\xspace}
\def\numub       {\ensuremath{\nub_\mu}\xspace}

The position of \nuprismlite, at 1~km from the neutrino source, as well as its huge fiducial mass makes this detector an excellent candidate for the studies of non-standard 
short-baseline neutrino oscillations. This section presents an initial, conservative sensitivity of nuPRISM to the so-called LSND anomaly. The LSND and MiniBooNE experiments detect an undetermined excess in their \nue and \nueb channels, which may be explained by sterile neutrino mixing with a 
$\stme \sim 10^{-3}$ and $ \dmsqfo \sim 2eV^2$ in the 3+1 model~\cite{miniboone-nue}. %\cite{steriles_whitepaper}.

Here we present the sensitivity studies for a two different layouts of the \nuprism detector: 3~m radius and 4~m 
radius. We performed our $\nue$ selection analysis considering an exposure of $4.6\times 10^{20}$ 
p.o.t. with a horn configuration enhancing neutrinos and defocusing anti-neutrinos. The possible $\nue$ disappearance due to sterile mixing is neglected as in 
the case of the LSND and MiniBooNE analyses. This is justified by the fact that we have only $1\%$ 
of $\nue$ in the beam and the $\numu \rightarrow \nue$ channel will be dominant. In the case where both 
$\nue$ disappearance and appearance are considered, our current results can be seen as 
lower limits for the mixing angle $\stme$.

We test the simplest sterile neutrino model by adding to the standard three-neutrino parametrization one additional mass state, mainly sterile, with a mass difference relative to the other states of $\Delta m^2_{41}$. 
Since the mixing with the sterile neutrino is dominant at short baselines, such as the \nuprismlite baseline, the new mass state is expected to be much larger ($\sim eV^2$) than the two standard neutrino mass splittings. In such conditions the two-neutrino approximation is valid and provides the following \nue
appearance probability,
\begin{equation}
\begin{split}
 P_{\numu\nue} & = P(\numu \rightarrow \nue) \\
 & = \stme\sin^2\left( 1.27 
\dmsqfo\mbox{[eV$^2$]}\dfrac{L\mbox{[km]}}{E\mbox{[GeV]}}\right),
\end{split}
\label{eq:app}
\end{equation}

where $L$ is the neutrino flight path fixed at $1km$ and $E$ the energy of the neutrinos. $\stme = 
4|U_{e4}|^2|U_{\mu4}|^2$ where $U$ are the new elements in the extended PMNS matrix. We consider an 
analysis on the reconstructed energy ($E_{rec}$) and off-axis angle (OAA) shape informations, so 
both rate and shape are taken into account by building bidimensional binned templates. The expected 
number of background and signal events entering in the $\nue$ selection are shown in Table~\ref{tab:steriles_events} for different oscillation hypothesis and both detector radius cases.

% and the $E_{Rec}$ distributions for 
% five cases of OAA's is shown in Fig. \ref{fig:steriles_erec}. \\

\begin{figure}[htpb]
\centering
\includegraphics[width=7cm]{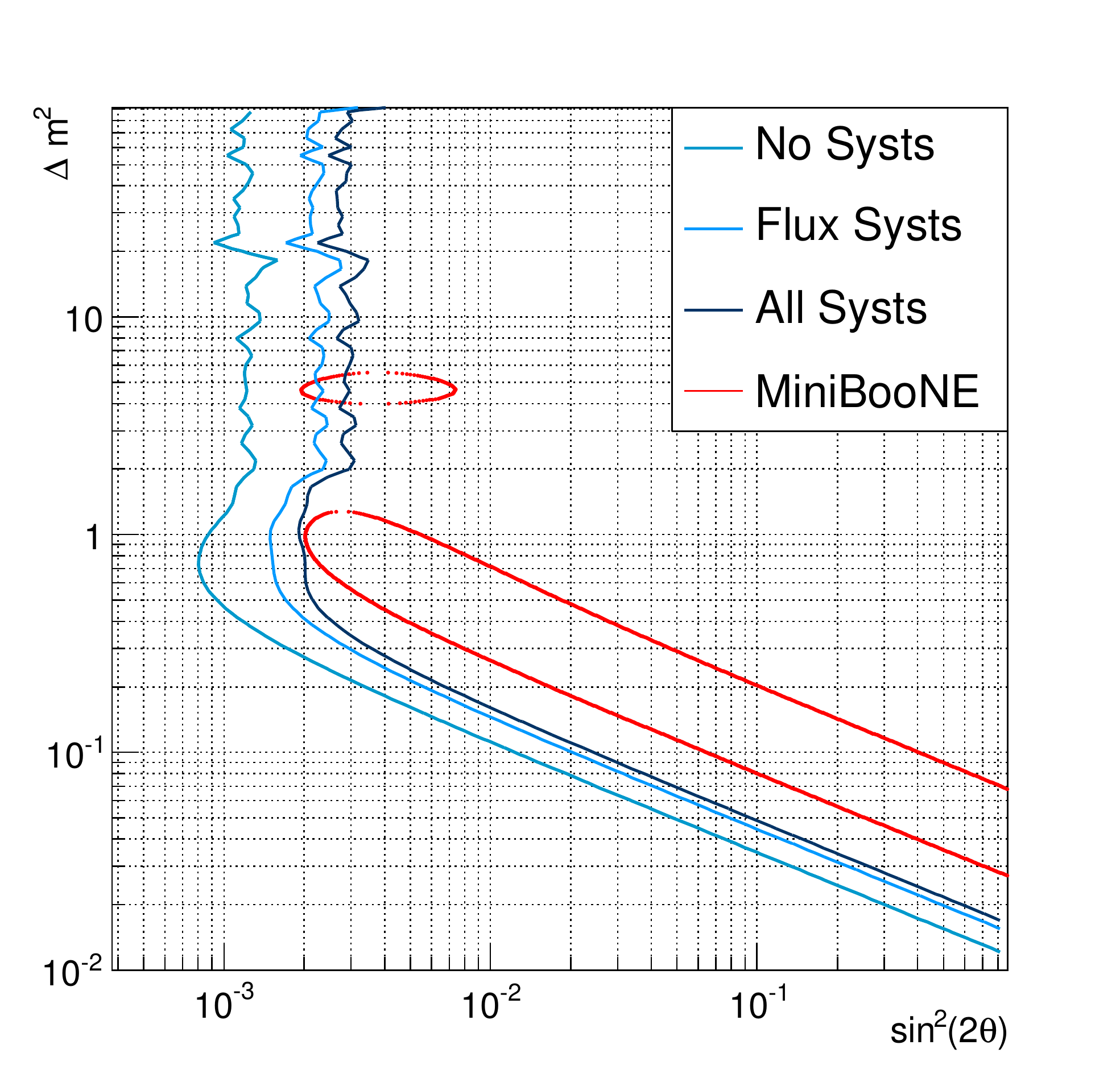}
\includegraphics[width=7cm]{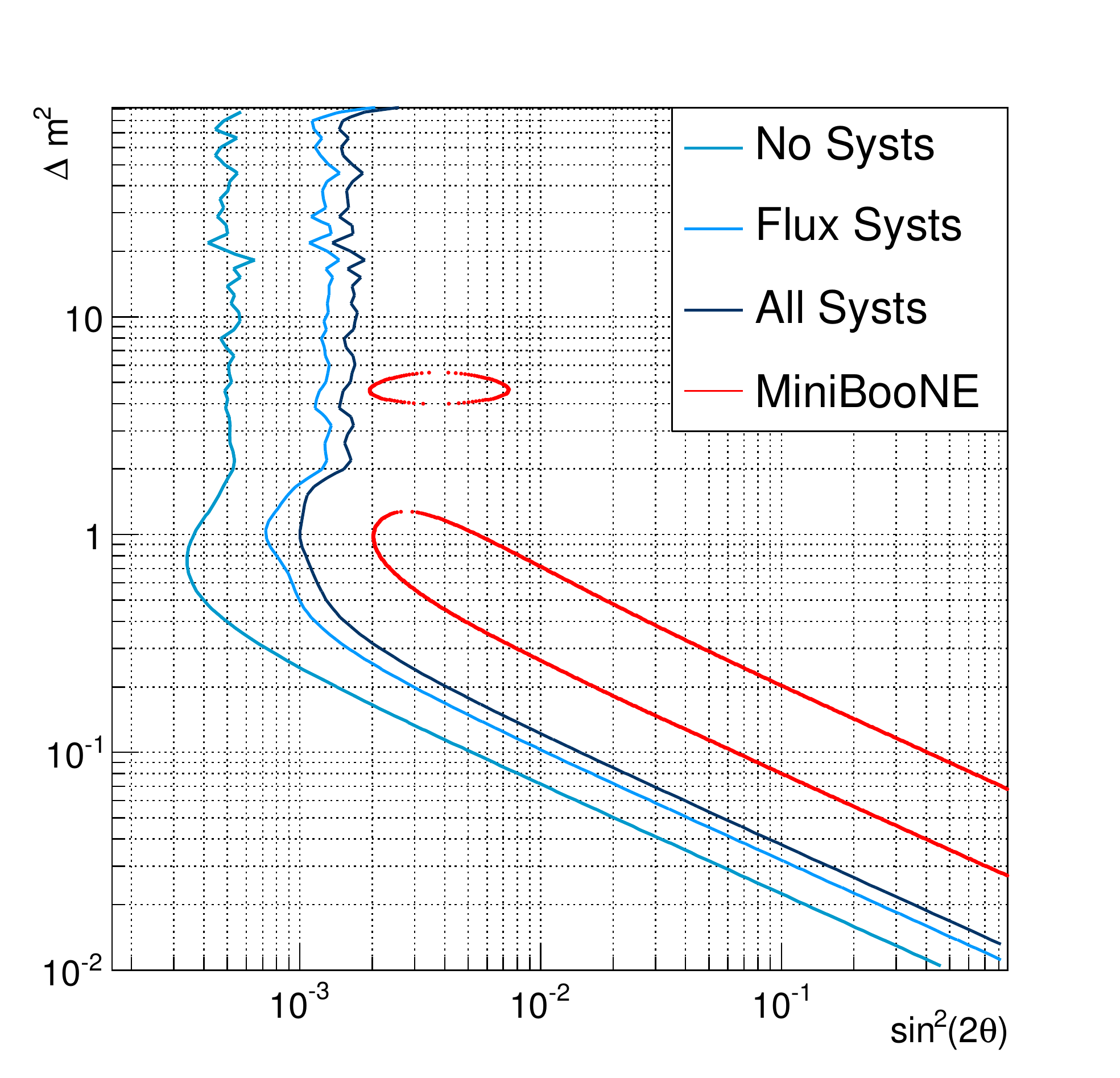}
\caption{90\% C.L. expected sensitivities for an exposure of $4.6 \times 10^20$ p.o.t. for three scenarios: statistical uncertainty only, both statistical uncertainties and flux systematic uncertainties, and statistical uncertainties with flux and cross-section systematic uncertainties. The sensitivity curves are shown for the two detector configuration considered: 3m (top) and 4m (bottom) inner detector radius. For comparison, the MiniBooNE allowed region at 90\% C.L. in antineutrino mode is shown in red.}
\label{fig:sterile_sensi}
\end{figure}

The systematic errors due to the flux and the cross-section uncertainties are included through a 
covariance matrix that is calculated using toy Monte Carlo throws. A $\chi^2$ test for a binned template 
of 10 $E_{Rec}$ bins and 10 OAA bins is performed between 0.2~MeV and 4~MeV, in order to obtain 
the expected sensitivity in the bidimensional oscillation parameter space ($\stme,\dmsqfo$). For 
each oscillation hypothesis the $\chi^2$ value is given by
\begin{equation}
\begin{split}
%\begin{align}
\chi^2 = \vec{n}_s \left(\stme, \dmsqfo \right)^T \times V^{-1} \\
 \times \vec{n}_s\left(\stme,\dmsqfo 
\right)
\end{split}
\end{equation}
%\end{align}

where $\vec{n}_s$ is the n-tuple of number of expected signal events due to $\numu \rightarrow 
\nue$ in $E_{Rec}$ and OAA bins, and $V$ is a $100\times100$ covariance matrix that includes the 
statistics and systematic errors. The $\chi^2$ is computed for each point of a bidimensional grid 
and the constant $\Delta \chi^2$ method is applied to determine the contours for the regions 
excluded at the 90\% C.L. The final sensitivity is shown in in Fig. \ref{fig:sterile_sensi} for the 
90\% C.L. along with a comparison with the MiniBooNE antineutrino results.

We observe that the final sensitivity, taking into account statistical uncertainties as well as flux and cross section systematic errors, contains, for the 4m inner detector radius case,
the full MiniBooNE allowed 
region at 90\% C.L. Regarding the 3m case, the detector is able to 
explore the whole low $\dmsqfo$ region allowed by MiniBooNE and it covers most of the high 
$\dmsqfo$ part. The sensitivity has been computed without using any constraints from ND280. In the nuPRISM analysis scenario, ND280 has the role of reducing 
model uncertainties in flux and cross sections, so the final errors for a full \nuprism+ND280 analysis are expected to be significantly reduced, but have not yet been computed.
Moreover, a \nue appearance analysis allows for the use of the nuPRISM \numu analysis to further constrain 
the flux and cross section systematics, which should further improve upon the sensitivity
predicted in this study.

\begin{table}
\begin{small}
 \caption{Expected number of events in the \nue selection for each oscillation hypothesis, and for the two detector inner diameters being considered.}
 \begin{center}
  \begin{tabular}{ c c c c }
\hline
\hline
& $(\stme, \dmsqfo)$  & 3~m radius & 4~m radius \\
\hline
$\numu \rightarrow \nue$ Signal &$(0.001,1~eV^2)$      &  87.6    &   484.3      \\
                                &$(0.005,1eV^2)$      &  437.8   &   2421.7     \\
				&$(0.01,10eV^2)$      &  635.2   &   3521.0     \\
				&$(0.001,10eV^2)$     &  63.5    &   352.1      \\
\hline
Background                      &$\nue$               &  1076.2  &   6695.5     \\
                                &$\numu$              &  983.8   &   4700.7     \\
\hline
\hline
  \end{tabular}
 \end{center}
\label{tab:steriles_events}
\end{small}
\end{table}

\subsection{$\bar{\nu}_\mu$ Measurements}

%Section goals:
%\begin{itemize}
%\item Reproduce wrong-sign flux in anti-nu mode with linear combination from nu-mode.
%\item Show flux plots, if available.
%\end{itemize}

In principle, the \nuprism technique of using multiple off axis angles to measure the oscillated $p_\mu$ and $\theta_\mu$ for each oscillated flux will work for anti-neutrinos as well. However, when running the T2K beam in anti-neutrino mode, there is a significant wrong-sign background from neutrino interactions. To disentangle these neutrino and anti-neutrino interactions, linear combinations of the neutrino-mode data can be used to construct the wrong-sign flux in anti-neutrino mode, analogous to the procedure used in Section~\ref{sec:disap} to construct the Super-K oscillated spectra and in Section~\ref{sec:nuexsec} to construct the electron neutrino spectrum. Hence, the neutrino flux in the anti-neutrino mode is described with the linear combination of neutrino mode fluxes:
\begin{equation}
\Phi^{\bar{\nu}mode}_{\nu_{\mu}}(E_{\nu},\theta_{oa}) = \sum c_i(\theta_{oa}) \Phi^{i,\nu mode}_{\nu_\mu}(E_\nu).
\end{equation}
$\Phi^{\bar{\nu}mode}_{\nu_{\mu}}(E_{\nu},\theta_{oa})$ is the anti-neutrino mode $\nu_{\mu}$ (wrong-sign) flux for a given off-axis angle $\theta_{oa}$.    $\Phi^{i,\nu mode}_{\nu_\mu}(E_\nu)$ is the neutrino mode $\nu_{\mu}$ (right-sign) flux for the $i^{th}$ off-axis bin and $c_i$ is the weight for the $i^{th}$ off-axis bin that depends on the off-axis angle for which the anti-neutrino mode wrong sign flux is being modeled.  

\begin{figure}[htpb]
\begin{center}
      \includegraphics[width=8cm] {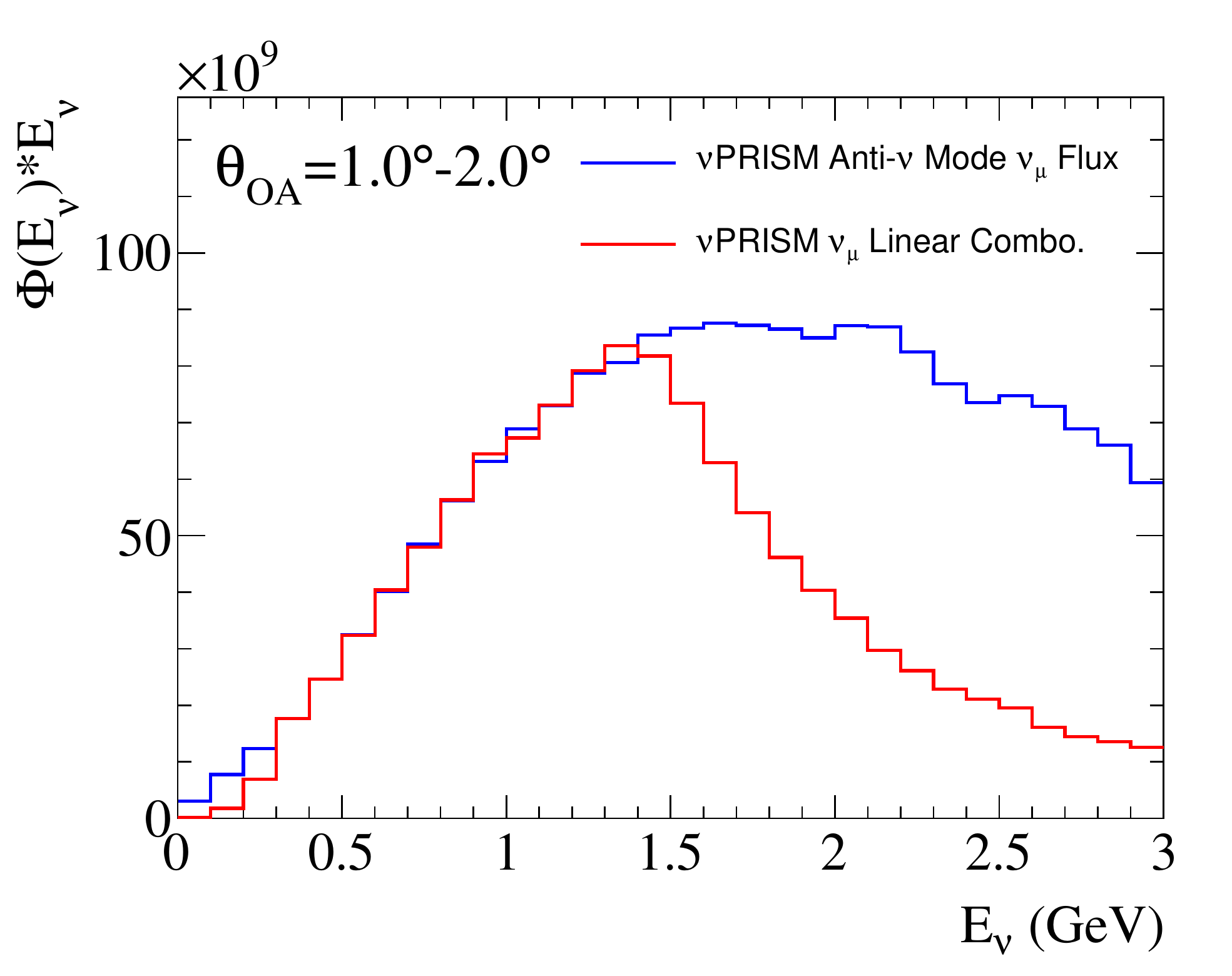}
      \includegraphics[width=8cm] {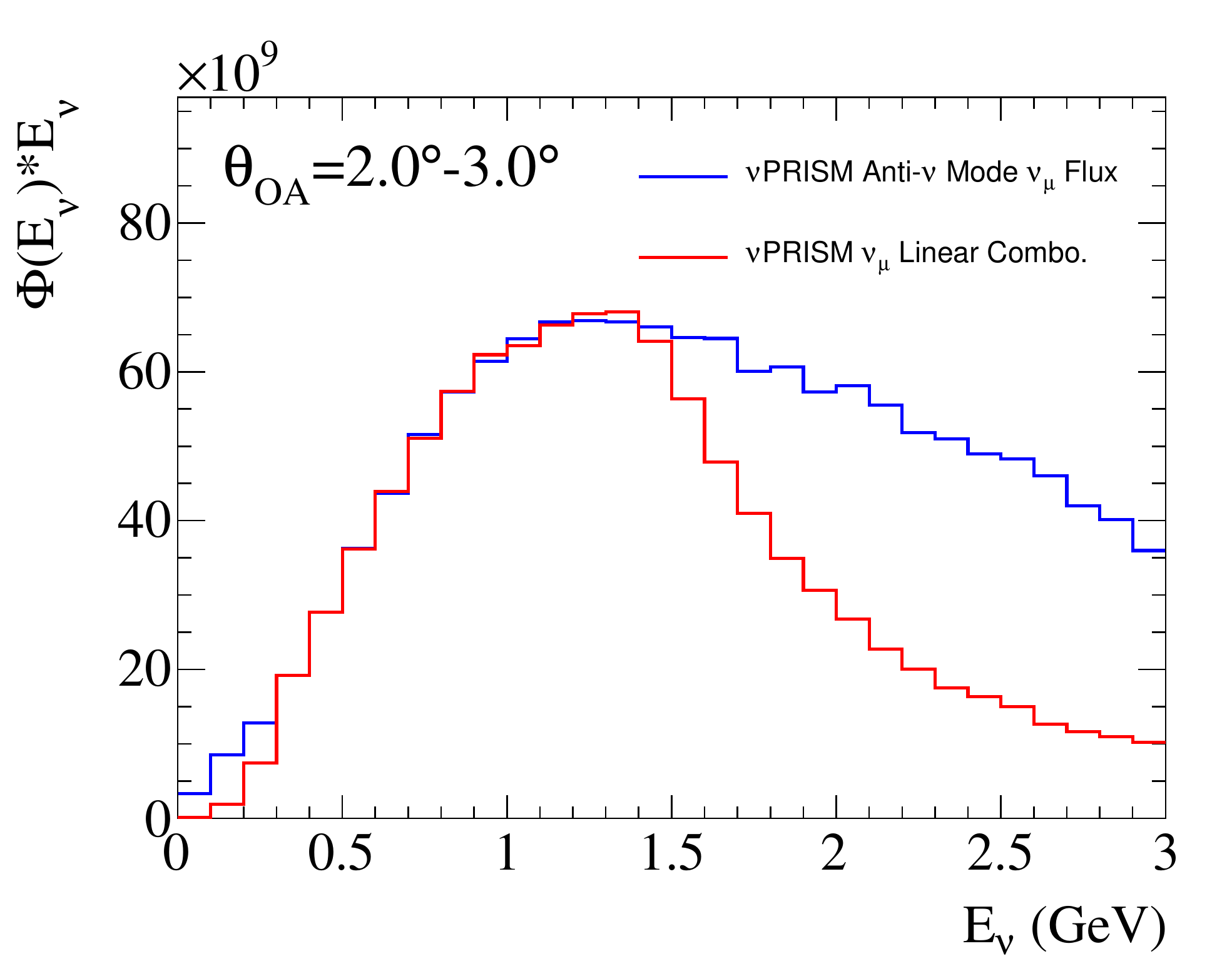}
      \includegraphics[width=8cm] {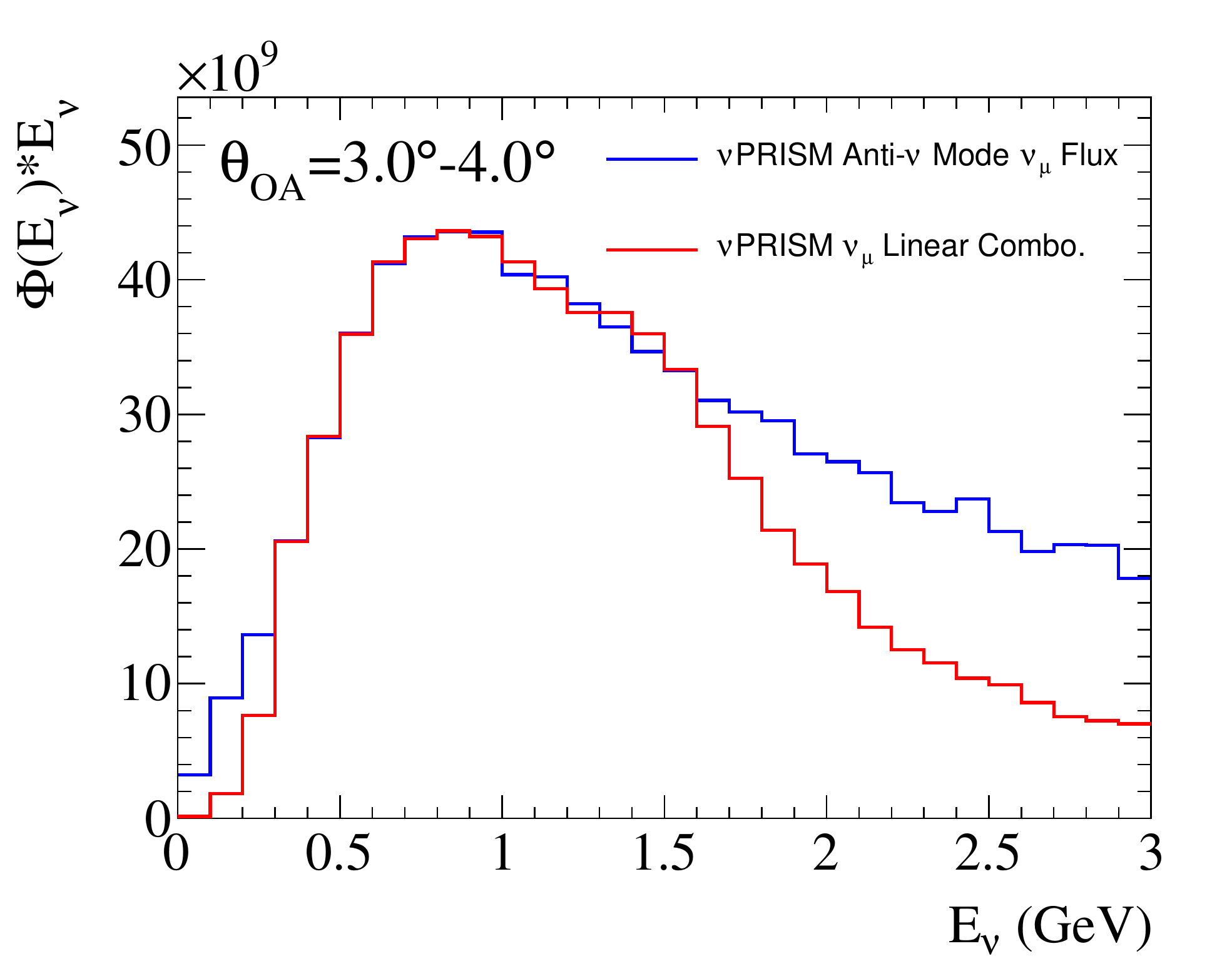}
\end{center}
\caption{The \nuprism anti-neutrino mode wrong-sign $\nu_{\mu}$ fluxes for $1.0-2.0^{\circ}$ (top), $2.0-3.0^{\circ}$ (middle) and $3.0-4.0^{\circ}$ (bottom), and the \nuprism linear combinations of neutrino mode $\nu_{\mu}$ fluxes.  }
\label{fig:wrongsignfit}
\end{figure}

Linear combinations to reproduce the wrong-sign $1.0-2.0^{\circ}$, $2.0-3.0^{\circ}$ and $3.0-4.0^{\circ}$ anti-neutrino mode fluxes are shown in Figure~\ref{fig:wrongsignfit}.  As with the combinations to produce the $\nu_{e}$ flux, the agreement is good up to about 1.5 GeV in neutrino energy.  As discussed in Section~\ref{sec:nuexsec}, it is less important to reproduce the high energy part of the flux since high energy interactions are suppressed by the event topology selected and the muon acceptance of \nuprism.

\begin{figure}[htpb]
\begin{center}
  \begin{minipage}[t]{.45\textwidth}
    \begin{center}
      \includegraphics[width=\textwidth] {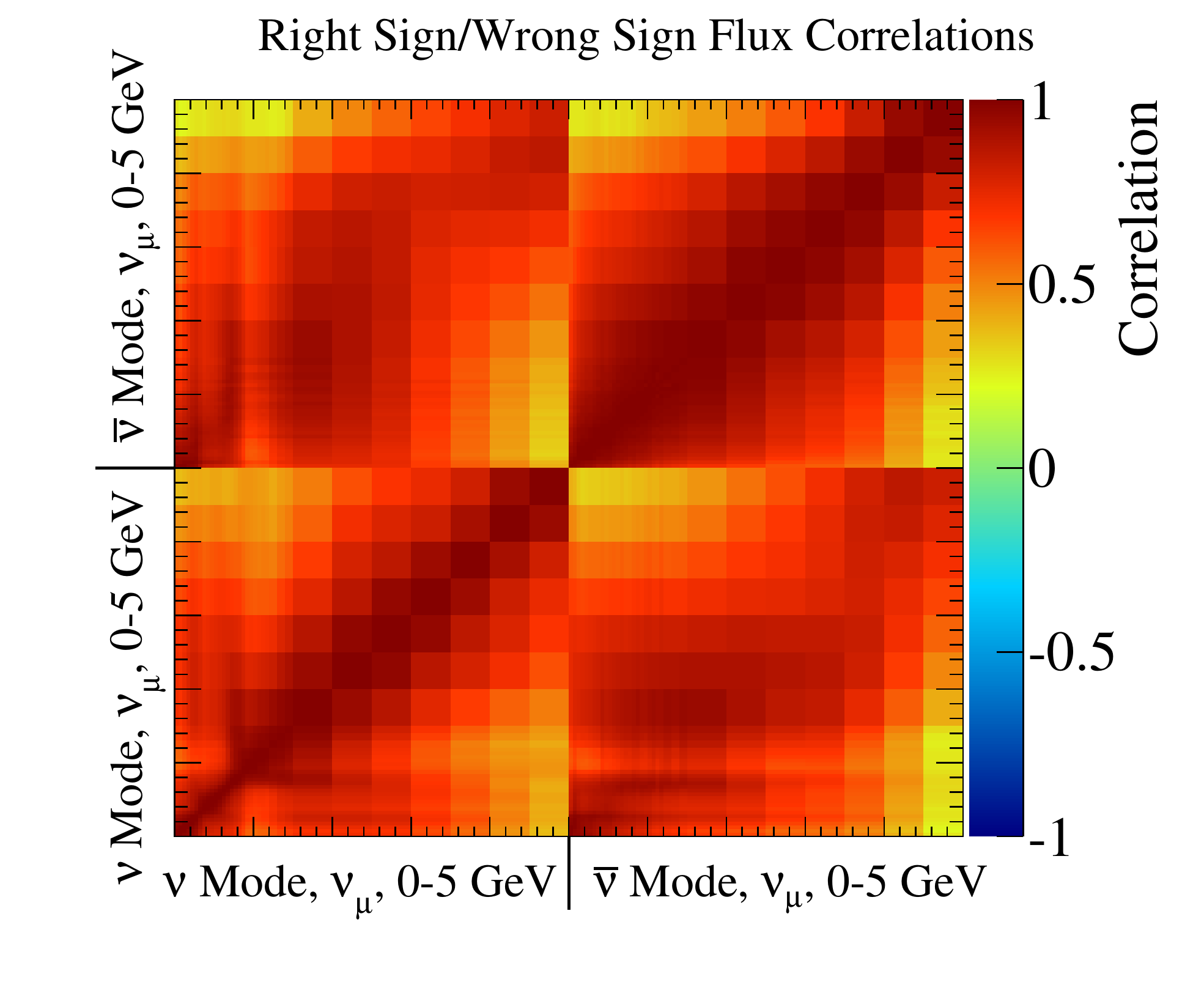}
    \end{center}
  \end{minipage}
\end{center}
\caption{The correlations between the flux normalization parameters for energy bins from 0 to 5 GeV for the neutrino mode and anti-neutrino mode $\nu_{\mu}$ fluxes. }
\label{fig:wrongsignflux}
\end{figure}

As shown Figure~\ref{fig:wrongsignflux}, there is significant correlation between the wrong-sign neutrino flux in anti-neutrino mode and the neutrino-mode flux, so the flux uncertainties will give some cancelation using this method. After subtracting the neutrino background, the remaining $\bar{\nu}_\mu$ events can 
then be combined as in the neutrino case to produce oscillated spectra at Super-K. 

\subsection{Cross Section Measurements}
\label{sec:xsec}

A unique feature of \nuprismlite is the ability to measure the true neutrino energy dependence of both CC and NC interactions using nearly monoenergetic beams. These measurements are expected to significantly enhance the reach of oscillation experiments, since the energy dependence of signal and background processes must be understood in order to place strong constraints on oscillation parameters. As explained in Section~\ref{sec:disap},  additional multinucleon processes, with a different energy dependences than the currently modeled CCQE and CC1$\pi$ cross sections can affect the T2K oscillation analysis. In the current disappearance analysis, there are also substantial uncertainties on NC1$\pi^+$ and NC1$\pi^0$ processes (for disappearance and appearance respectively). As a result, future proposed experiments which use water as a target (e.g. Hyper-Kamiokande and CHIPS) will directly benefit from the \nuprismlite cross section program; other programs benefit less directly through a critical validation of our assumptions of the energy dependence of the cross section on oxygen.
It is also not just long baseline oscillation programs which benefit, as cross section processes at T2K's flux peak are also relevant for proton decay searches and atmospheric neutrino oscillation analyses. Finally, should T2K run an antineutrino beam during \nuprismlite operation, all arguments made above equally apply for  antineutrino cross section measurements at \nuprismlite.
 
One should also consider the study of neutrino interactions interesting in its own right as a particle/nuclear theory problem. As an example, MiniBooNE's cross section measurements have received much attention from the nuclear theory community who predominantly study electron scattering data.

Some of the difficulties in improving our understanding of neutrino cross sections stems from the fact that we do not know, for a given interaction, the incident neutrino energy. Any given measurement is always averaged over the entire flux. The observed rate $N$ in a given observable bin $k$ depends on the convolution of the cross section, $\sigma$, and the flux, $\Phi$:

\begin{equation}
N^k  = \epsilon_k  \int \sigma(E_\nu) \Phi(E_\nu) dE_\nu
\end{equation}

where $\epsilon$ is the efficiency. Therefore, our understanding of the energy dependence of neutrino interaction for a particular experiment is limited by the flux width and shape. One then attempts to use different neutrino fluxes (with different peak energies) to try to understand the cross section energy dependence. As discussed later in this section, for CC interactions we have many examples of disagreements between experiments, and for NC, we have a limited number of measurements made, and the lack of information and conflicting information leaves unresolved questions about the true energy dependence of the cross section.

In addition to providing new measurements on oxygen, there are two main advantages of \nuprismlite over the current paradigm. First, we can directly infer the energy dependence of the cross section by combining measurements at different off-axis angles into a single measurement, as if we would have had a Gaussian neutrino flux source. Second, and equally important, we can fully understand the correlations between energy bins, in a way not possible previously when comparing across experiments with entirely different flux setups.  

In CC interactions, previous experiments use the muon and hadronic system to try to infer the neutrino energy dependence.  \nuprismlite has the capability to directly test if the neutrino energy dependence inferred from the lepton information is consistent with the energy information determined from the off-axis angle. \nuprismlite will also for the first time probe the energy dependence of NC cross sections within a single experiment.  
%In the case of NC interactions, we only can compare measurements made with different fluxes to understand the energy dependance of the NC cross section, and when only 1 or a few masurements exist, it is difficult to che

 Furthermore, there is no data for the kinematic information of pions out of NC\pip interactions. However, NC\pip is one of the backgrounds in the current T2K \rmu selection used for the disappearance analysis. A direct measurement of NC\pip, and a measurement of the pion momentum and angular distributions would reduce the substantial uncertainties on this process (in both cross section and detector efficiency) in the analysis.

%% This section may just be a repeat, so can consider removing
%If we want to measure the observed rate $N$, in a given observable bin $k$ is given by:

%\begin{equation}
%N^k  = \epsilon_k  \int \sigma(E_\nu) \Phi(E_\nu) dE_\nu
%\end{equation}

%where $\epsilon$ is the efficiency, $\sigma$ is the cross section, and $\Phi$ is the flux in that bin. Now we consider a carefully chosen sum over measurements in the bin taken with different fluxes denoted by index $i$:

%\begin{equation}
%\frac{\sum_i c_i N_i^k}{\epsilon_k }  =  \sum_i c_i \int \sigma(E_\nu) \Phi_i(E_\nu) dE_\nu =  \int \sigma(E_\nu) Ae^{\frac{-(E_\nu-E^{avg}_\nu)^2}{2s^2}} dE_\nu 
%\end{equation}

%This is equivalent to:
%\begin{equation}
%\frac{\sum_i c_i N_i^k}{\epsilon_k }  =   \int \sigma(E_\nu-E_\nu^{avg}) Ae^{\frac{-(E_\nu)}{s^2}} dE_\nu 
%\end{equation}

%While we cannot further simplify this expression without knowing the exact functional form of $\sigma$, we recognize that combinations of measurements would be the same as a single measurement according to a Gaussian flux. We then can divide out the flux assuming $\sigma(E_\nu-E_\nu^{avg})$ is approximately constant over that flux for a cross section measurement in that bin.
%% 

Oxygen is an interesting target material for studying cross sections because few measurements exist and it is a medium sized nucleus where the cross section is calculable. \nuprismlite will provide differential measurements in muon and final state pion kinematic bins. While these kinds of measurements will be done with the ND280 P0D and FGD2 detectors in the near term, \nuprismlite will have more angular acceptance than those measurements and so enhances the T2K physics program.

Possible cross section measurements, based on observable final state topologies, at \nuprismlite include:

\begin{itemize}
\item CC inclusive
\item CC0$\pi$ 
\item CC1$\pi^+$, $\pi^0$ (resonant and coherent)
%\item NC elastic (with a proton final state)
\item NC1$\pi^+$, $\pi^0$ (resonant and coherent) 
\item NC1$\gamma$ 
\end{itemize}

The above list is based on expected water Cherenkov detector capabilities from experience with MiniBooNE, K2K 1 kton and Super-Kamiokande (SK) analyses. All CC measurements can be done for \numu and \nue flavors due to the excellent e-$\mu$ separation at \nuprismlite.  Antineutrino cross section measurements are also possible with similar selections. A brief summary of each measurement follows.
Table~\ref{tab:rates} shows the number of events in the FV of \nuprismlite, broken down by interaction mode.

\begin{table}
 \caption{Expected number of events in the fiducial volume of \nuprismlite for $4.5\times 10^{20}$ POT, separated by true interaction mode in NEUT.}
 \begin{center}
  \begin{tabular}{ c c c c  }
\hline
\hline
Int. mode  & 1-2$^\circ$ & 2-3$^\circ$ & 3-4$^\circ$ \\
CC inclusive  & 1105454 & 490035 & 210408 \\
CCQE & 505275 &  271299 &  128198 \\
CC1\pip & 312997 & 111410 & 39942 \\
CC1\piz & 66344 & 23399 & 8495 \\
CC Coh & 29258 & 12027 & 4857 \\
NC 1\piz & 86741 & 32958 & 12304 \\
NC 1\pip & 31796 & 11938 & 4588 \\
NC Coh & 18500 & 8353 & 3523 \\
\hline
\hline
  \end{tabular}
 \end{center}
\label{tab:rates}
\end{table}

\subsubsection{CC Inclusive}

Inclusive measurements are valuable because they are the most readily comparable to electron scattering measurements and theory, as there is minimal dependance on the hadronic final state. Also, external CC inclusive neutrino data was used in the estimation of the T2K neutrino oscillation analyses to help determine the CCDIS and CC multi-$\pi$ uncertainties.

The CC \numu cross section has been measured on carbon by the T2K~\cite{Abe:2013jth} and SciBooNE~\cite{Nakajima:2010fp} experiments. MINERvA has produced ratios of the CC inclusive cross section on different targets (C,Fe,Pb) to scintillator~\cite{Tice:2014pgu}. In addition, the SciBooNE results include the  energy dependence of the CC inclusive cross section from the muon kinematic information. The CC \nue cross section on carbon is in preparation by T2K. 

\nuprismlite should be able to select CC \numu and \nue events with high efficiency and produce a CC inclusive measurement vs. true neutrino energy on water.  Using the latest T2K simulation tools, we estimate  a CC inclusive \numu (\nue) selection to be  93.7\% (50.4\%) efficient relative to FCFV and 95.9\% (39.5\%) pure based on observable final state. The low purity of the \nue selection is predominantly due to the small \nue flux relative to \numu.

%; MiniBooNE's result is in preparation. MiniBooNE proposes to also utilize scintillation light to verify the true neutrino energy dependance inferred from the muon, however the challenges of using this have delayed the result.  

\subsubsection{\label{sec:cc0pi}CC0$\pi$}

\begin{figure}[htpb]
\begin{center}
      \includegraphics[width=0.45\textwidth] {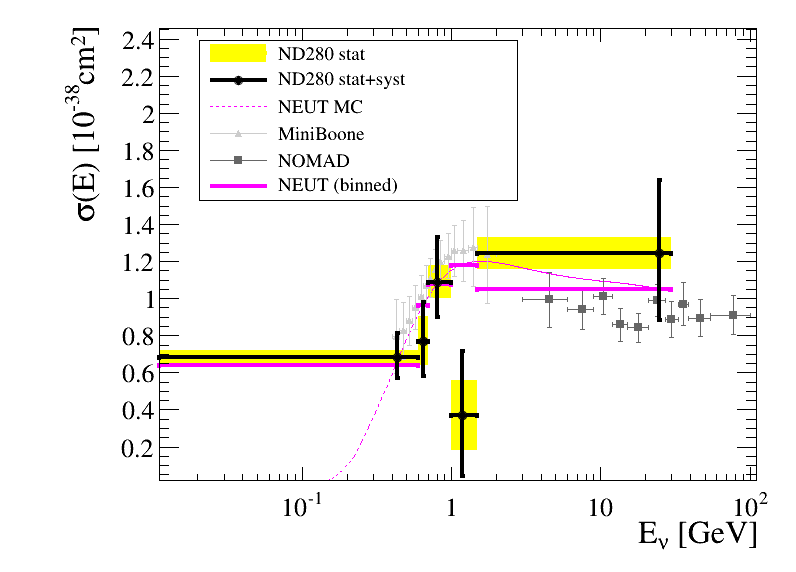}
\end{center}
\caption{The CCQE cross section as predicted by NEUT (pink dashed) vs. true neutrino energy. Also overlaid are results from MiniBooNE, NOMAD and T2K.}
\label{fig:ccqe}
\end{figure}

The CCQE \numu cross section has been measured on carbon by MiniBooNE~\cite{AguilarArevalo:2010zc} and is consistent with a larger cross section than expected which could correspond to an increased value of an effective axial mass ($M_A$) over expectation; SciBooNE's analysis was presented at NuInt2011~\cite{sciboone-nuint2011} but not published and is consistent with MiniBooNE. In addition, a measurement by NOMAD~\cite{Lyubushkin:2008pe} was done at higher neutrino energies which is not in agreement with MiniBooNE and SciBooNE. This is shown in Figure~\ref{fig:ccqe}, along with the recent T2K ND280 Tracker analysis results. An indirect measurement of the cross section was done with the K2K near detectors, where a higher than expected value of the QE axial mass, $M_A$, was also reported~\cite{Gran:2006jn}. There are also recent results from MINERvA~\cite{Fiorentini:2013ezn}.

MiniBooNE's selection was CC0$\pi$, that is 1 muon and no pions in the final state, and was 77.0\% pure and 26.6\% efficient; the \rmu selection at SK is 91.7\% pure  and 93.2\% efficient, based on observable final state.
It is postulated that the MiniBooNE selection, but not the NOMAD one, is sensitive to multinucleon processes, where a neutrino interacts on a correlated pair of nucleons and that this resulted in the higher cross section reported by MiniBooNE. However, the two experiments have very different flux, selection and background predictions and systematics.

By measuring the CC0$\pi$ cross section  at different vertex points in \nuprismlite, we should be able to infer the different energy dependence and constrain multinucleon and CC1\pip pionless $\Delta$ decay (PDD)  processes. This can be seen in Figure~\ref{fig:mono_beam_erec}, which shows the momentum of CCQE and MEC (Nieves' npnh) events for a particular angular range ($0.85<$cos($\theta$)$<0.90$) generated according to the T2K flux, and for a 1~GeV \nuprismlite flux. MiniBooNE and T2K have difficulty separating the MEC component of the CCQE cross section due to the shape of their neutrino energy spectra, but the \nuprismlite detector would give us additional information to separate out that component and characterize it, as demonstrated in Figure~\ref{fig:mono_beam_erec}.  Even though \nuprismlite is not a measurement on carbon, oxygen is of a similar density to carbon and so will be helpful in understanding the difference between the MiniBooNE and NOMAD results if it is indeed due to MEC.

%\begin{figure}[h!]
%\begin{center}
%  \begin{minipage}[t]{.45\textwidth}
%    \begin{center}
%      \includegraphics[width=\textwidth] {figures/t2kflux_ccqe_npnh.pdf}
%    \end{center}
%  \end{minipage}
%  \begin{minipage}[t]{.45\textwidth}
%    \begin{center}
%      \includegraphics[width=\textwidth] {figures/nuprism1000flux_ccqe_npnh.pdf}
%    \end{center}
%  \end{minipage}
%\end{center}
%\caption{The momentum of CCQE and MEC (Nieves' npnh) events for a particular angular range ($0.85<$cos($\theta$)$<0.90$) generated according to the T2K flux (left), and for a 1000 MeV \nuprismlite flux (right)}
%\label{fig:npnh}
%\end{figure}

\subsubsection{CC1\pip and CC1\piz}

The CC1\pip and CC1\piz cross sections have been measured on carbon by MiniBooNE~\cite{AguilarArevalo:2010bm},\cite{AguilarArevalo:2010xt}; K2K also produced measurements CC1\pip~\cite{Rodriguez:2008aa} and CC1\piz~\cite{Mariani:2010ez} with the SciBar detector. One may infer the coherent contribution to the CC1$\pi$ cross section from the angular distribution of the pion; this was done by K2K~\cite{Hasegawa:2005td} and SciBooNE. Improvements to the SK reconstruction could yield a similar efficiency and purity to the the MiniBooNE selections for CC1\pip (12.7\%, 90.0\%) and CC1\piz (6.4\%, 57.0\%) based on observable final state.

The CC1$\pi$ resonant cross section for the T2K flux is dominated by contributions from the $\Delta$ resonance~\cite{Lalakulich:2013iaa}, so \nuprismlite would provide clear information about the N$\Delta$ coupling and form factors. We can also compare the pion momentum produced out of CC1\pip interactions for different neutrino energies in order to better understand how final state interactions affect pion kinematics.

%{\it TODO: Add a plot of pion spectrum vs. offaxis angle} 

\subsubsection{NC1\pip and \ncpi}

\begin{figure}[htpb]
\begin{center}
      \includegraphics[width=0.45\textwidth] {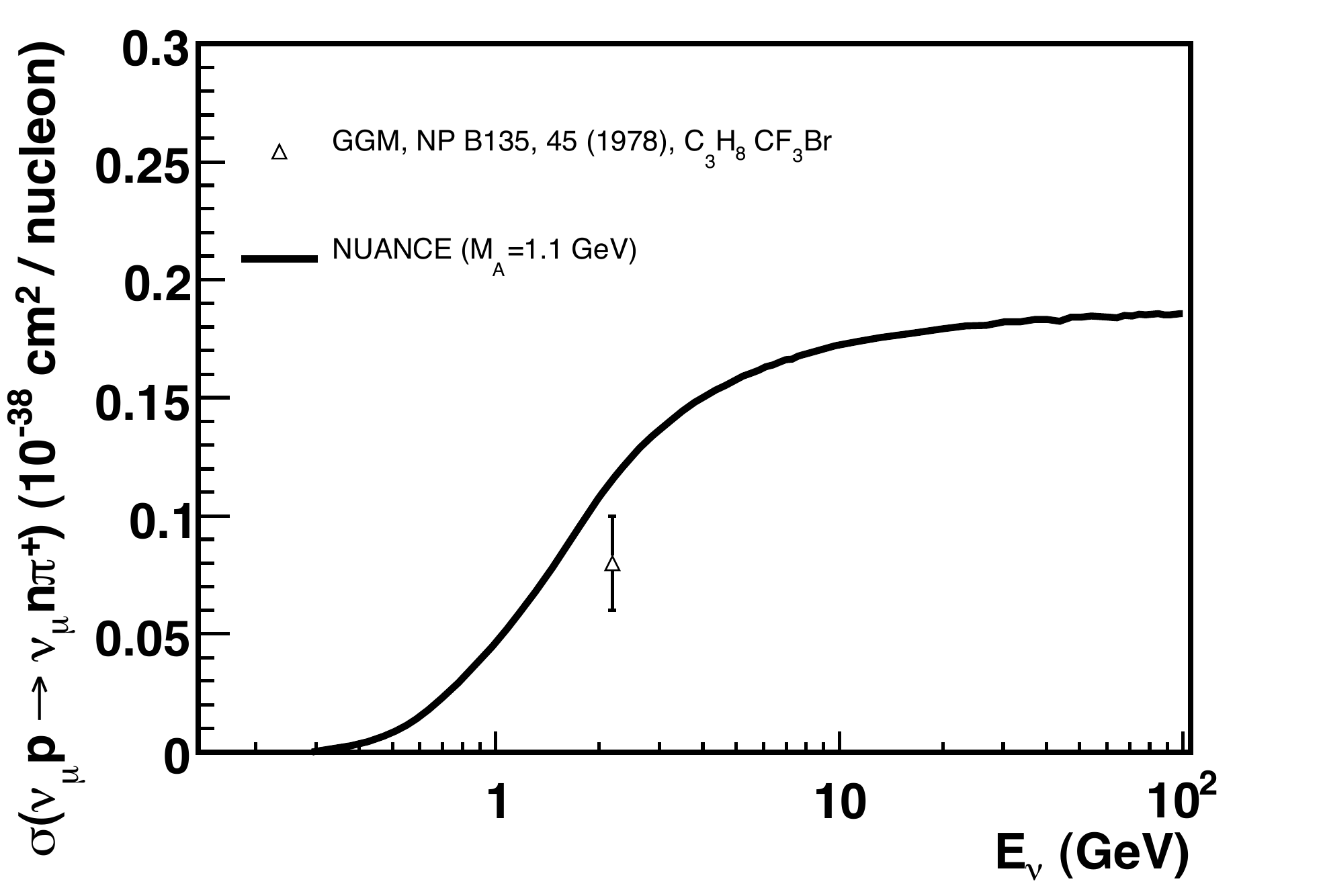}
\end{center}
\caption{The NC\pip  cross section as predicted by NUANCE vs. true neutrino energy overlaid with the only measurement (on C$_3$H$_8$CF$_3$Br). Figure from Ref.~\cite{Formaggio:2013kya}}
\label{fig:ncpip}
\end{figure}

The \ncpi  cross section has been measured on carbon by MiniBooNE~\cite{AguilarArevalo:2009ww} (36\% efficient, 73\% pure) and SciBooNE. A measurement of the ratio of \ncpi to the CCQE cross section has been done water by the K2K 1kton near detector~\cite{Nakayama:2004dp}. The efficiency and purity of the K2K selection is 47\% and 71\% respectively. A measurement of NC\pip exists~\cite{hawker} on a complicated target material (C$_3$H$_8$CF$_3$Br) but has no differential kinematic information. Figure~\ref{fig:ncpip} shows this measurement with a prediction from the NUANCE neutrino event generator.  

A measurement of NC\pip will be challenging but possible at \nuprismlite. T2K already has developed an ``NC'' enhanced selection for Super-K that is 24\% NC\pip, 14\% NC1proton, and 55\% CC\numu, by interaction mode. Recent developments in event reconstruction at Super-K include a dedicated pion ring finder, which should make possible a more pure selection of NC\pip from which the pion momentum and angular distribution can also be measured. Since \nuprismlite will allow for a first measurement of the energy dependence of the NC channels and like the CC channels, it will be particularly interesting to measure the outgoing pion spectra of these events in order to probe nuclear final state interactions.

To summarize, \nuprismlite's measurement of true neutrino energy dependence of the cross section is a unique and potentially critical input to our overall understanding of cross section processes around 1 GeV neutrino energy. In particular, \nuprismlite will help us understand for CC$0\pi$ events, if the shape and size of the PDD and mulit-nucleon components are modeled correctly. Furthermore, \nuprismlite can provide new information on the pion kinematics out of NC interactions relevant to the oscillation analysis and the energy dependence of those cross sections.

\clearpage

\section{Detector Design and Hardware}\label{sec:detector}

% T. Ishida, R. Henderson, A. Konaka, Y. Kudenko, T. Lindner, S. Nakayama, Y. Nishimura, F. Retiere, M. Shiozawa, H.-K. Tanaka, M. Ziembicki, TRIUMF, ICRR

\begin{figure}[htbp]
\centering\includegraphics[width=9cm,angle=0]{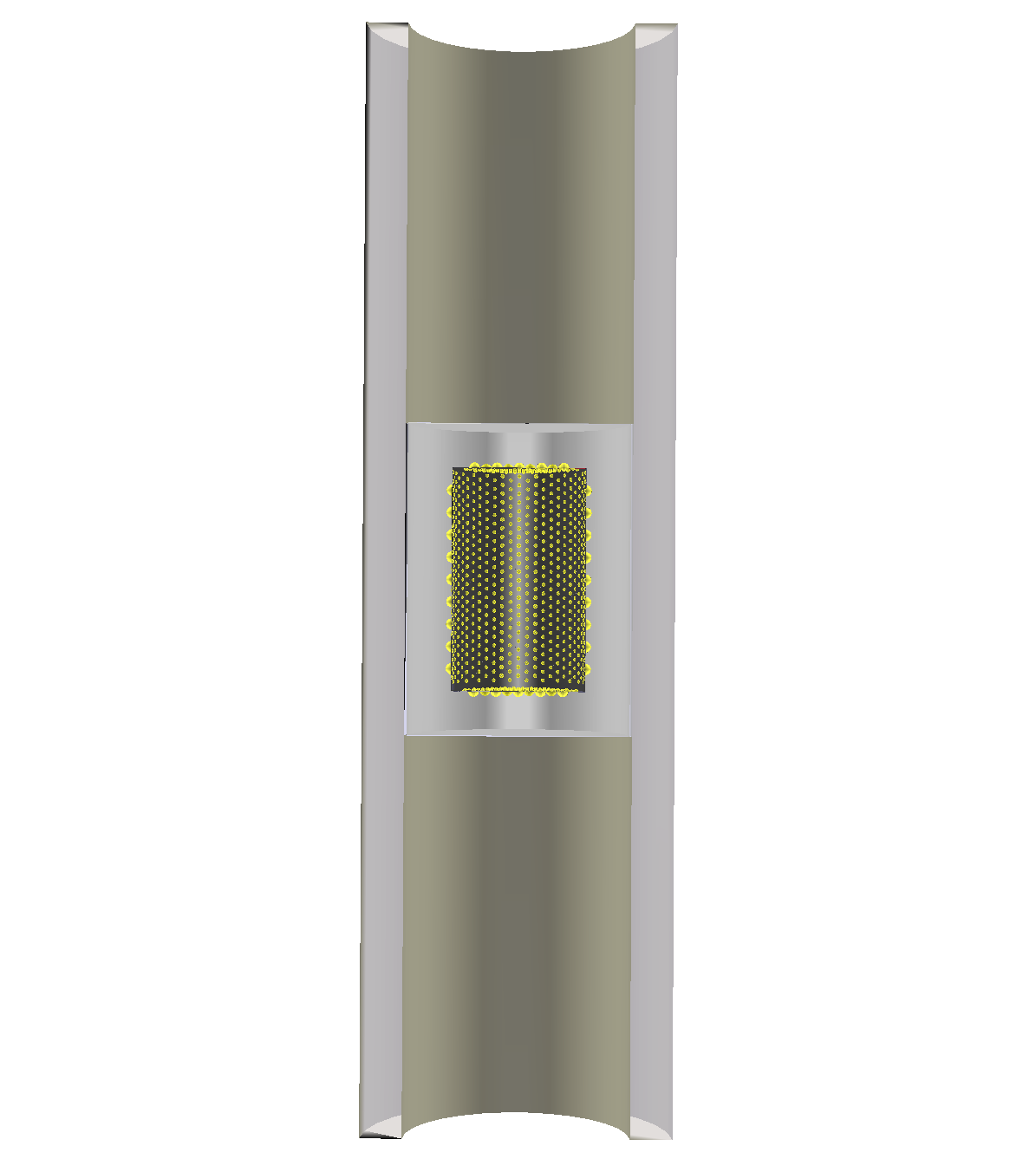}
\caption{The planned configuration of the nuPRISM detector within the water tank is shown. The instrumented portion of the tank moves vertically to sample different off-axis angle regions.}
\label{fig:nuprismpic} 
\end{figure}

The \nuprism detector uses the same water Cherenkov detection technology as Super-K with a cylindrical water volume that is taller than Super-K (50-100m vs 41m) but with a much smaller diameter (6-10m vs 39m). The key requirements are that the detector span the necessary off-axis range (1$^\circ$-4$^\circ$) and that the diameter is large enough to contain the maximum required muon momentum. The baseline design considers a detector location that is 1~km downstream of the neutrino interaction target with a maximum contained muon momentum of 1~GeV/c. This corresponds to a 50~m tall tank with a 6~m diameter inner detector (ID) and a 10~m diameter outer detector (OD), as shown in Figure~\ref{fig:nuprismpic}. A larger, 8~m ID is also being considered at the expense of some OD volume in the downstream portion of the tank. As the \nuprism analysis studies mature, the exact detector dimensions will be refined to ensure sufficient muon momentum, \nue statistics and purity, etc.

The instrumented portion of the tank is a subset of the full height of the water volume, currently assumed to be 10~m for the ID and 14~m for the OD. The novel feature of this detector is the ability to raise and lower the instrumented section of the tank in order to span the full off-axis range in 6 steps. The inner detector will be instrumented with either 5-inch or 8-inch PMTs to ensure sufficient measurement granularity for the shorter light propagation distances relative to Super-K. Also under consideration is to replace the OD reflectors with large SMRD-style scintillator panels, as discussed in Section~\ref{sec:scint}, although this has not yet been integrated into the overall detector design.

The remainder of this section describes the elements needed for \nuprism and corresponding cost estimates, where available. The cost drivers for the experiment are the civil construction and the cost of the PMTs, and, correspondingly, more detailed cost information is presented in those sections.

%Section goals:
%\begin{itemize}
%\item Overview of hardware issues (references to associated physics studies)
%\item Mention unique requirement to raise and lower detector
%\end{itemize}

\subsection{Site Selection}

The \nuprism detector location is determined by several factors, such as signal statistics, accidental pile-up rates, cost of digging the pit, and potential sites available.
At 2.5$^o$ off-axis position at 1~km with a fiducial volume size of 4~m diameter and 8~m high cylinder, the neutrino event rate at \nuprism is more than 300 times that of SK. At 2km, the number of events drops by a factor of 4, which yields 75 times more events than SK, for the same size of the detector. The impact of the number of events collected on the physics sensitivities is described in Section~\ref{sec:physics}. The event pile-up is dominated by sand muons, but at 1~km, the pile-up rate appears to be acceptable, which is explained in more detail in Section~\ref{sec:physics},  The detector depth and diameter scales with the distance to the \nuprism detector. In order to cover from 1-4$^\circ$ off-axis angles, the depth of the detector is 50~m at 1~km and 100~m at 2~km. There are standard Caisson-based excavation procedures available for pit depths of up to 65~m and diameters of up to 12~m. For deeper depth or larger diameter, more specialized construction may be required, and could increase the cost per cubic meter of excavation dramatically, as discussed in the next section.

%\begin{figure*}[htpb]
%\centering\includegraphics[width=15cm,angle=0]{figures/nuprismsites.pdf}
%\caption{Potential sites are shown for \nuprism if all rice field locations are excluded.}
%\label{fig:detsites}
%\end{figure*}
%
%
%The two far detector sites that must be considered are the Mozumi mine, where Super-K is located, and Tochibora, which is a candidate site for Hyper-K. There are four potential unused sites in the Tochibora and Mozumi directions, not including rice fields, a shown in Figure~\ref{fig:detsites}:
%\begin{itemize}
%\item 750m site near the Muramatsu community meeting centre:
%          This location is right next to R245 and owned by the local government. The space is limited but covers the 
%          Mozumi direction and the central line between Mozumi and Tochibora. This site would have the highest event pile-up rate.
%\item 1km site: a large un-cultivated private land covering both Tochibora and Mozumi directions
%\item 1.2km site: a large patch of private land at the foot of a forest covering both Tochibora and Mozumi directions
%\item 1.8km site: the originally considered 2km detector site owned by the local government covering Tochibora direction. This site would have the deepest detector, 90m deep.
%\end{itemize}
%If the rice field can be available, there are a lot more choices. At the time of the 2km detector site study, rice fields were also considered, although not selected in the end. Changing the land use for the rice field would require additional approval process, if it were possible. 

Potential sites for nuPRISM have been identified along the path from the neutrino beam to both the Mozumi mine, where Super-K is located, and to the Tochibora mine, which is a candidate site for Hyper-Kamiokande, and is positioned at the same off-axis angle as Mozumi. No specific sites are discussed in this public version of the document.
Land use will require consensus from the local community and involvement from one or more Japanese host institutions. There are existing facilities that are operated just outside J-PARC, such as the KEK-Tokai dormitory, KEK Tokai \#1 building at IQBRC, and  the dormitory of the Material Science Institute of Tokyo university.
%\clearpage

\subsection{Civil Construction}\label{sec:civil}

%Section goals:
%\begin{itemize}
%\item Discussion of considerations from original 2km investigation \red{(TI)}
%\item Discussion of rock condition at various sites and implications for making a pit \red{(TI \& HKT)}
%\item Discussion of different pit construction methods with cost estimates \red{(AK \& HKT)}
%\end{itemize}

Based on the current baseline design of the \nuprism detector described previous sections,
we have communicated with companies for the preliminary cost estimation of
\nuprism civil construction; the water tank construction and detector construction.
The \nuprism detector is also considered as a prototype detector of Hyper-Kamiokande (Hyper-K) for
testing new photo-sensors, readout electronics, and the water containment system design.

Two groups have been contacted to provide preliminary cost estimates for the civil construction associated with fabricating a 50~m deep cylindrical volume with a 10~m diameter. The first group consists of a general construction company and a heavy industrial company currently providing cost estimates for Hyper-K. The second group is a single general construction company that was associated with the cost estimates from the original T2K 2~km detector proposal~\cite{t2k2km}.

There are several techniques to construct the 10~m$\phi$ and 50~m long vertical ``tunnel'';
Pneumatic Caisson (PC) method, Soil Mixing Wall (SMW) method, New Austrian Tunneling (NAT)
method, Urban Ring (UR) method.
Each of the construction methods have pros and cons, and some of the methods are not applicable
depending on the actual geological condition. %Cost estimates from both construction groups are given in the appendix.

\subsection{Liner and Tank}

The \nuprism detector can be used for proof-testing various designs and components which
will be adopted in the Hyper-K detector. The \nuprism water tank will
have the same liner structure as that designed for Hyper-K.

\begin{figure}[htpb]
  \centering
  \includegraphics[width=0.5\textwidth]{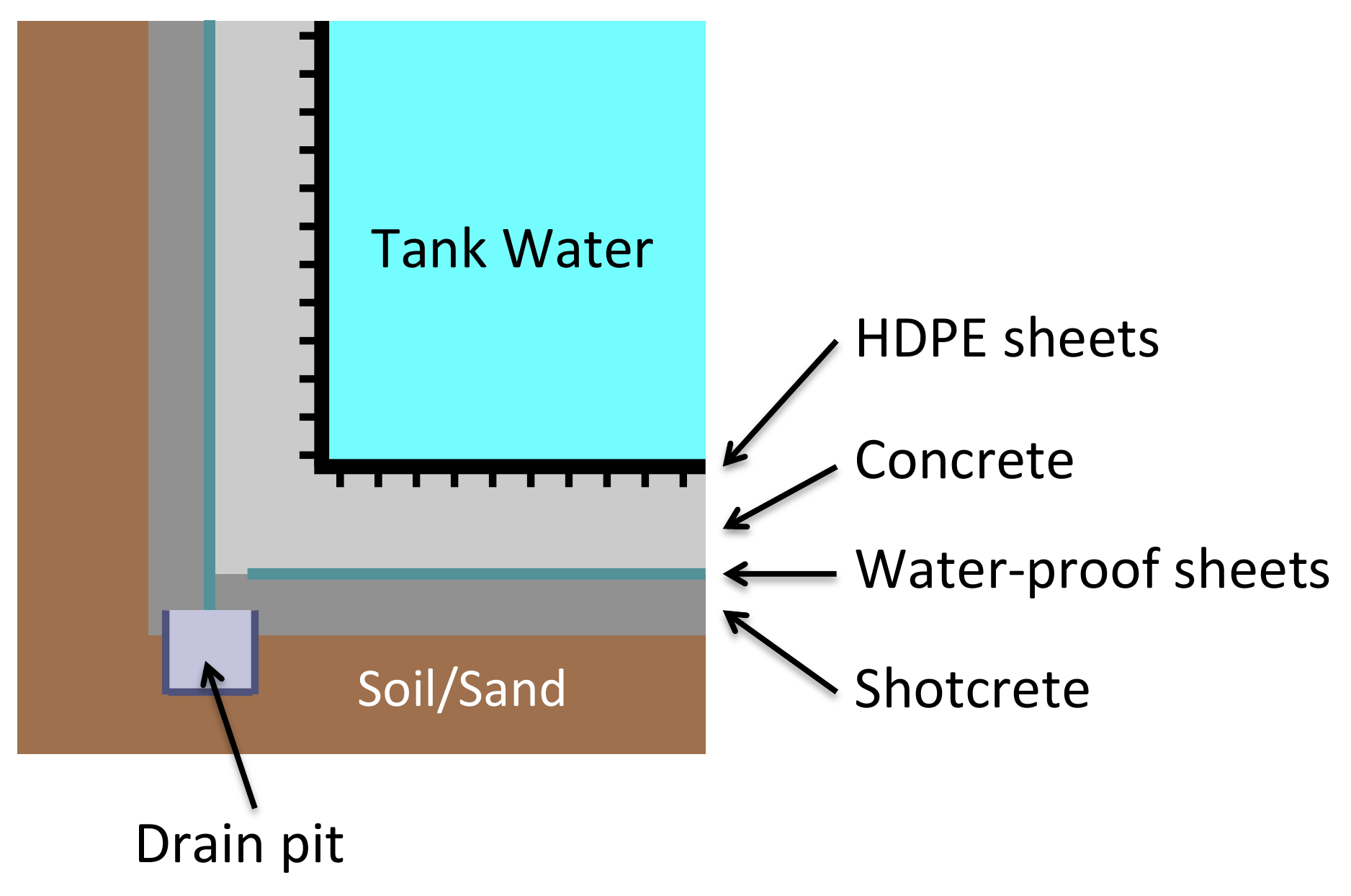}
  \caption{A schematic view of the \nuprism tank liner.}
  \label{fig:liner}
\end{figure}

The structure of the \nuprism tank liner is shown in Figure~\ref{fig:liner}. The innermost layer
contacting with the tank water must be a water-proofing component to seal the water within
the tank. We use High-Density Polyethylene (HDPE) sheets, which are commonly used as a
water-proofing tank liner material. The sheets have extremely low water permeability and also
are resistant to long-term damages from the ultra pure water. The adjoining sheets are
heat-welded, and the welded part also keeps the water-proof functionality.

We select the HDPE sheet with a number of studs protruding from one side. These studs work
for anchoring the sheet firmly on the backside concrete layer. To build this "HDPE on concrete"
liner, a HDPE sheet is fastened to the inside of a concrete form beforehand, then the concrete
is poured into the form for making the backfill concrete layer. While the thickness of the HDPE
liner is 5-10mm, the thickness of the backfill concrete layer is yet to be determined.

Though we aim to construct the HDPE sheet liner such that the tank water can not leak, an
additional water-proof layer is made between the backfill concrete layer and the shotcrete.
This layer works as a catcher and a guide for the water by the unexpected leakage through
the HDPE liner (and also the sump water through the shotcrete). This leaked water is drained
via pits placed under the water tank.

%The detailed information for the Hyper-K liner design (as a reference for the \nuprism liner)
%can be found elsewhere~\cite{hkmtg_tank_slides}.

\subsection{Detector Frame and Lifting Mechanism}

%Section goals:
%\begin{itemize}
%\item Describe PMT frame and rail system used to move the detector (RH talk from workshop)
%\end{itemize}

This section describes a proposed design for the frame that supports the \nuprism
PMTs and defines both the inner and outer detector.  We will also describe the system
by which this frame can be moved up and down in order to be able to make the \nuprism measurements.
Attention will be paid to the question of providing adequate water flow through the \nuprism
frame while maintaining optical separation.

\subsubsection{Detector Shape, Support and Positioning}
\begin{figure}[htpb]
\centering\includegraphics[bb=600 0 1450 820,clip,width=10cm]{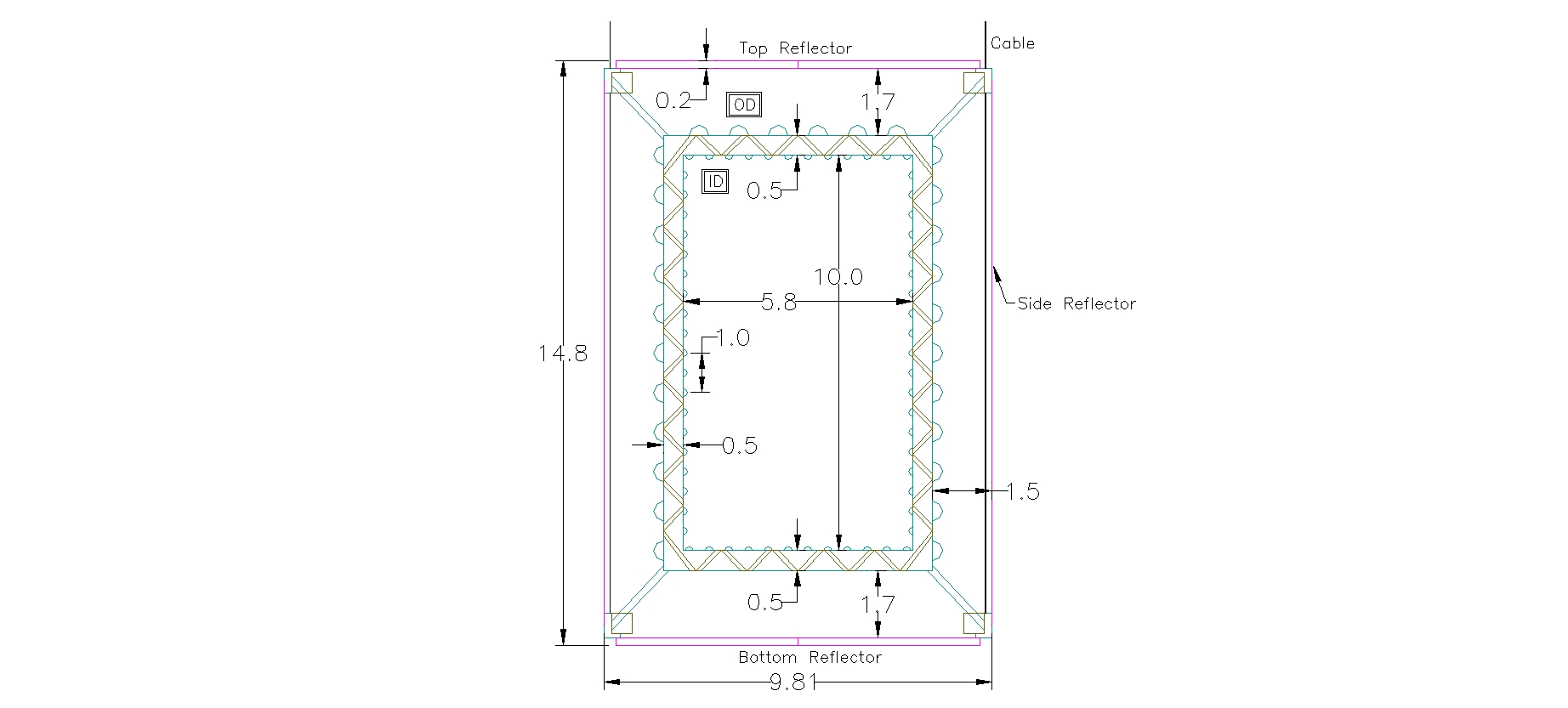}
\caption{The Detector is positioned on four rails inside shaft, and supported on top and bottom rings.
 Struts connect the detector to its rings. Four vertical cables support the assembly. Ballast can be 
added to the rings, if required. Distances are in meters.}
\label{fig:detdesign}
\end{figure}

Figure~\ref{fig:detdesign} shows a simple cylindrical design, the walls of the Inner Detector (ID) 
being 0.5 meters thick. The half circles represent the 20'' PMTs (0.5m) facing outward for the veto
region (OD). The smaller half circles represent 8'' PMTs (0.2m) facing inwards to the ID region.
% Is this actually the right spacing for the ID or OD PMTs?  Seems to be too much space between them?
 The 0.5m thickness of the detector wall is to contain the bodies of the PMTs (and PMT electronis)
 and, with internal
 stiffening braces, be stiff enough to accurately position the PMTs and not deform significantly 
under the weights and buoyancies. 

Figure~\ref{fig:detdesign} also shows a conceptual support
 and positioning system. The detector is positioned on four vertical rails fixed to the shaft walls,
 and supported on top and bottom rings. Struts connect the detector to these two rings. The struts 
are positioned
 at the corners of the detector where the structure is strongest, and angled so that
 the distance from the detector to the start of the reflector is 1.7m top and bottom, and 1.5m
 on the sides. The reflector encloses the OD region and is required to be optically isolated from
 the ID volume, and from the shaft water volumes above and below. We discuss the reflector
 in more detail below.

% Might be nice to have a top view of the detector as well, to understand where struts rails go.

\subsubsection{Water Flow and Optical Isolations}

Figure~\ref{fig:detcomp} shows views of the top-left corner of the detector and reflector.
 Two section views indicate the conceptual features and functions involved. The volume of the ID and OD are $\approx$264m$^3$ and $\approx$790m$^3$, respectively, with a combined volume of $\approx$1,190m$^3$ 
(including all
 wall volumes). If the apparatus is to traverse the shaft limits in $\approx$24 hours, the speed
 would be $\approx$1.5 meters/hour. Since the reflector side walls are close to the shaft, the 
displaced water needs to flow through the reflectors. This speed corresponds to a water
 flow of $\approx$118 m$^3$/hour = 2.0 m$^3$/min. Even if the water could flow past the sides of the
 reflector enclosure, ~1,190 tons of water would also be in motion, which would be difficult to accommodate.
With no water flowing through the sides of the reflector enclosure, the sides can be
 simple metal panels with a white inner surface to enhance the OD light collection.
As indicated in Figure~\ref{fig:detcomp}, these vertical reflector walls need to 
notch around the four rails and the associated couplings on the rings, and would be screwed to the top/bottom rings. With a height of 13.8m
 and circumference of 33.5m, it will need to be segmented with overlapping joints
 (or added joint strips). When the detector is out of the water, it would be useful
 to be able to easily remove the side reflector segments. Minimal segmentation would
 be four, with joints at the center of the `notches'.
This would allow the segments  to slid out past the rails and the support towers.
The top and bottom reflectors
 are also bolted to the top/bottom rings, but they have to be thicker to allow them
 to be strong and stiff due to the quantity of water flowing through them. The stiffness is achieved by
 making the top/bottom reflectors 0.2m thick and them having an internal bracing
 structure.  The top/bottom reflectors need an optical seal to the rest of the
 shaft, yet allow $\approx$2.0 tons/min of water to flow through. Figure~\ref{fig:detcomp}, 
shows two possible solutions:

\begin{figure}[hptb]
\centering\includegraphics[bb=0 350 900 1750,clip,angle=270,width=10cm]{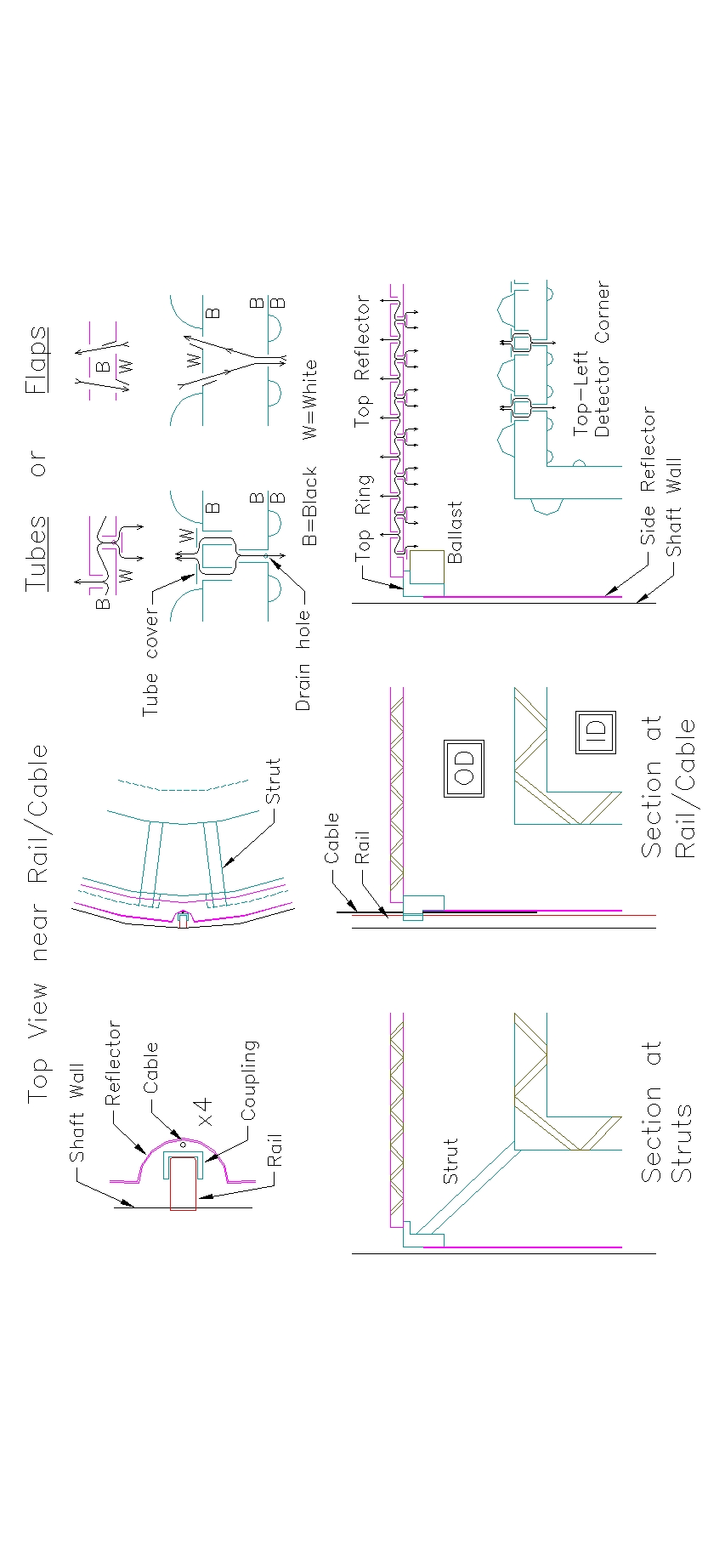}
\caption{This shows views of the top-left corner of the detector and reflector. Two 
sections views indicate the conceptual features and functions involved. A system of
 offset black pipes (or flaps) would allow water to flow through.}
\label{fig:detcomp}
\end{figure}

\begin{enumerate}
\item The first is a system of offset black pipes, so that water can flow through,
 but any light would need at least two reflections off black surfaces. The inner
 surface of the top/bottom reflectors would be white to enhance light collection.
 The tubes at the inner surface would have white `tube covers'. This is easily
 done by having the tube extend, with the tube cover fixed to it, but the tube
 having four side large slots, leaving webs of material to hold the cover. The
 cover and outer surface of tube extension would be white. To prevent water
 being trapped in the reflector wall when the detector is lifted out of the
 water, there would be `Drain holes' in the tubes, just inside the inner wall.
 Alternately, there could be some small drain tubes+covers extending slightly into OD
 volume. This scheme is a little complicated, but has the advantage of no moving parts. 
If the flow is fully distributed over the 55.0 m$^2$ area, the movement water flow would
 then be $\approx$36 liters/minute/m$^2$.
 
\item Another way solve this
 problem would to have a system of flaps that open only when the detector is moved,
 and close automatically when it stops. Half the flaps open when the detector moves
 down (water moves up), these flaps close under their own weight when movement stops.
 The other flaps open when the detector moves up (water down), these `down' flaps
 would need to be spring loaded or counterweighted to close when movement stops.
 In Figure~\ref{fig:detcomp}, we show both these flaps in the open position. The inner surfaces
 of the flaps would be white. In this scheme most of the water would drain through the spring
 loaded `down flaps', but would also require a system to small holes or pipes to drain out
 the last of the water. This system has many moving parts that cannot be lubricated, so
 binding and galling would be concerns, but it can probably be made to work. It would
 need to be made very reliable, a few flaps stuck closed wouldn't be a concern, but some
 stuck open could be a problem. This system has the disadvantage that it prevents lower
 levels of circulating water during data taking. This recirculating loop will probably
 be required for; the purification and temperature control of the water, cooling of
 electronics etc. For these reasons, we prefer the offset tubes option.
 % I think that clearly recirculating will be needed, so this is a little oddly phrased.

\end{enumerate}

When the detector is out of the water, the bottom reflector would need to be segmented to
 be removed between the support towers. With four towers (see Figure~\ref{fig:dettower}),
 the four bottom cover segments would be 4.6x4.6 meters. Higher segmentation (multiples of
 4) would also be possible. We imagine a scissor cart rolled under the detector, lifted to
 contact a segment. It could then be unbolted, lowered and rolled away. The segments would
 need to overlap on the inner surface for light seal, and on the outer surface for joining
 (or have extra joint strips). The top reflector would be craned out, in one piece or in
 segments.

\subsubsection{Walls of Inner Detector (ID)}
The top/bottom walls of the
 ID would also need to allow water flow, otherwise one would have to allow for
 the inertia of 400 tons of trapped water. The movement flows would be 42 tons/hour =
 0.7 tons/minute. Distributed, this is 27 liters/minute/m$^2$. This is somewhat less than
 the 36 liters/minute/m$^2$ of the reflector, but this wall has all the PMTs as well. In
 Figure~\ref{fig:detcomp}, I show the tubes and flaps options for this wall, similar
 to that for the top and bottom reflectors.

\subsubsection{Detector in the shaft}

\begin{figure}[htpb]
\centering\includegraphics[bb=750 0 1050 850,clip,width=7cm]{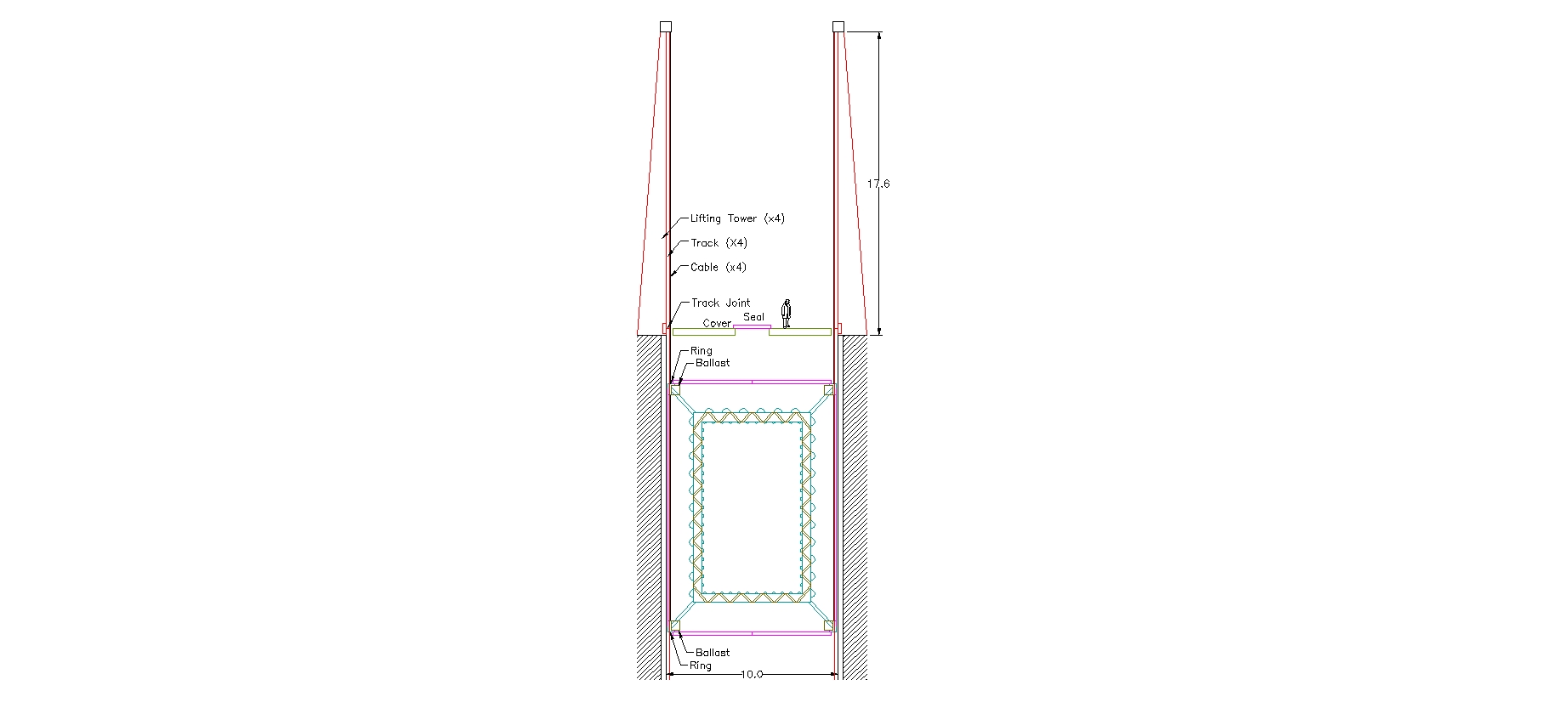}
\caption{This shows the detector in the shaft, the shaft covers and the external towers.
 Four vertical cables support the assembly. Above ground, there would be four towers upwards
 extending 17.6 meters. Four motors, acting together, lift or lower the detector in the shaft,
 or even lift it completely out of the water.}
\label{fig:dettower}
\end{figure}

The detector is guided within the shaft by a set of rails.
The current proposal has
 four rails and support cables but it could be three, five, etc. if dictated by other design considerations.
 It is important to understand that the ring connections to the rails do not need to be
 high precision rail bearings. Because the positioning accuracy required is only $\approx$1cm,
 they could be simple guides (see Figure~\ref{fig:dettower}). Similarly, the rails do not need to be
 complex. The loose tolerance makes it far less likely that the detector will jam on
 the rails. When the detector has been moved, there may be a system to lock two of the
 four guide locations to eliminate small position changes during data taking. Another
 reason for a looser coupling (before locking), is that then the rails do not need to
 be so precisely positioned on the shaft walls, i.e. several  millimeters versus
 0.1mm. 

Figure~\ref{fig:dettower} shows the detector in the shaft, the shaft covers
 and the external towers. Four vertical cables support the assembly. Ballast can be
 added to the rings, if required. Above ground, there would be four towers extending
 upwards ~17.6 meters. Four motors, acting together, lift or lower the detector in
 the shaft, or even lift it completely out of the water. The load will increase as
 it leaves the water (loss of buoyancy), if the load is too much, the top ballast
 can be removed by crane as it clears the water. Or, a lifting frame could be attached
 when the top ring clears the water, allowing the crane to raise it further, then it can
 be locked in the out position, freeing the crane.

In this concept, the signal and
 power cables for the detector would travel up out of the water beside the four support cables.
 They would nominally go up and over the towers, then down to the ground racks. With this
 scheme there would be no extra length in the water, wherever the detector was positioned
 in the shaft. When the detector is slowly lowered further down the shaft, the cables etc.
 should be cleaned before entering the water.

\begin{figure}[htpb]
\centering\includegraphics[bb=0 0 800 1000,clip,width=7cm,angle=270]{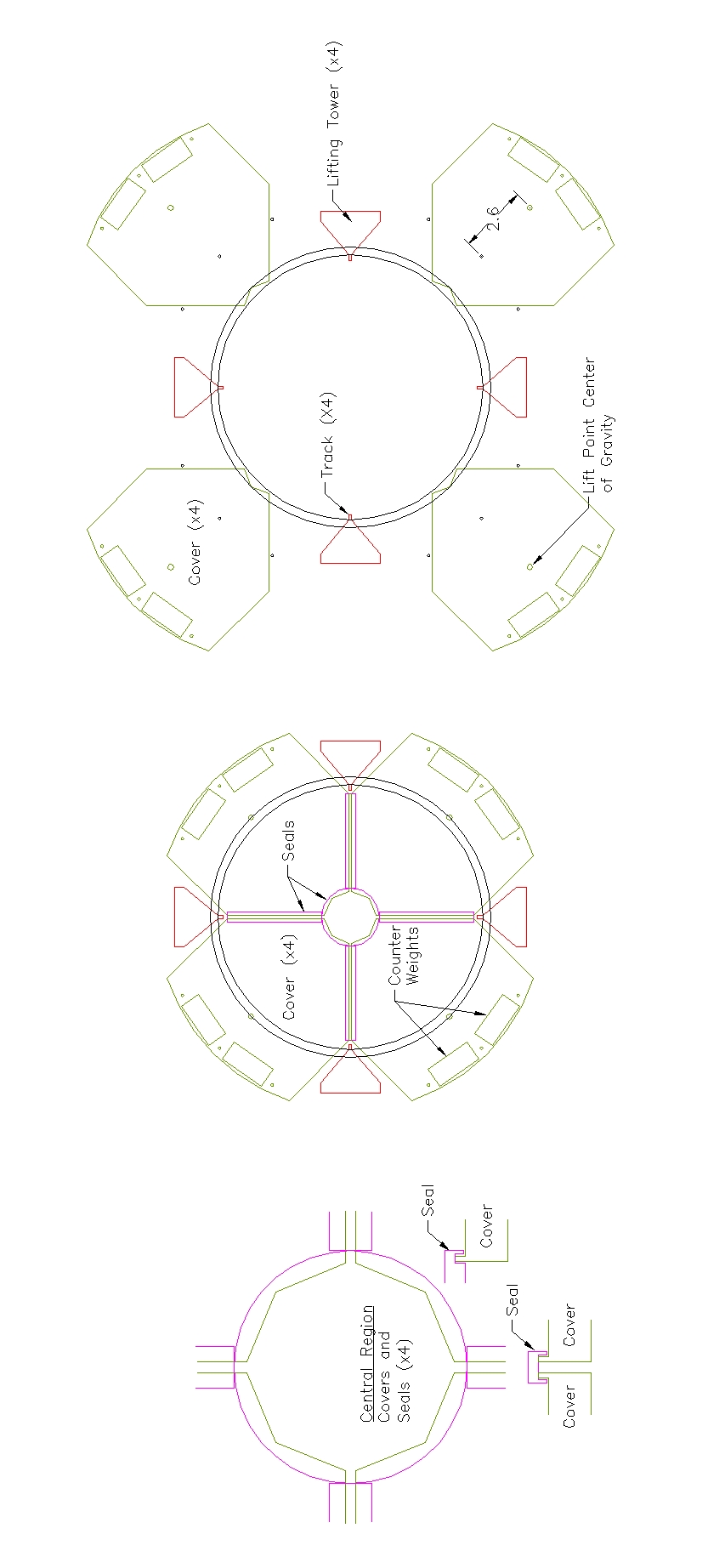}
\centering\includegraphics[bb=0 1000 800 1800,clip,width=7cm,angle=270]{figures/fig3-4.jpg}
\caption{The four covers can be craned in and out. Added counterweights make sure the center of
 gravity is beyond the detector shadow when in. The covers are bolted to the ground. Light weight seals
 cover the joints and the central region.}
\label{fig:detcover}
\end{figure}

Once the detector is entirely out of the water, the shaft covers can be craned back into
 position (see Figure~\ref{fig:detcover}). Adding counterweights will 
make sure the Center-of-Gravity (COG)
 of the covers are beyond the detector shadow when the covers are pushed in. 
The covers would be bolted to the ground.
 Lightweight seals cover the joints, the central region, and the four small areas where the support
 cables, signal and power cables exit the water.

\begin{figure}[htpb]
\centering\includegraphics[bb=0 300 800 900,clip=true,width=10cm,angle=270]{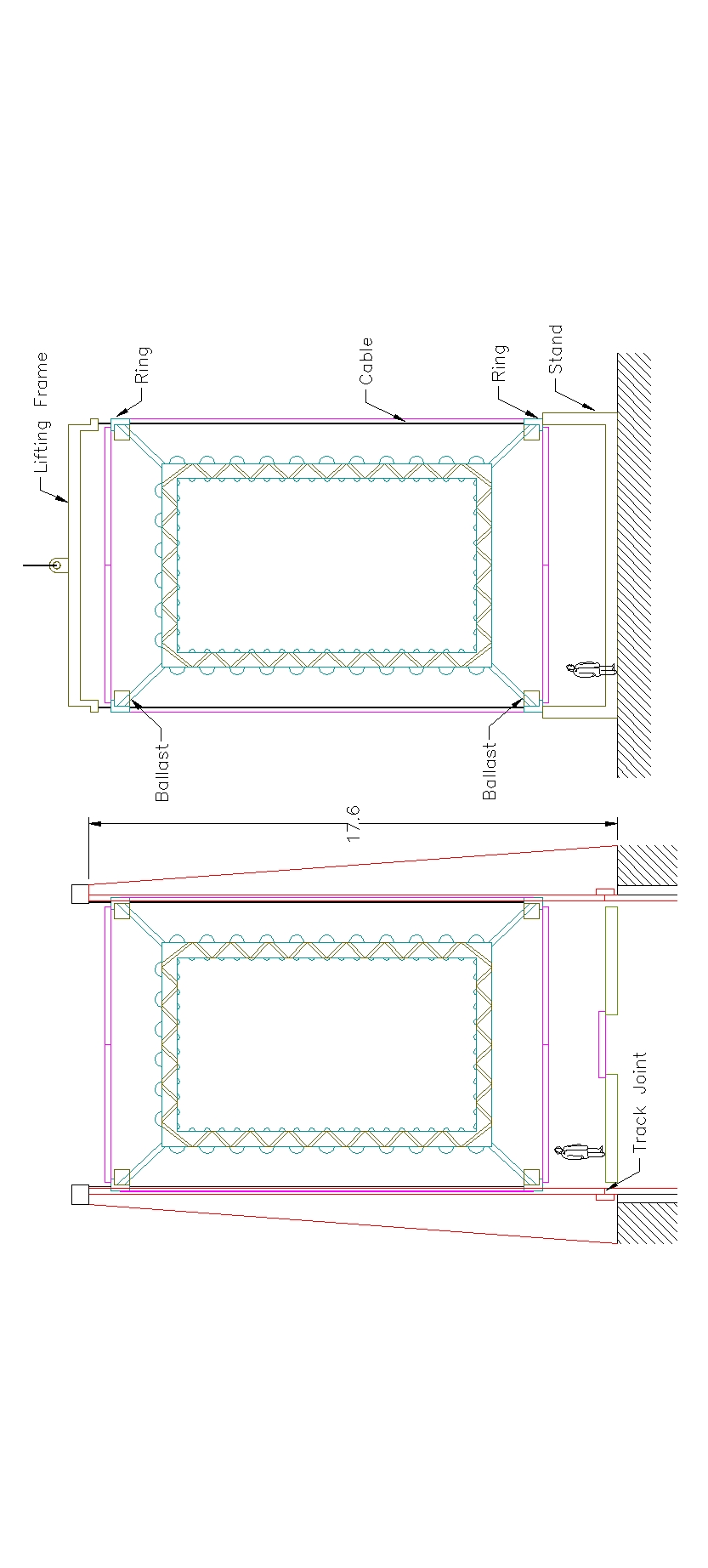}
\caption{The four towers allow the detector to be raised out of the water (units in meters). The covers  can be reinstalled under the detector, allowing people to work underneath it. A lifting frame can be craned over the detector, attached, and the towers removed. The detector can then be craned to a stand.}
\label{fig:detremoved}
\end{figure}

Figure~\ref{fig:detremoved} shows the detector out of the water and covers reinstalled. It is important that
 the covers and seals are safe for people and light equipment, so that the bottom of the detector can be
 worked on. Scaffolding can be erected to work on all parts of the detector. The figure also shows the
 detector moved to a stand. To move the detector, the lifting frame would be installed, the detector
 supported, then the eight ring guides removed and two of the towers removed (or laid down), opening
 a path for the detector move.

Whether above the shaft or on a separate stand, it would probably
 be useful to be able to remove the reflector sections and get access to parts of the ID.
 If the ID were bolted together sections, it might be possible to partially disassemble to make repairs and/or replacements.

\subsubsection{Detector Surveying}

As mentioned earlier, after the detector has been moved,
 there may be a system to lock two of the four guide locations to eliminate small position
 changes during data taking. A laser surveying system could be in place to look down through the
 water to periodically check the detector position at the four rail locations. The PMTs may have
 to be turned off during these times. The positioning of PMTs within the detector would be surveyed
 during its assembly (out of water) and then should only be subject to thermal expansion/contraction
 shifts in the water, plus deflections due to loads (primarily the top/bottom PMTs.

The thermal
 expansion/contractions of the detector will depend on its material. For a 10 meter Aluminum piece,
 the expansion would be ~2.2mm for a 10 OC change. 306 stainless steel would be ~1.6mm. The shaft, being reinforced concrete, should expand ~1.3mm for 10 degrees. Stainless steel is
 a better thermal match, the differential expansion being ~0.3mm for 10 degrees, compared to ~0.9mm 
for an Aluminum detector frame, but the difference is not likely to be significant.

%\afterpage{\clearpage}

\subsection{Scintillator panels}\label{sec:scint}

The veto system of the \nuprism detector can be composed of plastic scintillator detectors which completely surround the the water Cherenkov detector. The main purpose of the veto system is to identify backgrounds from beam neutrino interactions in the surrounding pit walls  and to provide a cosmic trigger signal for calibration purposes. The technology developed for the ND280 SMRD detector can be applied for this veto system.

\subsubsection{Scintillator counters with WLS/avalanche photodiode readout}

Scintillator counters with wavelength-shifting (WLS) fibers and opto-electronic readout are an established technology for neutrino detectors in  long-baseline neutrino oscillation experiments. ND280 consists of several subdetectors which use extruded plastic scintillators of various shape and dimensions~\cite{nd280}. Each of these subdetectors is comprised of plastic slabs and bars, wavelength shifting fibers and compact photosensors - multi-pixel avalanche photodiodes. The Kuraray  double-clad Y11 WLS fibers are used in all ND280 scintillator detectors for transportation of the reemitted light to photosensors.

{\it SMRD counter}. The SMRD detector was made of the polystyrene-based scintillator slabs, each with an embedded wave-length shifting  fiber. The slabs were produced at the Uniplast Factory (Vladimir, Russia). The scintillator composition is a polystyrene doped with 1.5\% of paraterphenyl (PTP) and 0.01\% of POPOP. The slabs were covered by a chemical reflector by etching the scintillator surface  in a chemical agent that results in the formation of a white micropore deposit over a polystyrene\cite{extrusion}.  The chemical coating is an excellent reflector, besides it dissolves  rough surface acquired during the cutting process. The WLS fiber was read out on both ends to increase light yield, improve uniformity and position accuracy, and provide redundancy.
%(see Fig.\ref{fig:smrd_slab}).
%\begin{figure}[h!]
%\centering\includegraphics[width=14cm,angle=0]{figures/slab.eps}
%\caption{SMRD scintillator slabs with a serpentine-routed Y11 WLS fiber.}
%\label{fig:smrd_slab} 
%\end{figure}

A key feature of   these counters is the usage of the one serpentine-shaped WLS fiber for readout of scintillating signal.   The serpentine geometry of a groove consists of 15 half-circles, each with a diameter of 58 mm and straight sections connecting the semi-circles. A 1 mm diameter Y11 (150) Kuraray WLS fibers of flexible S-type and with double-cladding was used for the SMRD counters. Fibers are bent into a serpentine-shape and glued into grooves with BC600 Bicron glue.
%The light yield distribution  of all SMRD counters is shown in Fig~\ref{fig:smrd_ly}.
%\begin{figure}[h!]
%\centering\includegraphics[width=12cm,angle=0]{figures/smrd_ly.eps}
%\caption{The light yield (sum of both ends) distribution for 2024 SMRD 
%counters.}
%\label{fig:smrd_ly}
%\end{figure}
%As seen from this figure, 
The mean light yield for sum of both ends  was about 40 p.e./MIP after subtraction of the MPPC cross-talk and after pulses. The high light yield allowed us to obtain the efficiency of more than 99.9\% for detection of  minimum ionizing particles.

\begin{figure}[htpb]
\centering\includegraphics[width=9cm]{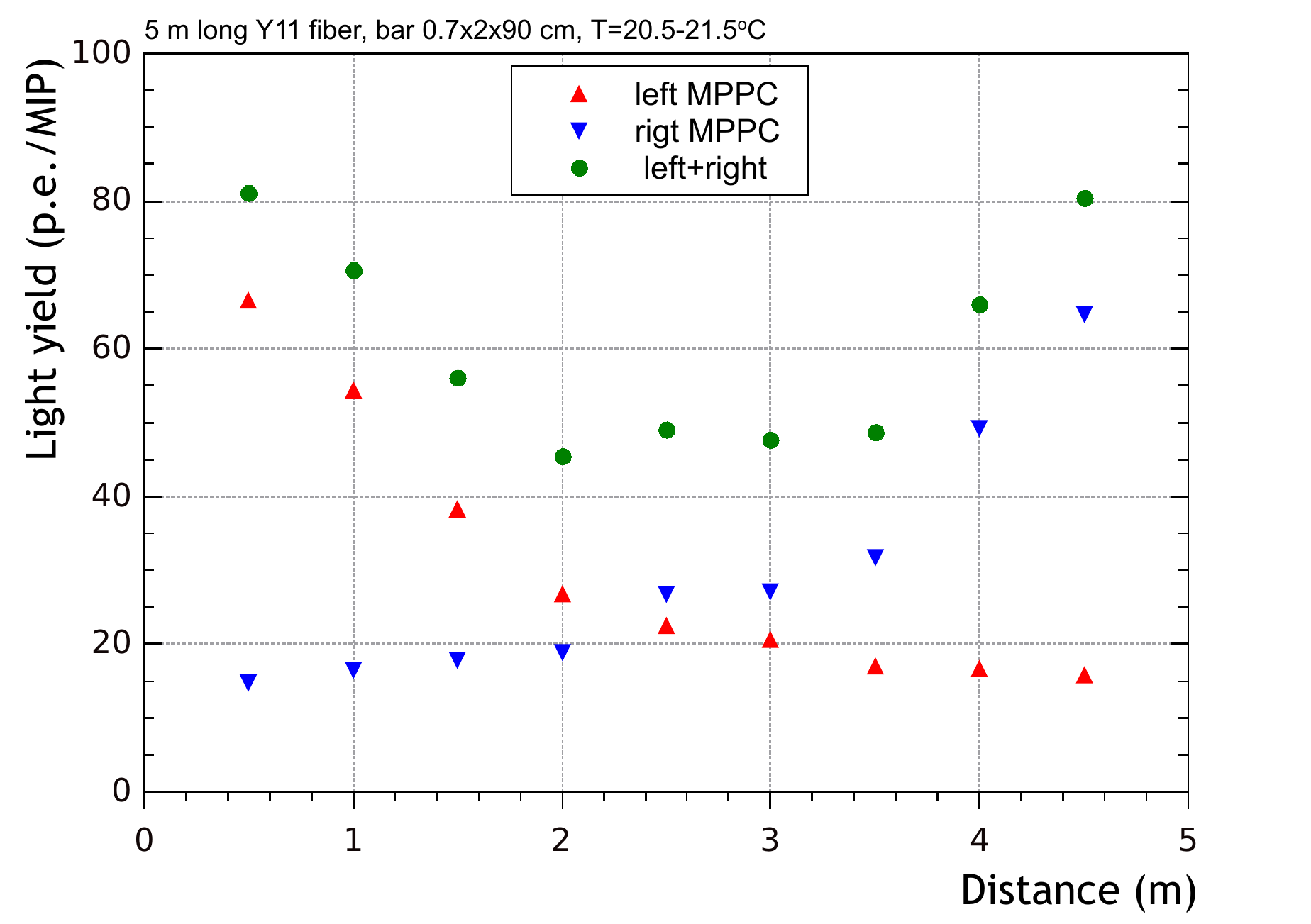}
\caption{Light yield of a scintillator counter with  5 m long WLS  fiber vs position along the fiber. The T2K 667 pixel MPPC's were used in this measurement.}
\label{fig:graph1} 
\end{figure}

%{\it Long fibers}. To test the performance of scintillator detectors with longer WLS fibers,  
%90~cm long and 0.7~cm thick scintillator slabs were extruded.   
%A 2~mm deep and 1.1~mm wide groove was machined along a bar central line to accommodate the 16~m long Y11 fiber. Since a tested bar is moved along the fiber the optical coupling between the fiber and the bar was implemented with an optical grease.
%Tests of scitillator bars  with embedded WLS fibers and MPPC's showed very good results, as seen in Fig.~\ref{fig:bar2}.
%\begin{figure}[h!]
%\centering\includegraphics[width=11cm,angle=0]{figures/bar2.eps}
%\caption{Total light yield from both fiber ends  vs position along the Y11 fiber for the 3~cm wide bar.}
%\label{fig:bar2} 
%\end{figure}
%
%For example, one-end readout provides the light yield  of about 12-15 p.e. for a MIP passing the detector at the distance of 6 meters from the photosensor.  
The light yield of about 14 p.e. per a minimum ionizing particle ($\sim 7$ p.e./MeV for 1 cm thick bar) provides the efficiency  for detection of minimum ionizing particles of more than  99\% in an individual scintillator bar  for a detection threshold of 1.5 p.e.  Time resolution depends on the  light yield as $\sim 1/\sqrt{N_{p.e.}}$ where $N_{p.e.}-$ is the number of photoelectrons. For the l.y. of 20 p.e. the typical resolution  is obtained to be $\sigma ~ 1$ ns. Detectors with shorter WLS fibers were also tested. 
Light yield of the detector with a 5 m long WLS  Y11 fiber is shown in Fig.~\ref{fig:graph1}

In this case, the minimum light yield of more that 40 p.e./MIP (sum of both ends) is obtained.

\subsubsection{Veto counters for \nuprism}

i%\begin{figure}[h!]
%\centering\includegraphics[width=10cm,angle=0]{figures/pde_new_mppc.eps}
%\caption{Photon detection efficiency  of the MPPC S12651-050C as a function of wavelength of the detected light~\cite{musienko_instr14}.}
%\label{fig:pde_new_mppc}
%\end{figure}

The excellent performance of the SMRD counters with one serpentine WLS fiber per counter gives a possibility to make a veto system using similar approach. One option is to construct the \nuprism veto system from scintillator counters,  each of 0.2 m$^2$.
%The geometry of one counter is shown in Figure~\ref{fig:nuprism_counter}.
%\begin{figure}[h!]
%\centering\includegraphics[width=13cm,angle=0]{figures/counter.eps}
%\caption{Drawing view of a scintillator counter for the \nuprism veto system.}
%\label{fig:nuprism_counter} 
%\end{figure}
One WLS Y11 S-type fiber is embedded in the extruded plastic slab of $2000 \times 200 \times 7 {\rm mm}^3$. Half-circles have the radius of 3 cm that allows to keep the performace of the fiber without loosing the transmission of the reemitted light along the fiber. A 6 m long Y11 fiber is readout on both ends by MPPC's. Taking into account the improved parameters of new MPPC's, for exmple, higher PDE, 
%as shown in Figure~\ref{fig:pde_new_mppc}, 
we can expect to obtain miminum light yield of 20-30 p.e./MIP and time resolution of about 1 ns for these detectors. More accurate information can be obtained after tests of the conter prototypes. 

\subsection{Photomultiplier Tubes}

%Section goals:
%\begin{itemize}
%\item 20 inch Hyper-K HPDs for OD
%\item 5 or 8 inch PMTs for ID (could be high QE or HPDs)
%\item discussion of cost
%\end{itemize}

The original T2K 2~km detector proposal used 8" PMTs to better match the granularity of the 20" PMTs used in the much-larger Super-K detector. The baseline design for the \nuprism detector is only 6~m in diameter and 10~m tall, which corresponds to 3,120 PMTs for 40\% photocathode coverage. This is significantly smaller than the 11,129 PMTs used at Super-K, so to improve the granularity of the detector, 5" PMTs are also being investigated, of which 7,385 PMTs would be required for 40\% coverage. Additional options such as avalanche photodiodes and high quantum efficiency coating are also being explored. %Initial cost estimates from Hamamatsu for a wide variety of PMT configurations are given in the appendix.

\subsection{Electronics}

%\subsubsection{Electronics Overview}

Part of the goal of the \nuprism is to serve as a prototype for the Hyper-K.
We therefore want \nuprism to use a set of electronics that is as close as possible to the
electronics being proposed for Hyper-K.  Some of the key features of the Hyper-K electronics 
are the following:

\begin{itemize}
\item Front-end electronics will be placed in the water, as close as possible to the PMTs.
\item Front-end electronics are expected to find all hits above 0.25 PE and send all information 
about hits up to back-end electronics.  In back-end computers trigger decisions will be made using software.
No global triggers will be propagated to the front-end electronics.  
\item PMT digitization should provide 0.05 PE charge resolution, 0.25 ns timing resolution (for 1PE hits) and
0.1-1250 PE dynamic range.
\end{itemize}

We shall note various aspects of the \nuprism electronics where we may differ from
the default HK electronics plan.
In particular, one clearly different aspect of \nuprism will be the much higher rate of 
`pile-up' events during beam spills.  The rate of sand muon events entering the ID
may be as high as 0.19 per bunch.  
At minimum we therefore need electronics that can cleanly distinguish 
between PMT hits in different bunches; ie, hits with separation of order $\approx$ 600ns.  We may also 
want to have some capacity to distinguish between hits within a single bunch; ie hits that differ by 10s of ns.
This would be a more challenging requirement.

\subsubsection{FADC Digitization}\label{fadc_digitization}

Given this requirement for inter-bunch and intra-bunch hit resolution we propose using 
FADC digitization with basic digital signal processing in the front-end electronics.  
The basic scheme is as follows:  

\begin{enumerate}
\item The stretched/shaped PMT signal is fed into the FADC.
Use a standard commercial FADC, with sampling frequency between 80-500 MHz and 
12-16 bit resolution.
\item The digital output of FADC is fed into an FPGA (on the front-end electronic card), 
where we do basic digital pulse processing
(on the fly, at same rate as original digitization).  Digital pulse processing would involve the 
following:
\begin{itemize}
\item Finding PMT hits (for instance, by using simple threshold comparison).
\item Calculating the pulse time and charge.  
\end{itemize}
\item The digital pulse information is then transferred to the back-end electronics.  
We send different types of data depending on the pulse charge.

\end{enumerate}

It is worth emphasizing that the 
expected timing resolution using FADCs is not intrinsically limited by the sampling.  For instance, if you appropriately 
stretch and shape a PMT pulse you can easily achieve 0.25 ns timing resolution using a 100 MHz digitizer (ie a sample each 10 ns), as long
as you have high signal to noise ratio and a reasonable number of ADC samples on the leading and falling edges. 
We will explore the trade-offs involved in optimizing the performance of such a system in Section \ref{fadc_performance}.

\begin{figure*}[hptb]
\centering
%\resizebox{0.85\textwidth}{!} {
  \includegraphics[width=0.7\linewidth]{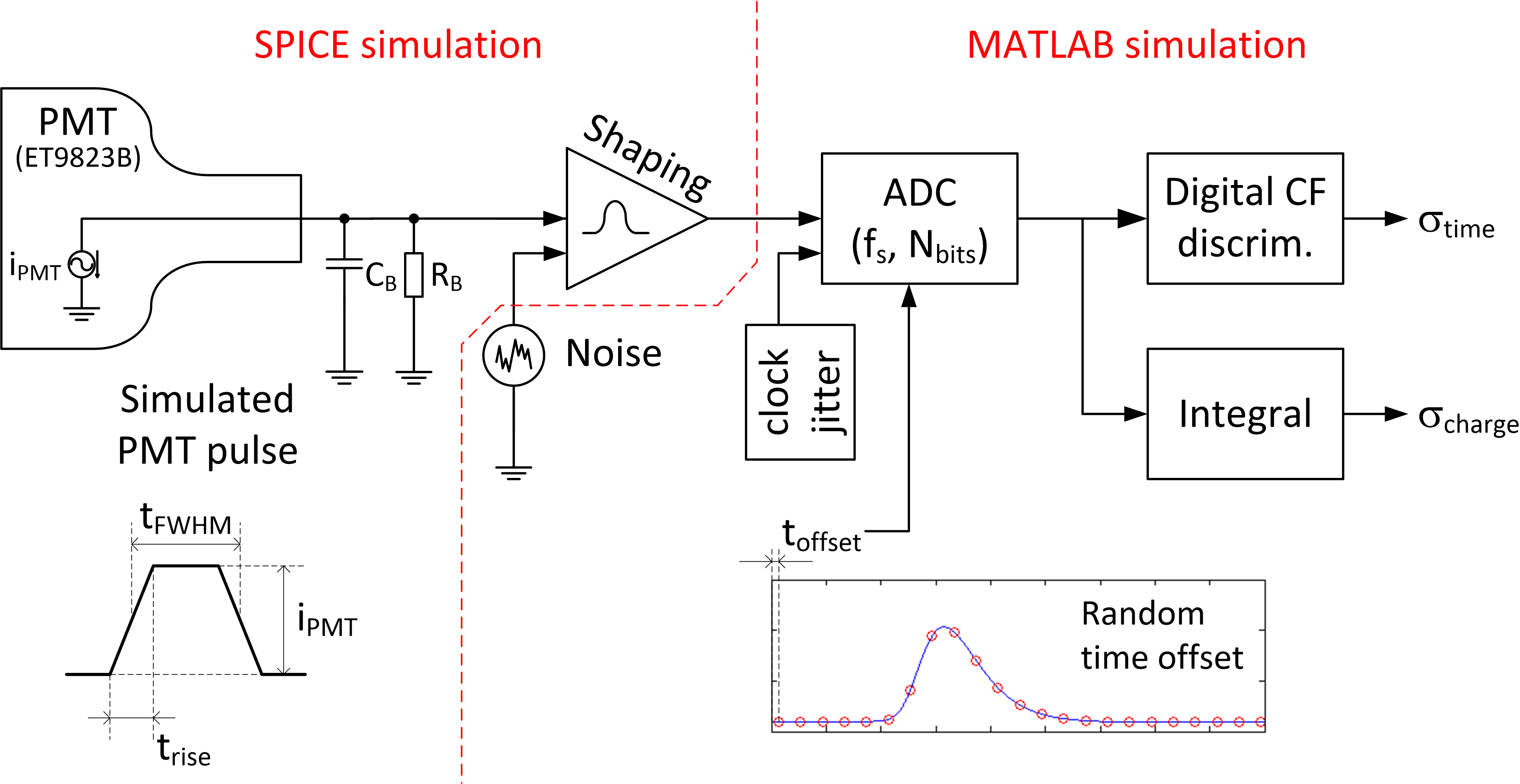}
%}
  \caption{Simulation setup for the study of the FADC performance.}
\label{fig:fadcSimSetup}
\end{figure*}

\subsubsection{Signal Conditioning And PMT HV}\label{fadc_signal_conditioning}

We propose to use differential transmission in order to deliver signals from the PMT bases to the digitization board. An advantage of such a solution is that, in principle, it would allow us to use a standard unshielded twisted pair cable, while still maintaining fairly good immunity to pickup of electromagnetic interference. The base of the PMT would contain shaping circuitry, which would stretch PMT signals, limiting their bandwidth to match FADC requirements and converting them into a symmetric form, suitable for transmission via a twisted pair cable. Preliminary studies show that signal shaping using a 5-th order Bessel-type low pass filter should provide satisfactory results. 

One of the design goals for the \nuprism is minimization of the amount of necessary cables. As such, it would be advisable to use a single cable to provide both high voltage to the PMT and to transmit the signal from the PMT base to the digitization board. Therefore, the preferable solution would be to synthesize the high voltage directly on the PMT base, from a 48-200 V DC supply, using either a commercial high voltage module or a custom designed voltage multiplier structure. This way, power to the PMT base could be delivered via an additional twisted pair of the same cable that would be used to transmit the shaped PMT signal. The slow control link necessary to tune the high voltage for specific PMT could be realized via a DC power line, thus avoiding the need to use additional cables. In any case, it should be emphasized that the details of the PMT HV implementation will depend strongly on the exact PMTs that are chosen.

\subsubsection{Digitization Performance/Optimization}\label{fadc_performance}

There is a strong inter-dependance of the digitization performance on the signal conditioning, type and
parameter of the chosen analog-to-digital converter (ADC) and the applied signal processing algorithms. The key parameters here are the speed and accuracy of the ADC\ as well as signal to noise ratio (SNR) of the whole system. Cost-wise, it would be best to use as slow and as least accurate ADC as possible while still meeting the performance requirements. Therefore, a Monte-Carlo study has been performed in order to estimate impact of the electronics chain on the overall system performance.  

\begin{figure}[htpb]
\centering
%\resizebox{0.99\textwidth}{!} {
  \includegraphics[width=9cm]{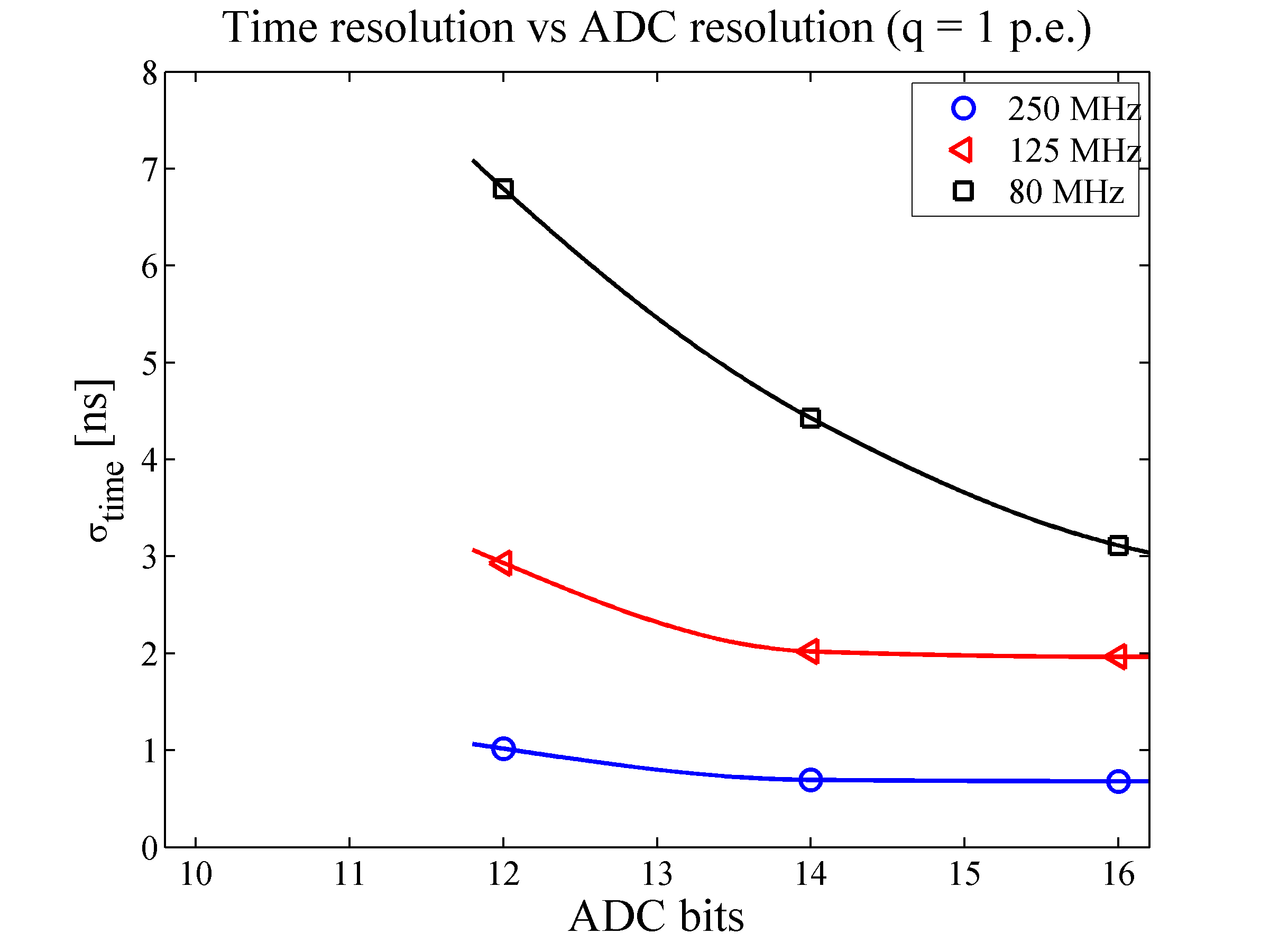}
  \includegraphics[width=9cm]{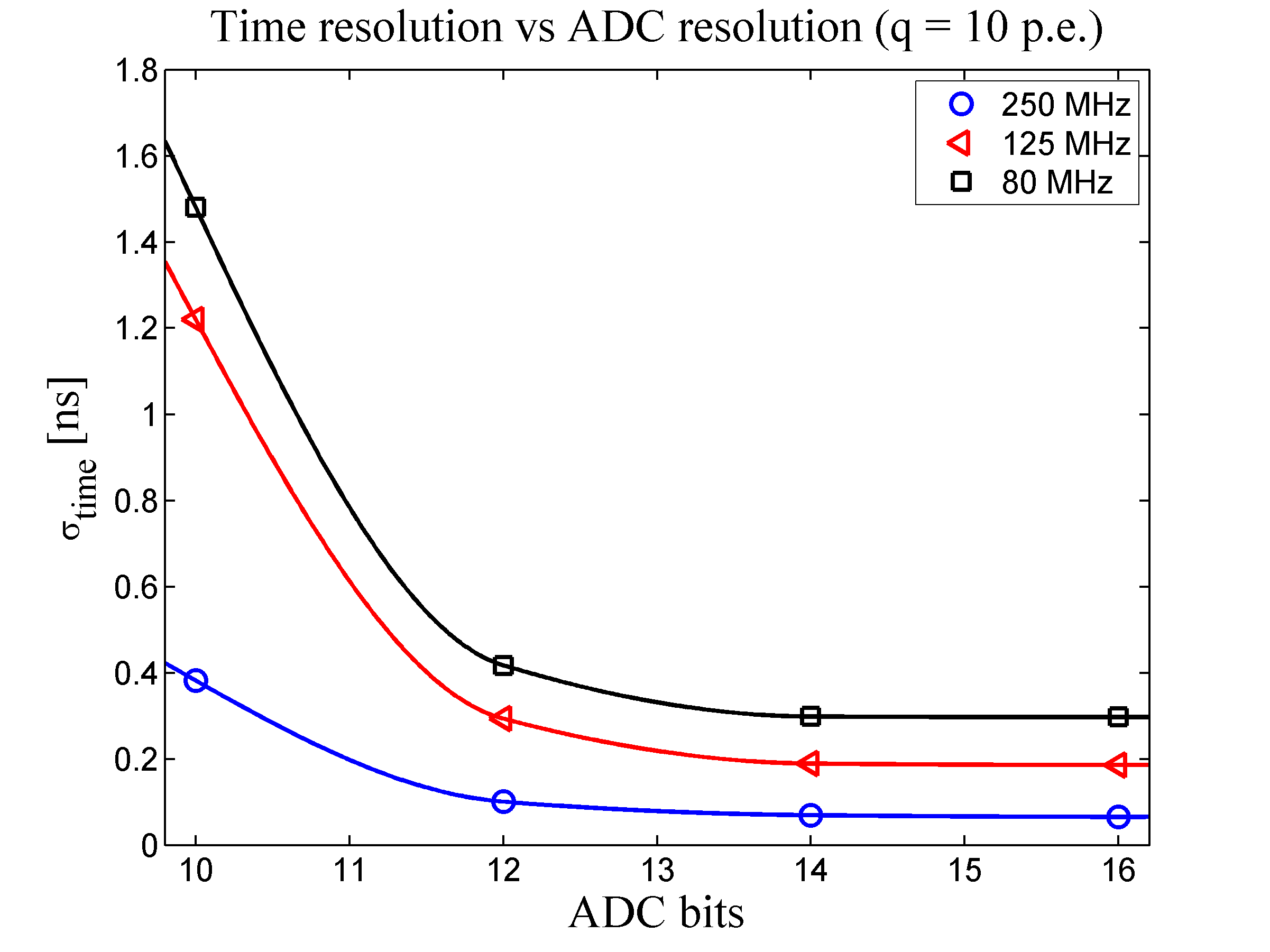}
%}
  \caption{Estimated timing resolution for FADC digitization, as a function of sampling frequency and ADC precision.  The top plot is for 1PE pulses; bottom is for 10PE pulses.}
\label{fig:timing_vs_dyn}
\end{figure}

\begin{figure}[htpb]
\centering
%\resizebox{0.99\textwidth}{!} {
  \includegraphics[width=9cm]{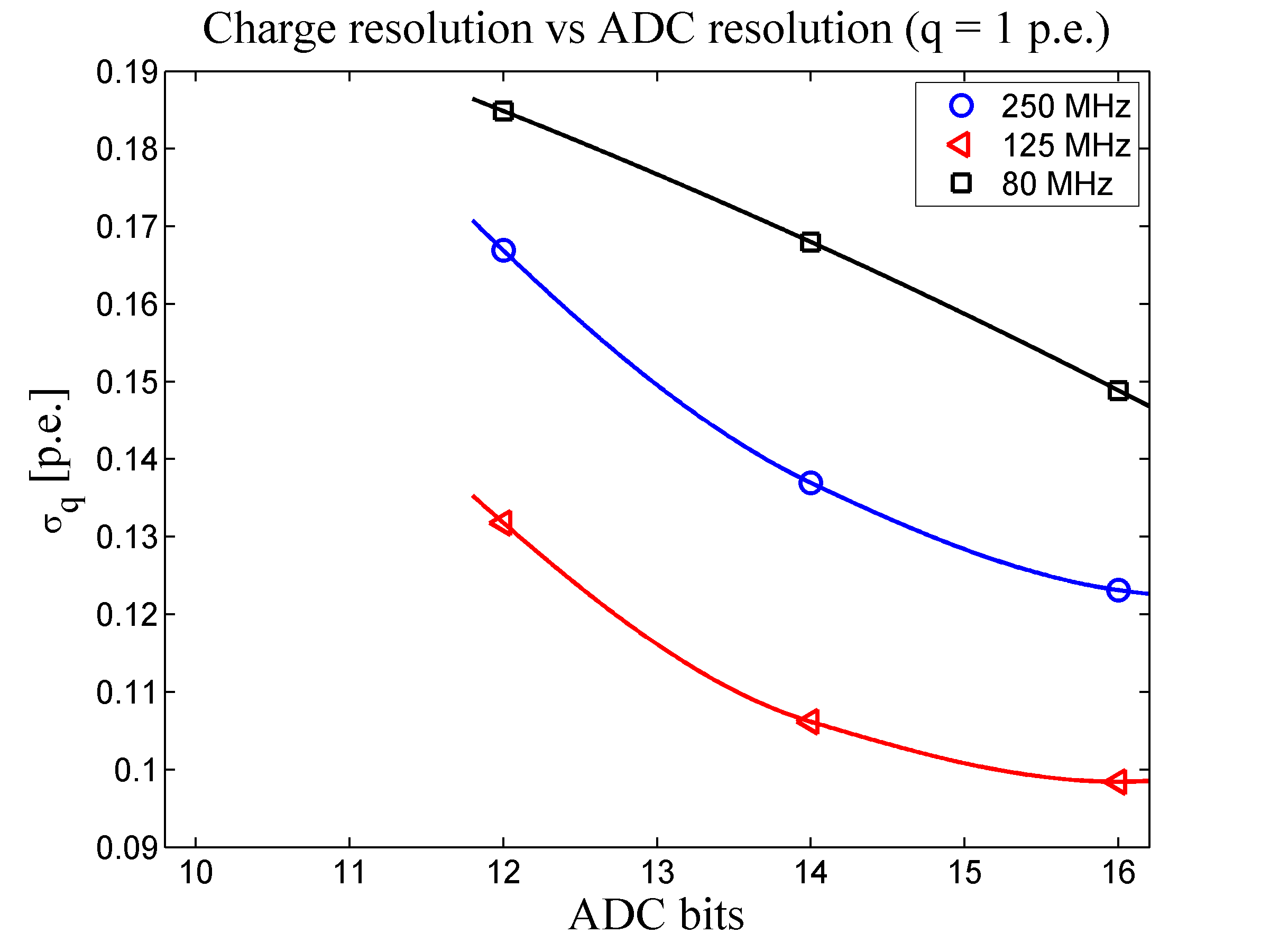}
  \includegraphics[width=9cm]{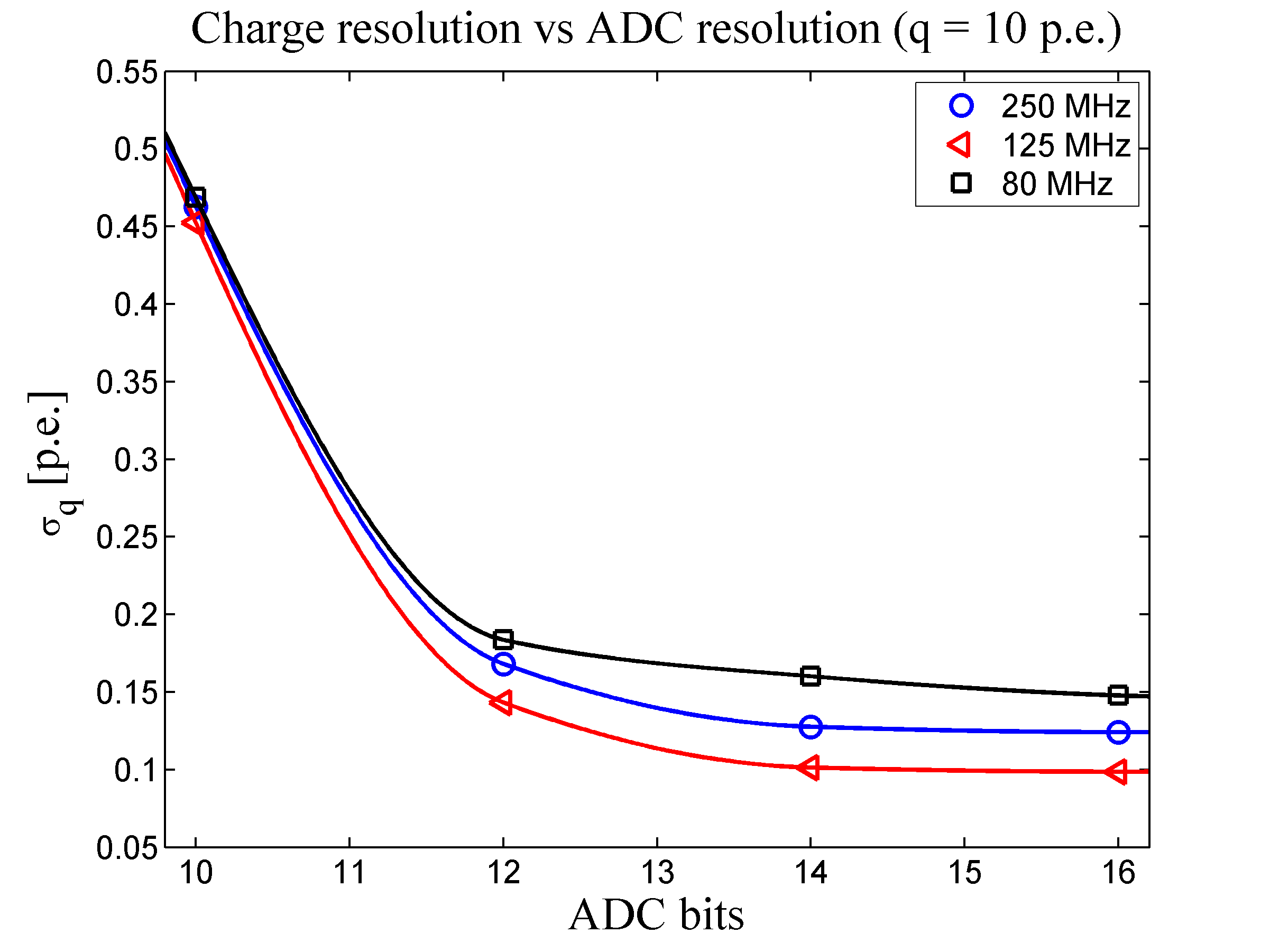}
%}
  \caption{Estimated Charge resolution for FADC digitization, as a function of sampling frequency and ADC precision.  The top plot is for 1PE pulses; bottom is for 10PE pulses.}
\label{fig:charge_vs_dyn}
\end{figure}

Simulation setup is presented in Fig.~\ref{fig:fadcSimSetup}. The photomultiplier has been simulated as a current source (\(i_{PMT}\)), connected in parallel with a base resistor \(R_{B}\)  and a capacitor \(C_{B}\), which together form the first pole of the shaping low-pass filter. Both the \(R_{B}\) and the \(C_{B}\) were chosen to fulfil the dynamic range requirement while maintaining the best possible signal to noise ratio,  i.e. to provide the highest possible PMT signal for maximum pulse charge (2000 p.e.) without saturating the amplifiers. The PMT current pulse waveform was approximated using a trapezoid pulse, with timing parameters (\(t_{rise}\), \(t_{FWHM}\)) corresponding to manufacturer specification given in the PMT datasheet. The rise and fall times were assumed to be equal. Given the time constants of the shaper, the PMT pulse can be treated as a delta function. 

The output of the shaper's response simulated in the SPICE   program was then sampled, quantized and subsequently analyzed using a digital Constant Fraction Discrimator (CFD), modeled in MATLAB.
Using the result of the CFD,
the difference between the calculated time and the real time was calculated, as well as the difference of calculated pulse charge and the real pulse charge.
A summary of the results is presented in Figures \ref{fig:charge_vs_dyn} and \ref{fig:timing_vs_dyn}.  As can be seen, there is some
difficulty in achieving the desired timing and charge resolution for the 1 p.e. pulses, which is due to poor signal to noise ratio. In particular, even with a 16-bit, 250MHz sampling we can only achieve approx. 0.8 ns timing resolution for the single p.e. (compared to the desired 0.25ns resolution).

As such, further studies are ongoing in order to find a working solution. The considered options include splitting the signal from the PMTs to separate high and low gain
branches which would then be digitized by their own ADCs. Other possibilities include dropping the linearity requirement for the PMT response to large number of photons and running it at higher gain. A significant effort is also foreseen to optimize signal processing algorithms for poor SNR conditions, in particular an adoption of matched filtering approach is planned.

\subsection{Water System}

Starting with the very first large-scale Water Cherenkov detector -- the
Irvine Michigan Brookhaven [IMB] proton decay experiment, which began
taking data in the early 1980's -- exceptional water clarity has been of
key importance for massive devices of this kind.  There is
little benefit in making a very large detector unless the target mass
contained within the detector can be efficiently observed.  Good water
quality has two main advantages: the light generated by physics
interactions in the water can propagate long distances with minimal
attenuation until it is collected by photomultiplier tubes or other
technologies, aiding accurate energy reconstruction, and the light can traverse
these distances (10's of meters) with minimal scattering, which aids
in the precise reconstruction of event vertices.

The strategy employed to create
kilotons of extremely clear water has been to remove all suspended solids,
dissolved gases, ions, and biologics from solution via a series of
filtration elements.  These include microfiltration filters, degasifiers
(vacuum and/or membrane type), reverse osmosis membranes [RO],
de-ionization resins [DI], and exposure to intense ultraviolet light [UV].

These water systems typically run in one of two modes: fill or
recirculation.  During the fill mode, water supplied by the local
municipality or ground water in the vicinity of the experiment is first brought
up to ultrapure levels and then injected into the detector.  The capacity of the water 
system, along with availability of water, defines how long it will take to fill the detector.  
During recirculation mode, already high-quality water from the detector is
continuously passed through the filtration system and returned to the
detector after being cleaned even further. This is necessary as transparency-impairing
materials are steadily leaching into the chemically active ultrapure water.  In addition, 
during the process of filtration the water is typically chilled to further
impede biological growth, with the added benefit of simultaneously reducing PMT 
dark noise which is typically strongly temperature dependent.

In the current baseline design, \nuprism will have interior dimensions ten times smaller than Super-K.
It is therefore possible that a commensurately less powerful water filtration system would be able to
provide sufficient water transparency.  Nevertheless, for now we will base our initial system design 
and flow rates on water systems known to have worked and produced useful physics in the past.

\begin{figure}[htpb]
     \begin{center}
       \includegraphics[width=9cm]{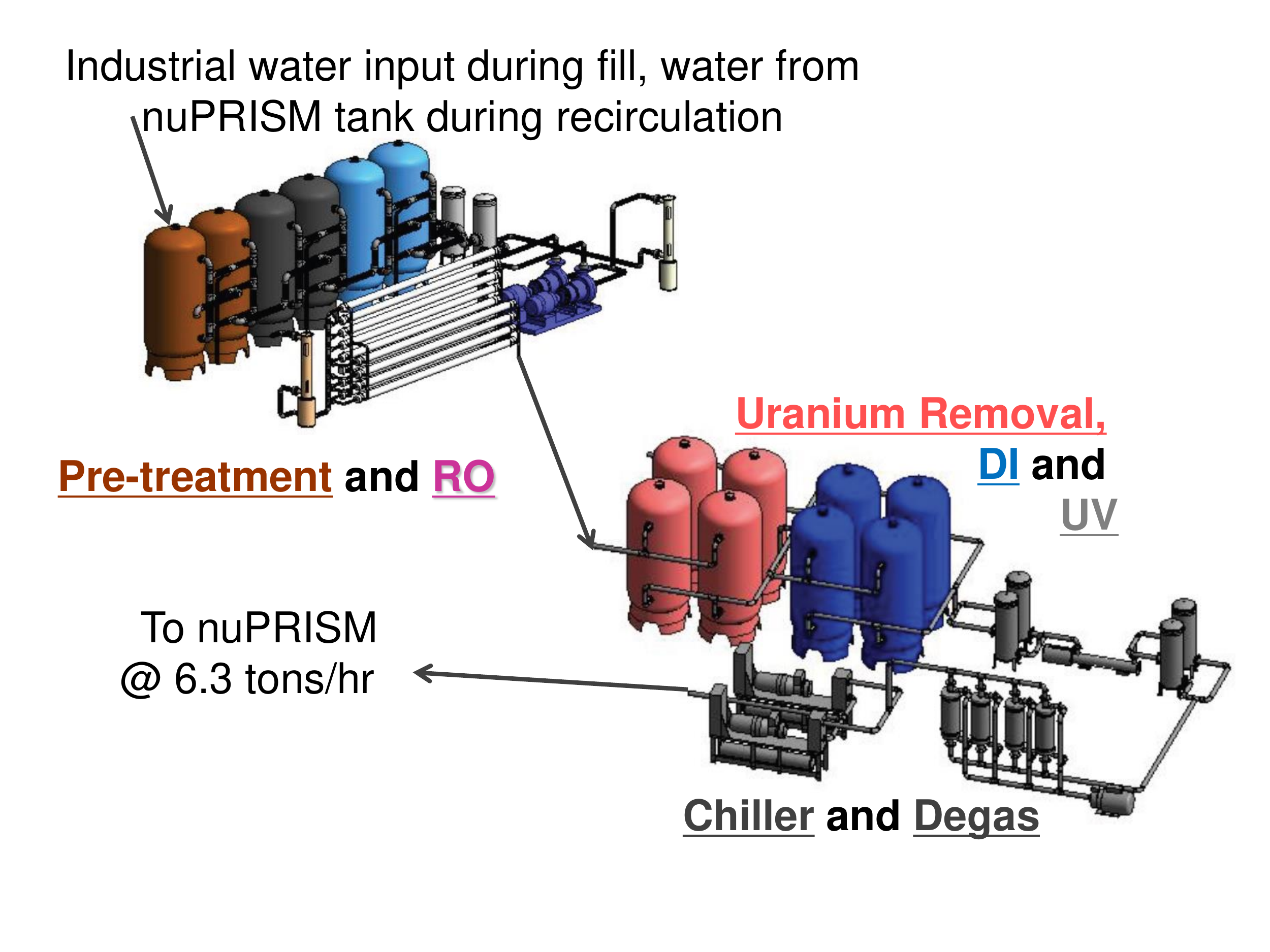}
       \caption{A preliminary baseline design of the \nuprism water system.}
       \label{water:water}
     \end{center}
\end{figure}

Following this approach, a baseline design and cost estimate for the \nuprism water system 
has been prepared.  The primary components described above are represented graphically in 
Figure~\ref{water:water}. This system will be capable of filling the detector at a rate of 
6.3 tons/hour, such that a complete fill can be completed in one month of operations.  
It will be capable of recirculating the water at a rate of  6.3 tons/hour through the entire system 
plus an additional 22.8 tons/hour through what is known as a secondary "fast recirculation" 
path which trades some filtration components for faster overall flow. The combination of 
complete cleaning and fast recirculation has been shown at previous experiments (including 
the K2K one kiloton near detector) to be the most cost-effective way of achieving the desired 
water transparencies.  A preliminary cost estimate for this baseline water system from South Coast Water in the is \$350,000, including shipping, duties, and installation at the detector site.

\subsubsection{Gd option}

If it is decided to add 0.2\% gadolinium sulfate by mass to 
Super-Kamiokande in order to provide efficient tagging 
of neutrons in water, it will likely be useful for a near detector at Tokai to also be 
Gd-loaded such that the responses of both detectors are as similar as possible.  As a large 
water Cherenkov detector, \nuprism is a natural candidate for eventual Gd-loading.  Therefore, 
the implications this has on the water system design must be taken into account.

Over the past decade there have been focused R\&D programs both in the US and Japan aimed 
at devising a method capable of maintaining the exceptional water transparency discussed 
above, while at the same time maintaining the desired level of dissolved gadolinium in solution.  
In other words, somehow the water must be continuously recirculated and cleaned of 
everything {\em except} gadolinium sulfate.

\begin{figure}[htpb]
     \begin{center}
       \includegraphics[width=9cm]{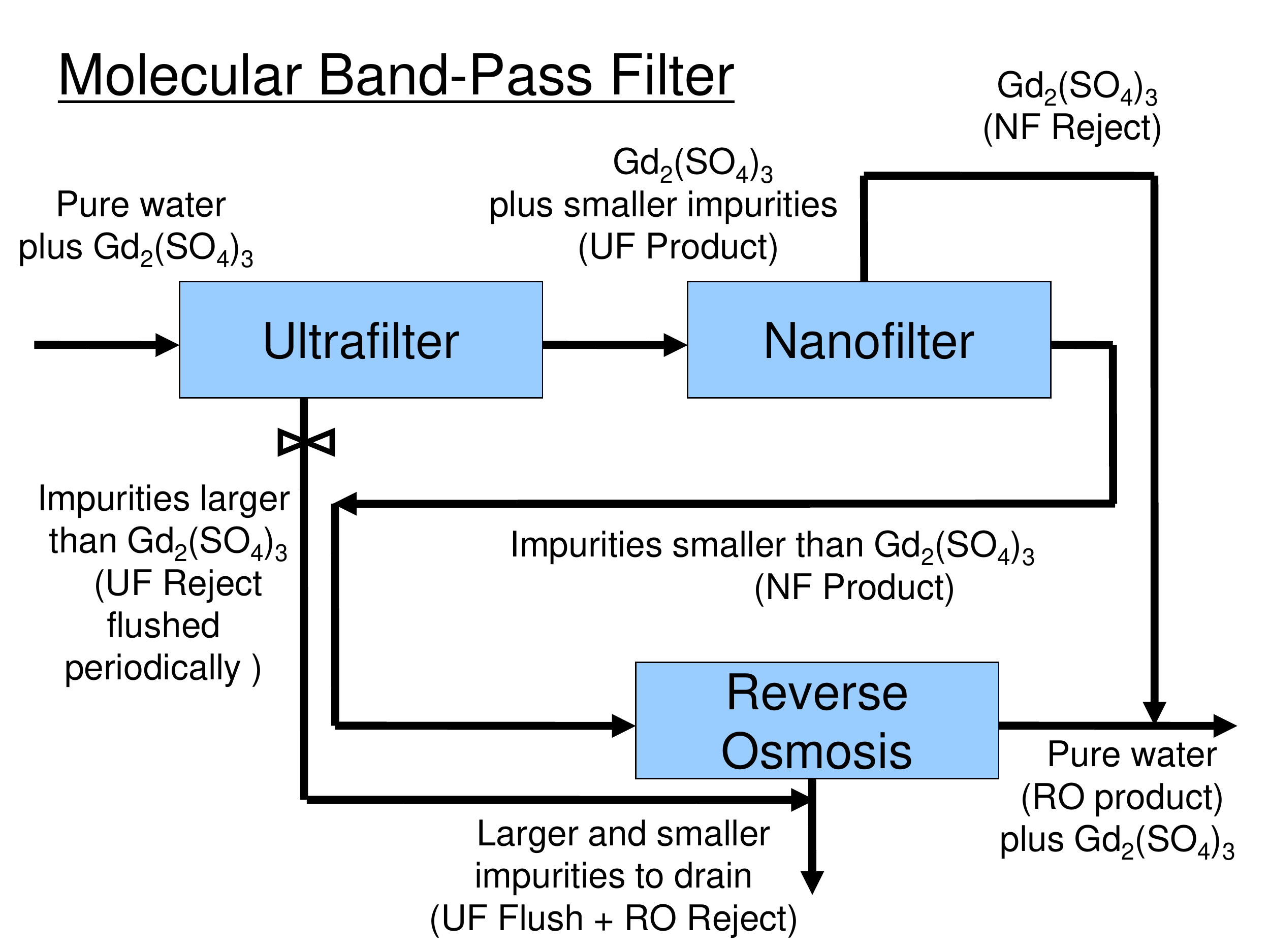}
       \caption{A schematic illustration of the principle of the "molecular band-pass filter".  
       Successively fine filter elements isolate the dissolved gadolinium sulfate ions and return
       them to the main tank, bypassing water system elements which would be fouled if they 
       were to trap gadolinium.}
       \label{water:bandpass}
     \end{center}
\end{figure}

Starting in 2007 with a 0.2~ton/hour prototype at the University of California, Irvine, since 2009 
the Kamioka-based EGADS (Evaluating Gadolinium's Action on Detector Systems) project 
has shown that such a selective water filtration technology -- known as a ''molecular band-pass
filter''  and schematically shown in  Figure~\ref{water:bandpass} -- is feasible at  3~tons/hour.  
As the EGADS design is modular and uses off-the-shelf and readily available equipment, albeit 
in novel ways, scaling it up from the current 3~tons/hour to 60~tons/hour for Super-Kamiokande, 
is straightforward, while scaling to \nuprism's 6.3 tons/hours would be trivial.

\clearpage

%\subsection{GPS \red{(Y. Hayato?)}}
%
%Section goals:
%\begin{itemize}
%\item Discussion of needed GPS system based on ND280 and recent GPS work
%\end{itemize}

\section{Detector Calibration}

%Section goals:
%\begin{itemize}
%\item Discussion of unique calibration challenges in \nuprismlite (Hiro's workshop talk) \red{(HT)}
%\begin{itemize}
%\item moving detector
%\item deployment of sources in the ID
%\end{itemize}
%\end{itemize}

The calibration systems for \nuprismlite will largely borrow from the existing Super-K calibration systems. However, \nuprismlite will also face some unique challenges:
\begin{itemize}
\item The PMT frame will move within the water volume.
\item Accessing the inner detector is more difficult when the position of the top of the detector is not fixed.
\end{itemize}
To address these issues, \nuprismlite will consist of calibration sources that are fixed within the ID (e.g. laser balls, LEDs, and scintillation cubes), as well as sources that can be lowered though remote-controlled access portals (e.g. radioactive sources). It is expected that each time the detector is moved, all of the PMTs will need to be recalibrated. This can be accomplished using the fixed light sources within the ID, and additional calibration runs with radioactive sources will be taken for each new detector position.

In addition to the detector response, it will also be necessary to precisely determine both the relative position of the PMTs within the ID, as well as the absolute position of the PMT frame within the water volume. This will be accomplished with a laser calibration system. An R\&D program is planned to demonstrate the effectiveness of such a system when operated in water.

As \nuprism will essentially reuse many of the established Super-K calibration techniques, the remainder of this section will provide a brief description of Super-K calibration systems. Further details can be found elsewhere~\cite{Fukuda:2002uc, SK_calib_paper}.

\subsection{Overview of Super-K Calibration Systems}

%Section goals:
%\begin{itemize}
%\item Description of relevant SK detector systems -- Hide's talk at the collaboration meeting  \red{(HKT)}
%\end{itemize}

This section overviews Super-K detector calibrations.
For further details, reader can also refer to \cite{Fukuda:2002uc, SK_calib_paper}.

The Super-K detector calibration can be divided into two steps;
the detector hardware calibrations and the calibrations for physics analyses.
The first step is common over all physics analyses, but the second step is
designed for each physics analysis goal.

\subsubsection{Detector hardware calibrations}

The detector hardware calibrations (measurements) consist of several parts:
\begin{itemize}
  \item Geometrical surveys: tank geometry, PMT positions
  \item Geomagnetic field
  \item PMT calibration: gain, photo-detection efficiency
  \item Readout channel (PMT and electronics) calibrations: linearity, timing, timing resolution
  \item Optical properties: water, PMT glass, black sheet, etc (for detector MC tuning)
  \item Water temperature
\end{itemize}
All of these calibrations and measurements are indispensable to understand the detector and
to model the detector in the simulation.
This section focuses on the PMT calibrations and readout channel calibrations, which
will be most relevant to \nuprismlite.

The PMT calibration procedure can be divided into three large steps;
1) pre-calibration, 2) post-installation calibration, 3) detector monitoring.
At the stage of `pre-calibration', a fraction of all Super-K PMTs have been
calibrated prior to the installation, e.g. a tuning of PMT gain.
The pre-calibrated PMT, called {\it standard PMTs}, were used to calibrate all
other PMTs {\it in-situ} after installed, at the stage of post-installation calibration.
Once all PMT are calibrated, the stability of the PMTs is monitored continuously
for the lifetime of the experiment.
The following sections discuss our ideas for each of the PMT
calibration steps.

\paragraph{Pre-calibration}

\begin{figure}[htb]
  \centering
  \includegraphics[width=9cm]{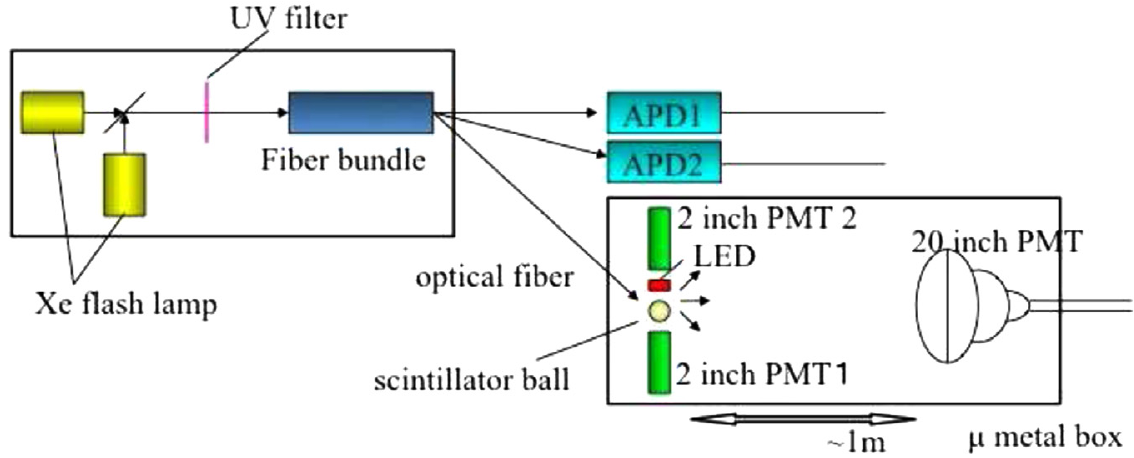}
  \caption{SK pre-calibration set-up. (Figure quoted from \cite{SK_calib_paper})}
  \label{fig:sk_precalib_setup}
\end{figure}

SK has 420 standard PMTs, which corresponds to about 4\% of all SK PMTs.
The SK standard PMTs were calibrated prior to the installation by adjusting
HV values to have identical charge ($\sim30$ p.e.) over the standard PMTs.
For the pre-calibration, SK employed a xenon lamp and scintillator ball.
Figure~\ref{fig:sk_precalib_setup} shows a schematic diagram of the pre-calibration
set-up.

\begin{figure}[htb]
  \centering
  \includegraphics[width=9cm]{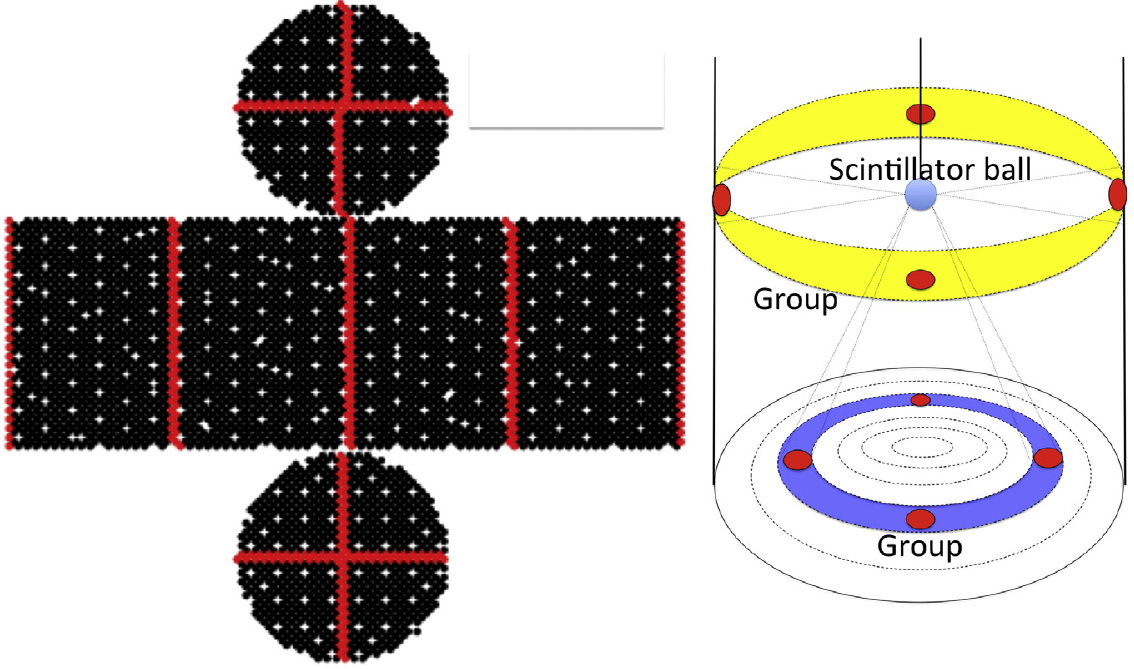}
  \caption{Layout of SK standard PMTs. (Figure quoted from \cite{SK_calib_paper})}
  \label{fig:sk_precalib_PMT_layout}
\end{figure}

The SK standard PMTs were installed in the tank in a geometrically symmetric configuration.
Figure~\ref{fig:sk_precalib_PMT_layout} shows the location of the standard PMTs
in SK inner detector.
%In SK pre-calibration, HV values of the standard PMTs were determined to obtain
%the same pulse height (30 p.e.) between them.

\paragraph{Post-installation calibrations}

In the post-installation calibration, all PMTs other than the standard PMTs
were calibrated {\it in-situ} after installed.
At this stage, all PMT parameters were determined and measured.
We will discuss the following items in this section,
\begin{itemize}
% --------------------------
\item HV (gain) tuning\\
  Tune HV for all PMTs, referencing to the standard PMTs by using the Xe lamp and deploying
  a scintillator ball in the tank (the same light source used in the pre-calibration).
  Move the scintillator ball along Z-axis (height direction), and tune HV group-by-group,
  where the group is defined by Fig.~\ref{fig:sk_precalib_PMT_layout}.

% --------------------------
\item Charge to photo-electron conversion\\
  Conversion factor of charge (pC) to photo-electron (p.e.) were obtained by measuring 1~p.e.
  distribution.
  SK deployed ``nickel source'' in the tank, that generate 1~p.e. level of light, where the
  nickel source is nickel-californium source; Ni(n,$\gamma$)Ni, E$_{\gamma}\sim$9~MeV.
  %Figure~\ref{fig:sk_1pe_dist} shows 1~p.e. distribution of SK.
  %\begin{figure}[htb]
  %  \centering
  %  \includegraphics[width=0.3\textwidth]{sk_1pe_dist.pdf}
  %  \caption{SK 1~p.e. distribution. Note that this distribution is accumulated over all PMTs.
  %    (Figure quoted from \cite{SK_calib_paper})}
  %  \label{fig:sk_1pe_dist}
  %\end{figure}
  Figure~\ref{fig:sk_nickel_source} shows the SK nickel source.
  \begin{figure}[htb]
    \centering
    \includegraphics[width=9cm]{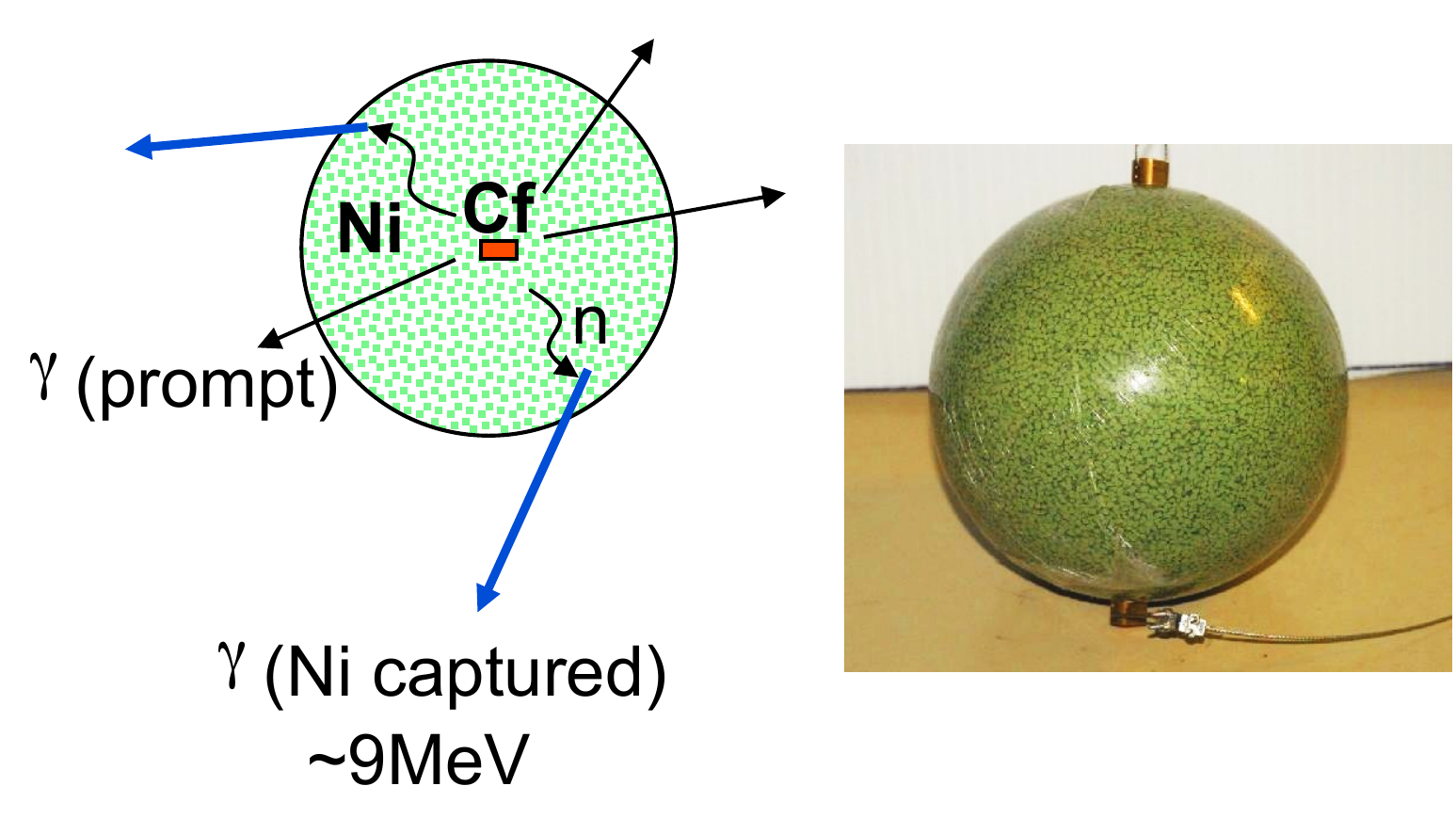}
    \caption{SK ``Nickel source'' (Figure quoted from \cite{SK_calib_paper})}
    \label{fig:sk_nickel_source}
  \end{figure}

% --------------------------
\item Photo-detection efficiency\\
  The photo-detection efficiency, $\epsilon$, is defined by Quantum Efficiency times Collection
  Efficiency (CE).
  Hit rate (Nhit) for 1 p.e. level of light is proportional to the photo-detection efficiency;
  $N_{hit} \propto N_{photon} \cdot \epsilon$.
  For this measurement, SK used the Nickel source to evaluate the hit rate, and compare with MC
  to evaluate {\it relative} efficiency over all PMTs.
  
% --------------------------
\item Timing calibration\\
  Calibration for time response of readout channel (PMT and electronics), e.g. {\it time-walk}
  effect.
  SK employed N$_2$-dye laser and deployed diffuser-ball in tank, that light source can generate
  0.1$\sim$1000~p.e. level light and covers the entire dynamic range of electronics.
  Evaluate TQ-maps for every single PMTs, and evaluate detector timing resolution (for MC input).

\end{itemize}

\subsubsection{Calibrations for physics analyses}

The calibrations for physics analyses need to be designed for physics goal basis.
This section describes the calibrations used for SK atmospheric neutrino and T2K
analyses, that relevant to \nuprismlite physics goals.

\paragraph{Photon yield and charge scale}

\begin{figure}[htb]
  \centering
  \includegraphics[height=6cm]{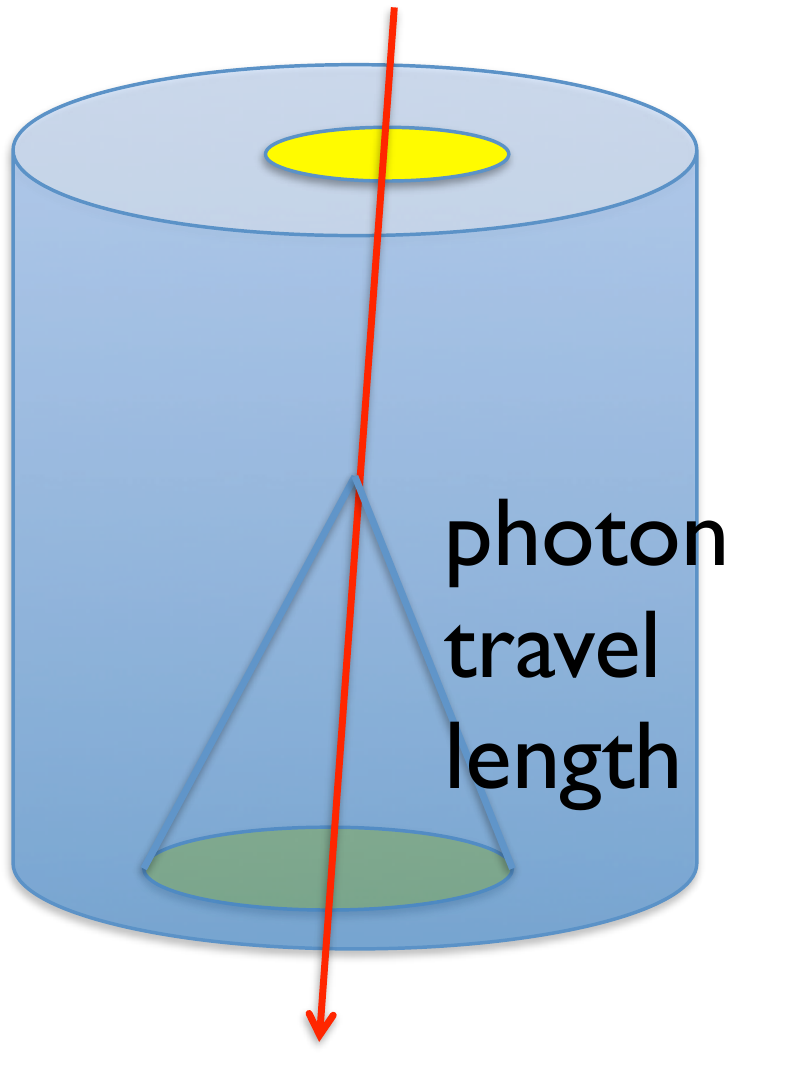}
  \caption{Schematic of SK vertical through-going muon events.}
  \label{fig:sk_thru_muon}
\end{figure}

Although several detailed detector calibrations have been carried out, there are
uncertainties on the photon propagation and photon detection of the detector,
that need to be tuned in the detector simulation using a well known control samples.
For that, SK uses cosmic-ray muons, called ``vertical through-going muons''.
Figure~\ref{fig:sk_thru_muon} shows a schematic of vertical through-going muon event
of SK.
The absolute photon yield and charge scale in the detector simulation have been tuned
to data using the vertical through-going muon events that provide known muon track length
and Cherenkov photon travel distance.

\paragraph{Momentum and energy scale}

SK event reconstruction algorithm uses a conversion table that translate the observed total
charge in the Cherenkov ring to the particle (muons and electrons) momentum.
The conversion table is called ``momentum table'' have been evaluated using the detector
simulation by generating particles in momentum range of 10's MeV/c to GeV/c.
Based on all detector calibrations and the simulation tuning, the detector and simulation
are ready to use for physics analyses.
Absolute energy scale is checked using natural sources; decay electron, $\pi^{0}$ mass,
sub-GeV stopping muons, and multi-GeV stopping muons, these sources cover the energy range of
10's MeV to 10~GeV.
SK detector simulation reproduces data within $\sim2$\% and that have been continuously
monitored.
SK defines the energy scale uncertainty as the data-MC difference.
If the simulation does not reproduce the data reasonably well, the detector calibrations
and simulation tuning need to be revised.

%\subsection{Additional Calibration Systems \red{(R. Tacik \& K. Mahn)}}
%
%Section goals:
%\begin{itemize}
%\item Discussion of possible additional calibration systems to be used in \nuprismlite \red{(RT)}
%\item MiniBooNE cubes? \red{(RT)}
%\item LEDs? \red{(RT or Asher?)}
%\item ...
%\item PMT position calibration \red{(KM)}
%\end{itemize}

\section{Conclusion}

The proposed \nuprismlite detector has the potential to address the remaining systematic uncertainties that are not well constrained by ND280. In particular, this detector can constrain the relationship between measured lepton kinematics and incident neutrino energy without relying solely on rapidly-evolving neutrino interaction models. Since \nuprismlite is a water Cherenkov detector, the neutral current backgrounds with large systematic uncertainties at Super-K, particularly NC$\pi^+$ and NC$\pi^0$, can be measured directly with a nearly identical neutrino energy spectrum. The ability to produce nearly monoenergetic neutrino beams also provides the first ever ability to measure neutral current cross sections as a function of neutrino energy. Finally, \nuprismlite provides a mechanism to separate the many single-ring e-like event types to simultaneously constrain $\nu_e$ cross sections, neutral current background, and sterile neutrino oscillations.

The main long-baseline oscillation analysis presented in this note was a $\nu_\mu$ disappearance measurement, since the effects of various cross section models on this measurement had already been well studied, which provided a useful basis for comparison. However, it is also expected that \nuprismlite will provide a significant improvement to the ultimate T2K constraint on $\delta_{CP}$ by constraining neutral current backgrounds and electron-neutrino cross sections. Initial studies have also been presented that demonstrate the impact \nuprismlite can have on both $\nu_e$ appearance measurements and anti-neutrino oscillation measurements. Other planned improvements to the analysis include a realistic detector simulation and event reconstruction. Thanks to the work done on event simulation and reconstruction in Hyper-K, these tools already exist and can be quickly incorporated into the current analysis to perform more detailed studies of event pileup and detector performance for various detector configurations and PMT sizes and coverage.

%\bibliographystyle{unsrt}
%\bibliography{nuprismLOI}% Produces the bibliography via BibTeX.

%\newpage
%
%\appendix
%
%\input{detectorcost.tex}

\end{document}